\documentclass[fleqn,usenatbib]{mnras}

\usepackage{fix-cm}
\usepackage{newtxtext,newtxmath}

\usepackage[T1]{fontenc}

\DeclareRobustCommand{\VAN}[3]{#2}
\let\VANthebibliography\thebibliography
\def\thebibliography{\DeclareRobustCommand{\VAN}[3]{##3}\VANthebibliography}

\makeatletter 
  \patchcmd{\NAT@citex}
    {\@citea\NAT@hyper@{%
      \NAT@nmfmt{\NAT@nm}%
      \hyper@natlinkbreak{\NAT@aysep\NAT@spacechar}{\@citeb\@extra@b@citeb}%
      \NAT@date}}
    {\@citea\NAT@nmfmt{\NAT@nm}%
    \NAT@aysep\NAT@spacechar\NAT@hyper@{\NAT@date}}{}{}

  \patchcmd{\NAT@citex}
    {\@citea\NAT@hyper@{%
      \NAT@nmfmt{\NAT@nm}%
      \hyper@natlinkbreak{\NAT@spacechar\NAT@@open\if*#1*\else#1\NAT@spacechar\fi}%
        {\@citeb\@extra@b@citeb}%
      \NAT@date}}
    {\@citea\NAT@nmfmt{\NAT@nm}%
    \NAT@spacechar\NAT@@open\if*#1*\else#1\NAT@spacechar\fi\NAT@hyper@{\NAT@date}}
    {}{}
\makeatother

\usepackage{graphicx}	
\usepackage{amsmath}	
\usepackage{orcidlink}

\title[Predicting ionizing photon escape]{From \textsc{thesan-zoom} to JWST: Predicting ionizing photon escape and the rise of UV-bright reionization sources}

\author[Summerfield, McClymont, Tacchella et al.]{%
Zebedee Summerfield $\orcidlink{0009-0009-9994-4557}$,$^{1,2}$\thanks{E-mail: \href{mailto:zns22@cantab.ac.uk}{zns22@cantab.ac.uk} (ZS)}
William McClymont $\orcidlink{0009-0009-5565-3790}$,$^{1,2}$\thanks{E-mail: \href{mailto:wjm50@cam.ac.uk}{wjm50@cam.ac.uk} (WM)}
Sandro Tacchella $\orcidlink{0000-0002-8224-4505}$,$^{1,2}$\thanks{E-mail: \href{mailto:st578@cam.ac.uk}{st578@cam.ac.uk} (ST)} 
Aaron Smith $\orcidlink{0000-0002-2838-9033}$,$^{3}$
\newauthor
Rahul Kannan $\orcidlink{0000-0001-6092-2187}$,$^{4}$
Enrico Garaldi $\orcidlink{0000-0002-6021-7020}$,$^{5}$
Ewald Puchwein $\orcidlink{0000-0001-8778-7587}$,$^{6}$
Xuejian Shen $\orcidlink{0000-0002-6196-823X}$,$^{7}$
Josh Borrow  $\orcidlink{0000-0002-1327-1921}$,$^{8}$
\newauthor
A. Lola Danhaive$\orcidlink{0000-0002-9708-9958}$,$^{1,2}$
Laura Keating $\orcidlink{0000-0001-5211-1958}$,$^{9}$
Gabriel Maheson $\orcidlink{0009-0005-8978-8598}$,$^{1,2,10,11}$
Parth Mehrotra $\orcidlink{0009-0001-4403-892X}$,$^{12}$ 
\newauthor
Charlotte Simmonds $\orcidlink{0000-0003-4770-7516}$,$^{13}$
Amanda Stoffers,$^{1,2}$
Mark Vogelsberger $\orcidlink{0000-0001-8593-7692}$,$^{7}$
and Oliver Zier $\orcidlink{0000-0003-1811-8915}$$^{14,7}$
\\
\\
$^{1}$ Kavli Institute for Cosmology, University of Cambridge, Madingley Road, Cambridge CB3 0HA, UK\\
$^{2}$ Cavendish Laboratory, University of Cambridge, 19 JJ Thomson Avenue, Cambridge CB3 0HE, UK\\
$^3$ Department of Physics, The University of Texas at Dallas, Richardson, TX 75080, USA \\
$^4$ Department of Physics and Astronomy, York University, 4700 Keele Street, Toronto, ON M3J 1P3, Canada \\
$^5$ Kavli IPMU (WPI), UTIAS, The University of Tokyo, Kashiwa, Chiba 277-8583, Japan \\
$^6$ Leibniz-Institut f\"ur Astrophysik Potsdam, An der Sternwarte 16, 14482 Potsdam, Germany \\
$^{7}$ Department of Physics, Kavli Institute for Astrophysics and Space Research, Massachusetts Institute of Technology, Cambridge, MA 02139, USA \\
$^{8}$ Department of Physics and Astronomy, University of Pennsylvania, 209 South 33rd Street, Philadelphia, PA 19104, USA \\
$^{9}$ Institute for Astronomy, University of Edinburgh, Blackford Hill, Edinburgh, EH9 3HJ, UK \\
$^{10}$Dipartimento di Fisica e Astronomia, Università di Bologna, Via Gobetti 93/2, I-40129, Bologna, Italy\\
$^{11}$INAF, Astrophysics and Space Science Observatory Bologna, Via P. Gobetti 93/3, I-40129 Bologna, Italy\\
$^{12}$Department of Physics, Indian Institute of Science Education and Research, Dr. Homi Bhabha Road, Pune, 411008, Maharashtra, India\\
$^{13}$Departamento de Astronomía, Universidad de Chile, Camino El Observatorio 1515, Las Condes, Santiago, Chile\\
$^{14}$ Center for Astrophysics $|$ Harvard $\&$ Smithsonian, 60 Garden Street, Cambridge, MA 02138, USA\\
}

\date{Accepted XXX. Received YYY; in original form ZZZ}

\pubyear{\the\year{}}

\begin{document}
\label{firstpage}
\pagerange{\pageref{firstpage}--\pageref{lastpage}}
\maketitle

\begin{abstract}
Understanding the sources and evolution of cosmic reionization remains a central challenge in astrophysics, with the escape of ionizing Lyman-continuum (LyC) photons from early galaxies representing a major uncertainty. In this work, we use more than 35,000 galaxy realisations from the \textsc{thesan-zoom} cosmological radiation-hydrodynamic simulations to identify indirect diagnostics of the LyC photon escape fraction ($f_\mathrm{esc}$) and the LyC photon escape rate ($\dot{N}_\mathrm{ion,esc}$) across the redshift range $z=3-16$. We train random forest regression models using these diagnostics to predict both quantities. We present four models: two trained with the full set of simulation-derived indicators to predict $f_\mathrm{esc}$ and $\dot{N}_\mathrm{ion,esc}$, and two restricted to observables accessible to  JWST photometric surveys. We find the 10-to-100$\,$Myr star-formation rate ratio ($\mathrm{SFR}_{10} / \mathrm{SFR}_{100}$) and the gas-to-stellar mass ratio ($M_\mathrm{gas} / M_*$) to be the strongest diagnostics of $f_\mathrm{esc}$, suggesting a strong relationship between ionizing photon escape and gas clearing through bursty star formation. In contrast, rest-frame UV ($1500 \, \textup{\AA}$) absolute magnitude ($M_\mathrm{UV}$) dominates $\dot{N}_\mathrm{ion,esc}$ prediction. Motivated by the strong predictive power of $M_\mathrm{UV}$, we combine observed UV luminosity functions with derived $\dot{N}_\mathrm{ion,esc} - M_\mathrm{UV}$ relations to construct histories of reionization. These are consistent with observational constraints, avoiding the recently reported crisis in the ionizing photon budget. Our analysis suggests that the bulk of reionization occurred rapidly after $z \approx 8$, driven by UV-bright galaxies, with the $M_\mathrm{UV} < -17$ populations providing the dominant contribution.
\end{abstract}




\section{Introduction} \label{sec:introduction}

The Epoch of Reionization (EoR) represents a major phase transition of the Universe, where the neutral hydrogen of the intergalactic medium (IGM) was ionized by the Lyman-continuum radiation (LyC; with energies in excess of $13.6 \,$eV) emitted by the first stars and galaxies. While observational evidence indicates that reionization reached completion in the redshift range of  $5 < z < 6$ \citep{becker_mean_2021, bosman_hydrogen_2022, davies_constraints_2024, zhu_damping_2024}, the onset and evolution of the transition remain uncertain. 

Yet, in recent years, a wide variety of observational probes have enabled increasingly tight constraints.  The early stages of reionization are indirectly constrained by Cosmic Microwave Background (CMB) measurements \citep{heinrich_reionization_2021},  while recent surveys from the James Webb Space Telescope (JWST) have identified luminous galaxies at $z \geq 13$ exhibiting Lyman alpha (Ly$\alpha$) emission  \citep{carniani_spectroscopic_2024, witstok_witnessing_2025}, seemingly implying patchy reionization had already started at this early epoch. Additionally, the evolution of the hydrogen neutral fraction with cosmic time has been inferred from the decreasing abundance of Ly$\alpha$ emitters at higher redshift \citep{tang_jwstnirspec_2024, kageura_census_2025}, measurements of the Ly$\alpha$ and Ly$\beta$ forests \citep{jin_nearly_2023, zhu_damping_2024}, and damping-wing analyses of high-redshift quasar spectra \citep{davies_quantitative_2018, wang_significantly_2020}. However, there is still much debate regarding the nature of the sources responsible for reionization \citep{robertson_galaxy_2022}.

It is widely agreed that the ionizing photons produced by young, massive stars in star-forming galaxies play an important role in reionization \citep{madau_cosmic_2017, rosdahl_sphinx_2018, yeh_thesan_2023, choustikov_inferring_2024}. However, it remains unclear whether active galactic nuclei (AGN) make a significant contribution. JWST has revealed an emergent population of broad-line AGNs at $4 \lesssim z \lesssim 13$ \citep{maiolino_jades_2024, fujimoto_uncover_2024, madau_cosmic_2024}, and it is known that galaxies harbouring AGNs can leak a large fraction of their produced LyC radiation \citep{cristiani_spectral_2016, iwata_ionizing_2022, madau_cosmic_2024}. Although other studies argue that AGN must have a subdominant impact on reionization \citep{parsa_no_2018, puchwein_consistent_2019, trebitsch_modelling_2020, shen_bolometric_2020}, as their contribution is limited by number densities that only become relevant at $z \lesssim 7$ \citep{dayal_uncovering_2025}, and by constraints on the thermal history of the IGM \citep{upton_sanderbeck_models_2016, puchwein_consistent_2019}. The community also remains divided over whether reionization is driven primarily by numerous faint galaxies or by the most luminous populations \citep{kakiichi_role_2018, naidu_rapid_2020, dayal_reionization_2020, larson_searching_2022, atek_most_2024, mascia_new_2024, simmonds_ionizing_2024}.

The evolution of the EoR is characterised by the cosmic ionizing emissivity, $\dot{n}_{\mathrm{ion}}$, defined as the ionizing photons emitted per unit time per unit comoving volume. This quantity can be expressed as a function of redshift in terms of the mean galactic LyC photon escape rate, $\dot{N}_\mathrm{ion,esc}$, using:
\begin{equation} \label{integral_magnitude}
\dot{n}_{\mathrm{ion}}(z) = \int_{M_\mathrm{UV,min}}^{M_\mathrm{UV,max}} \, \Phi(M_\mathrm{UV}, z) \, \left\langle \dot{N}_\mathrm{ion,esc} \,\middle|\, M_\mathrm{UV}, z \right\rangle \, \mathrm{d}M_\mathrm{UV} \, , 
\end{equation}
where $\Phi$ is the ultra-violet (UV) luminosity function, defined as the number density of galaxies as a function of redshift and rest-frame UV absolute magnitude. The integration limits $M_\mathrm{UV,min}$ and $M_\mathrm{UV,max}$ correspond to the faint- and bright-end magnitude limits of the galaxy population contributing to reionization, and are chosen on a case-by-case basis depending on observational constraints and modelling assumptions. Often, at the galactic level, $\dot{N}_\mathrm{ion,esc}$ is decomposed into:
\begin{equation} \label{n_ion}
    \dot{N}_\mathrm{ion,esc} = \dot{N}_\mathrm{ion} \, f_\mathrm{esc}
    = L_{\mathrm{UV}} \, \xi_{\mathrm{ion}} \, f_\mathrm{esc}
\end{equation}
for a galaxy of monochromatic UV luminosity $L_\mathrm{UV}$, where $f_\mathrm{esc}$ is the fraction of LyC photons that escape to the IGM and  $\xi_{\mathrm{ion}}$ is the ionizing photon production efficiency. The latter is defined as the ratio of the ionizing photon production rate, $\dot{N}_\mathrm{ion}$, to the monochromatic UV luminosity, $L_{\mathrm{UV}}$: $\xi_{\mathrm{ion}} = \dot{N}_\mathrm{ion} \, / \, L_{\mathrm{UV}}$. In the absence of a direct estimate for $\dot{N}_\mathrm{ion,esc}$, which is unobservable during the EoR, all three quantities $L_{\mathrm{UV}}$ (or a UV luminosity density function $\rho_{\mathrm{UV}}$), $\xi_{\mathrm{ion}}$, and $f_\mathrm{esc}$ must be specified to study $\dot{n}_{\mathrm{ion}}$. Among these parameters, the escape fraction has historically been the most challenging to constrain, as it arises from complex, highly non-linear processes within galaxies, such as those governing the opacity of the interstellar medium (ISM) \citep{jaskot_ionizing_2025}. More importantly, the neutral IGM is opaque to ionizing radiation and thus direct measurements are not feasible at high redshifts, with no clear lines of sight above $z \sim 4$ \citep{inoue_monte_2008}. Consequently, characterising $f_\mathrm{esc}$ remains a focus of active observational \citep{chisholm_far-ultraviolet_2022, mascia_new_2024, jaskot_multivariate_2024, mascia_little_2025, stoffers_challenge_2025} and simulation-based research \citep{yoo_origin_2020, rosdahl_lyc_2022, katz_nature_2022, yeh_thesan_2023, kostyuk_ionizing_2023, choustikov_physics_2024, choustikov_great_2024}.

Attempts to investigate LyC emitters observationally have been made through low-redshift analogues \citep{saxena_strong_2022, flury_low-redshift_2022, stoffers_challenge_2025}. However, it is unclear how representative these are of higher redshift reionization sources \citep{price_reconstructing_2016, katz_first_2023}. Furthermore, recent JWST-based studies infer an enhanced abundance of UV-bright galaxies in the early Universe \citep[$z \gtrsim 9$;][]{robertson_identification_2023, robertson_earliest_2024, carniani_spectroscopic_2024, donnan_jwst_2024, tacchella_star_2024, whitler_z_2025}; when combined with escape fractions inferred from low-redshift analogues \citep{chisholm_far-ultraviolet_2022}, these results could imply an overproduction of ionizing photons during the EoR, leading to a ``photon budget crisis'' \citep{munoz_reionization_2024, simmonds_ionizing_2024}.

Thus, due to the inability of observations to directly investigate ionizing photon emission during the EoR, theoretical models, and in particular numerical simulations, have become increasingly important. Fortunately, the fidelity of cosmological simulations has advanced significantly in recent years, driven by rapid developments in computational power and numerical methods \citep{vogelsberger_cosmological_2020}. The study of reionization in a cosmological context requires the use of intermediate- or large-volume simulations that emulate regions spanning tens to hundreds of megaparsecs \citep{gnedin_cosmic_2014, rosdahl_sphinx_2018, ocvirk_cosmic_2020, kannan_introducing_2022, puchwein_sherwood-relics_2023}. While these capture the large-scale statistical process of reionization, the combination of limited resolution and simplified ISM and feedback physics hampers a detailed treatment of radiative transfer through the ISM, hindering robust predictions of $f_\mathrm{esc}$. Some studies instead employ idealised galactic simulations \citep{jensen_machine-learning_2016, yoo_origin_2020}; however, their representativeness of LyC sources is ambiguous due to the absence of realistic feedback processes and a cosmological environment. Many recent works use a zoom-in simulation technique, where the resolution of a targeted halo is enhanced within a broader cosmological environment, enabling realistic modelling of radiation propagation \citep{ma_simulating_2018, pallottini_deep_2019, rosdahl_lyc_2022, bhagwat_spice_2024, choustikov_great_2024, kannan_introducing_2025}. Indirect measurements of $f_\mathrm{esc}$ can therefore be obtained using models calibrated with these zoom-in mock galaxies \citep[e.g.,][]{choustikov_physics_2024}.

In this work, we develop predictive models, using the random forest (RF) regression machine learning technique for both the widely studied galactic LyC photon escape fraction, $f_\mathrm{esc}$, and the galactic LyC photon escape rate, $\dot{N}_\mathrm{ion,esc}$, which provides a more direct probe of a source's contribution to reionization. To do so, we employ the recently developed \textsc{thesan-zoom} simulations \citep{kannan_introducing_2025}, a suite of zoom-in radiation-hydrodynamic simulations that resolve the multi-phase ISM with high spatial and mass resolution at $z > 3$. We apply our models to a sample of over 40,000 high-redshift JWST/NIRCam-detected JADES galaxies, deriving relations for $\dot{N}_\mathrm{ion,esc}$ dependent on redshift and rest-frame UV ($1500 \, \textup{\AA}$) absolute magnitude, $M_\mathrm{UV}$, for both the JWST sample and \textsc{thesan-zoom} itself. Using these relations in Eq.~\eqref{integral_magnitude}, we infer implications for the EoR, finding a history of reionization in which the majority of the neutral hydrogen in the IGM is ionized rapidly after $z \sim 8$, driven predominantly by LyC emission from relatively UV-bright galaxies.

This paper is structured as follows. In Section~\ref{sec:simulation}, we introduce the \textsc{thesan-zoom} simulations. We explore how potential diagnostics are identified and how they are used to train RF predictive models in Section~\ref{sec:prediction}, where we also present and discuss our resulting best-performing and observationally applicable models. In Section~\ref{sec:implications}, we explore the implications for reionization, calculating histories of $\dot{n}_{\mathrm{ion}}$, the hydrogen ionized fraction, $Q_\mathrm{HII}$, and the Thomson optical depth, $\tau$, before presenting our conclusions in Section \ref{sec:conculsions}.

\section{Simulation Methodology} \label{sec:simulation}

\subsection{The \textsc{thesan-zoom} project} \label{sec:thesan-zoom}

We employ the \textsc{thesan-zoom} simulations \citet{kannan_introducing_2025}, a state-of-the-art set of high-resolution zoom-in simulations that extends the original \textsc{thesan} project \citep{kannan_introducing_2022, smith_thesan_2022, garaldi_thesan_2022, garaldi_thesan_2024}.

The original \textsc{Thesan} project comprises a suite of large-volume ($\mathrm{L_{box}} = 95.5\,$cMpc) radiation-magneto-hydrodynamic simulations aimed at modelling the EoR and the early Universe, with galaxy formation following the  IllustrisTNG model \citep{springel_first_2018, marinacci_first_2018, nelson_illustristng_2021}. The \textsc{thesan} simulations follow the prevailing $\Lambda$CDM cosmological framework and assume the cosmological parameters from the Planck Collaboration \citep{ade_planck_2016}: $H_0 = 100 \, h \, \mathrm{km\,s}^{-1}\mathrm{Mpc}^{-1}$ with $h=0.6774$, $\Omega_\text{m} = 0.3089$, $\Omega_\Lambda = 0.6911$, $\Omega_\text{b} = 0.0486$, $\sigma_8 = 0.8159$, and $n_\text{s} = 0.9667$, where all symbols have the usual meaning. The simulations were initialised at $z_{\mathrm{in}} = 49$ \citep{kannan_introducing_2022} with homogeneous expanding backgrounds and imposed Gaussian perturbations to instigate the formation of structure. \textsc{thesan} was designed to capture the large-scale properties of the IGM and the galaxies driving reionization \citep[e.g.,][]{yeh_thesan_2023, borrow_thesan-hr_2023, shen_thesan-hr_2024, neyer_thesan_2024, zhao_thesan_2025, jamieson_thesan_2025, almualla_thesan_2025}, however, the effective equation-of-state treatment of the ISM limits \textsc{thesan}'s ability to model galactic processes, motivating the need for more advanced ISM and stellar feedback models to resolve escape at galactic scales.

Full details of \textsc{thesan-zoom} are presented in \citet{kannan_introducing_2025}, although we provide a brief overview here. \textsc{thesan-zoom} uses a zoom-in technique to enhance the resolution and physics of 14 regions from the original \textsc{thesan} simulations, enabling detailed studies probing galaxy-scale phenomena and properties during the EoR \citep{mcclymont_thesan-zoom_2025-1, mcclymont_thesan-zoom_2025, mcclymont_overmassive_2025, mcclymont_thesan-zoom_2025-2,  wang_thesan-zoom_2025, zier_thesan-zoom_2025, zier_thesan-zoom_2025-1, pruto_thesan-zoom_2026, shen_thesan-zoom_2026}. The suite includes three distinct resolution tiers achieving enhancements in spatial resolution by factors of 4, 8, and 16 and corresponding increases in mass resolution by factors of 64, 512, and 4096 compared to the original \textsc{thesan} runs. The corresponding baryonic mass resolutions are $9.09 \times 10^3 \, \mathrm{M_\odot}$, $1.14 \times 10^3 \, \mathrm{M_\odot}$ and $1.42 \times 10^2 \, \mathrm{M_\odot}$, respectively \citep{kannan_introducing_2025}. \textsc{thesan-zoom} inherits its initial conditions and cosmological parameters from \textsc{thesan}, ensuring large-scale structures and cosmological properties are preserved. 
 
The propagation of radiation is tracked across the boundaries of the zoom-in regions, using the time-varying radiation field from the parent volume as a boundary condition, to ensure consistency within the context of the large-scale environment \citep{kannan_introducing_2025}. The radiation-hydrodynamics solver \textsc{arepo-rt} \citep{kannan_arepo-rt_2019}, built on the \textsc{arepo} moving mesh code \citep{springel_e_2010}, and employing a novel node-to-node communication strategy \citep{zier_adapting_2024}, is used to self-consistently model LyC photon–gas interactions and radiative transfer ``on-the-fly''. It adopts a moment-based scheme with a reduced speed-of-light approximation \citep{kannan_arepo-rt_2019}, accounting for both internal (zoom-in) and external radiation fields. This approach is crucial for accurately capturing the coupling between ionizing radiation and the ISM. A dust model is incorporated which tracks dust production, growth and destruction across five species \citep[][Garaldi et al. in prep.]{mckinnon_simulating_2017,kannan_simulating_2020,kannan_introducing_2025}. The stellar feedback combines multiple channels, including photoionization, radiation pressure, stellar winds, supernovae, and an additional early feedback channel to account for unresolved and unaccounted for physics \citep{kannan_simulating_2020, kannan_introducing_2025}. Stellar winds and star formation are implemented using the \textsc{SMUGGLE} multi-phase ISM model \citep{marinacci_simulating_2019}. These feedback processes regulate star formation and drive outflows, shaping early galaxy properties and facilitating radiation escape. 

In this work, we limit our focus to simulations at the lowest resolution tier, with spatial resolution enhanced by a factor of four, ensuring consistent physical modelling across all galaxies and allowing us to include simulations which contain relatively massive galaxies at high redshift. We consider only subhaloes resolved with at least 100 stellar particles, which corresponds to a stellar mass cut-off of $9.09 \times 10^5 \, \mathrm{M_\odot}$. Haloes were identified using the friends-of-friends (FOF) algorithm \citep{davis_evolution_1985}, with self-gravitating haloes being identified within the FOF groups using the SUBFIND-HBT algorithm \citep{springel_populating_2001, springel_simulating_2021}. Due to the computational expense of our analysis, we only consider the largest zoom region in the suite that was run down to $z=3$, which is region m12.6 in \citet{kannan_introducing_2025}. We include both central and satellite galaxies across $3 < z < 16$ in this analysis, which leaves us with a sample of 43\,076 subhaloes. Stellar masses, star-formation rate, and luminosities are calculated based on particles within the virial radius. Gas masses are calculated within the galaxy (defined as within twice the UV half-light radius) in order to trace the gas within the ISM.

 \subsection{Post-processing of \textsc{thesan-zoom}} \label{sec:catalogue}

 \begin{figure*}
	\includegraphics[width=\textwidth]{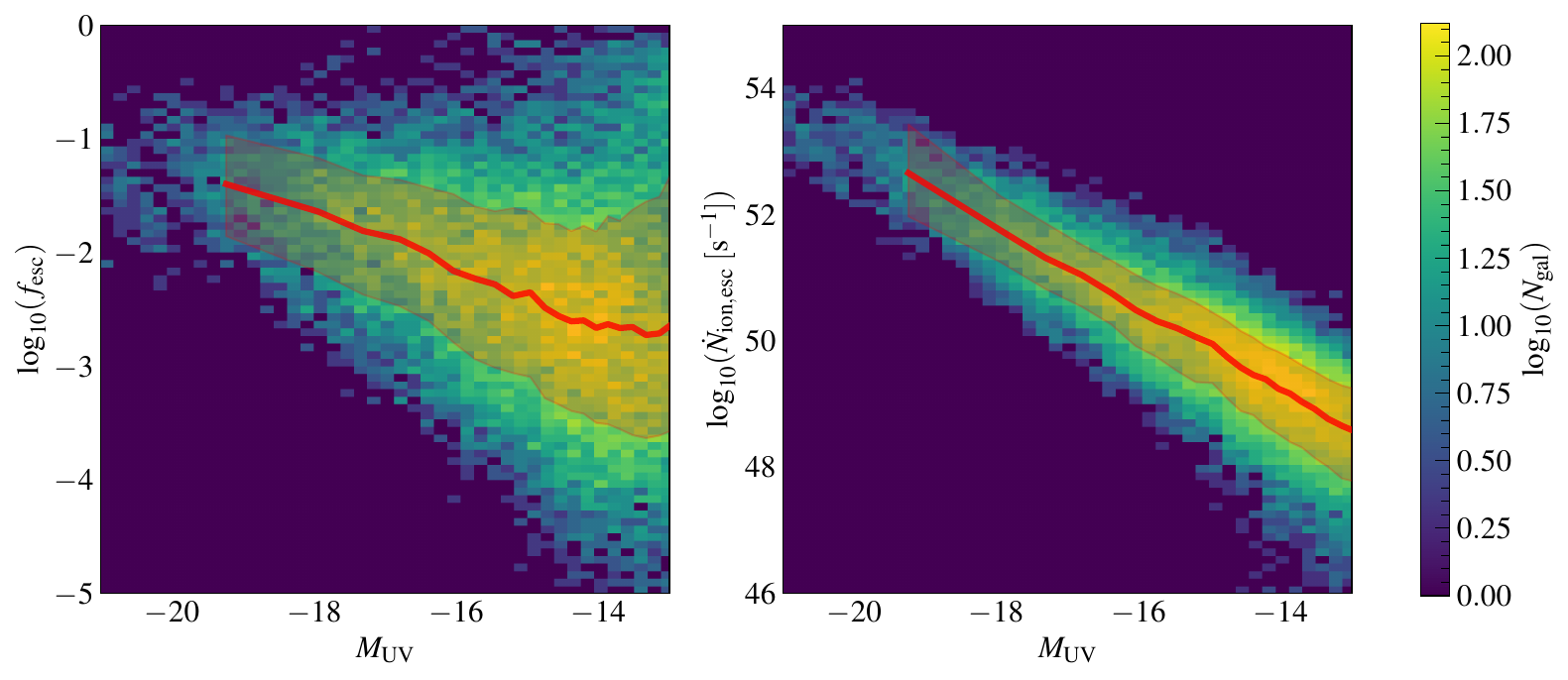}
    \caption{Ionizing photon escape fraction ($f_\mathrm{esc}$; \textit{left panel}) and ionizing photon escape rate ($\dot{N}_\mathrm{ion,esc}$; \textit{right panel}) as a function of rest-frame UV absolute magnitude ($M_\mathrm{UV}$) for the 35,512 galaxies selected from the \textsc{thesan-zoom} simulations across $3 < z < 16$. The colour scale indicates the number of galaxies per bin. The red lines trace the median relations, with the red shaded regions denoting the $16^\mathrm{th} - 84^\mathrm{th}$ percentile range. The mean half-width of these intervals corresponds to $1\sigma$ scatters of 0.829 dex and 0.628 dex, respectively. Brighter galaxies exhibit both higher escape fractions and higher escape rates. However, the $\dot{N}_\mathrm{ion,esc} - M_\mathrm{UV}$ relation displays reduced scatter relative to that of $f_\mathrm{esc}$. Therefore, we expect $M_\mathrm{UV}$ to serve as a good indirect diagnostic for both these escape parameters, while providing a particularly strong tracer of $\dot{N}_\mathrm{ion,esc}$.}
    \label{fig:uv_mag_vs_targets}
\end{figure*}

The COsmic Lyman-$\alpha$ Transfer (\textsc{colt}) code is a Monte Carlo Radiative Transfer (MCRT) solver for post-processing hydrodynamical simulations \citep[][for public code access and documentation see \href{colt.readthedocs.io}{colt.readthedocs.io}]{smith_lyman_2015, smith_physics_2019, smith_physics_2022, mcclymont_modelling_2025}. \textsc{colt} can be run on simulation outputs to trace the paths of photons through the ISM and circumgalactic medium (CGM) \citep{tacchella_h_2022, smith_physics_2022, mcclymont_nature_2024,mcclymont_thesan-zoom_2025-1, mcclymont_thesan-zoom_2025}. The code treats radiation as individual photon packets, initialised with positions, directions and frequencies, that propagate through a 3D adaptive Voronoi tessellation. At each step, an optical depth for the packet is randomly selected and the path length is calculated to determine whether it scatters or escapes. \textsc{colt} models many physical processes: photoionization, recombination cascades, collisional excitation, and both absorption and scattering interactions with dust. \textsc{colt} can therefore accurately model the anisotropic escape, scattering, absorption, and (re-)emission of ionizing, continuum, and line photons.

We have applied \textsc{colt} to the \textsc{thesan-zoom} simulations to generate a catalogue of galaxy realisations spanning $3 < z < 16$. LyC escape is considered based on photons which escape the virial radius. When calculating both LyC emission/escape and observable emission, we use the contribution of both the live dust model and an additional contribution of 40\% of the metals to dust. This is motivated by the attenuation due to the live dust model being too small to match observed UV luminosity functions \citep[UVLFs;][]{kannan_introducing_2025}. $M_\mathrm{UV}$ is the observed rest-frame $1500 \, \textup{\AA}$ UV absolute magnitude calculated with \textsc{colt}. However, we use the intrinsic value for $L_\mathrm{H\alpha}$ (assuming that reasonably accurate dust correction would be available).

Most of a galaxy's luminosity is produced by massive O-type and B-type stars, which have lifetimes of less than 10\,Myr \citep{sternberg_ionizing_2003}. Consequently, high-redshift surveys struggle to identify galaxies with low star-formation rates, as these objects tend to be intrinsically faint. For consistency with such surveys, to which we aim for our models to be applicable, we exclude galaxies with no star formation over the preceding 50\,Myr of their evolution ($\mathrm{SFR}_{50} = 0$).  Following this selection, the catalogue contains 35\,512 mock galaxy realisations.

\section{LyC Escape Prediction} \label{sec:prediction}

\subsection{Identifying potential diagnostics} \label{sec:identifying_indicators}

\begin{figure*}
	\includegraphics[width=\textwidth]{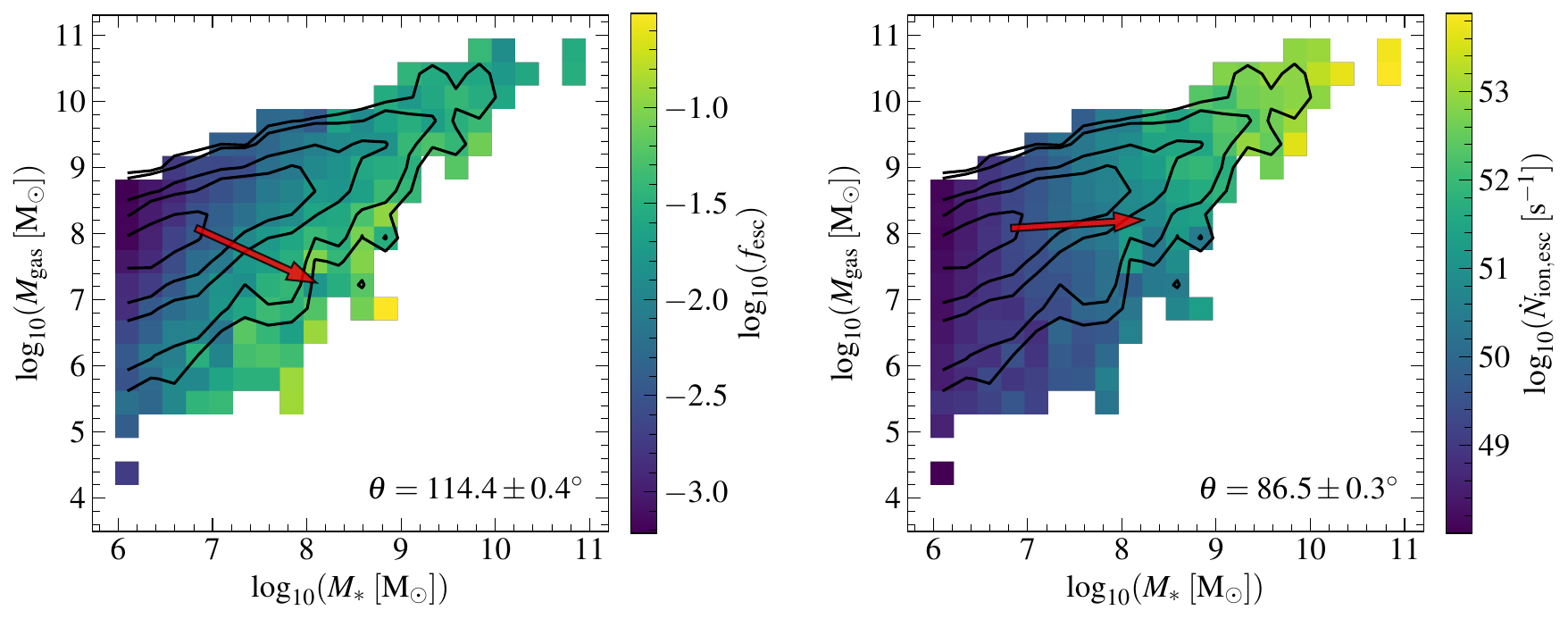}
    \caption{Gas mass ($M_\mathrm{gas}$) versus stellar mass ($M_*$) for \textsc{thesan-zoom} galaxies, colour-coded by median ionizing photon escape fraction ($f_\mathrm{esc}$; \textit{left-panel}) and median ionizing photon escape rate ($\dot{N}_\mathrm{ion,esc}$; \textit{right panel}). Only bins containing at least five galaxies are plotted. The contours indicate the density of galaxies in the plots, with the innermost contours encompassing 34.0\% of the sample and the outermost contours encompassing 97.5\%. The red arrows indicate the direction of the strongest linear positive colour gradient, with their angles, $\theta$, calculated from the partial correlation coefficients and defined clockwise from the positive $y$-direction. $\theta = 114.4\pm 0.4 ^\circ > 90^{\circ}$ for the left plot, demonstrating that $f_\mathrm{esc}$ tends to increase as a galaxy's $M_\mathrm{gas}$ decreases and its $M_*$ increases, implying that gas-to-stellar mass ratio is a good indicator for $f_\mathrm{esc}$. Conversely, $\theta = 86.5 \pm 0.3 ^\circ < 90^{\circ}$ for the right plot which indicates that both $M_\mathrm{gas}$ and $M_*$ are positively correlated with $\dot{N}_\mathrm{ion,esc}$, with $M_\mathrm{gas}$ exhibiting a very weak correlation, which suggests this ratio is not an important diagnostic of $\dot{N}_\mathrm{ion,esc}$. This highlights how $f_\mathrm{esc}$ and $\dot{N}_\mathrm{ion,esc}$ are distinct quantities governed by different physical processes.}
    \label{fig:histograms_star_mass_vs_gas_mass}
\end{figure*}

We assess how LyC escape fractions and escape rates correlate with galactic properties in the \textsc{thesan-zoom} simulations. Any indirect diagnostic must strongly correlate with $f_\mathrm{esc}$, such that $f_\mathrm{esc}$ admits a clear dependence on the diagnostic; the same applies for diagnostics of $\dot{N}_\mathrm{ion,esc}$. In the left panel of Fig.~\ref{fig:uv_mag_vs_targets} we consider the correlation of $f_\mathrm{esc}$ with $M_\mathrm{UV}$. As highlighted by the red line of median $f_\mathrm{esc}$ against median $M_\mathrm{UV}$, we find a predominantly monotonic-decreasing relationship down to $M_\mathrm{UV} \approx -13$, implying high UV luminosity correlates with a high escape fraction for \textsc{thesan-zoom} galaxies in the range $-20 < M_\mathrm{UV} < -13$. The relationship we find is different to that of \citet{rosdahl_lyc_2022} who investigate $f_\mathrm{esc}$ in the \textsc{SPHINX} cosmological simulations \citep{rosdahl_sphinx_2018}, and find a peak in $f_\mathrm{esc}$ at $M_\mathrm{UV}\sim-17$. It is also opposite to the trend found in the Low-redshift Lyman Continuum Survey by \citet{chisholm_far-ultraviolet_2022}, who find high escape fractions are correlated with low UV luminosities.

We also consider the correlation of $\dot{N}_\mathrm{ion,esc}$ with $M_\mathrm{UV}$ in the right panel of Fig.~\ref{fig:uv_mag_vs_targets}, likewise finding that escape rate increases with luminosity over the same range. Comparing the plots in Fig.~\ref{fig:uv_mag_vs_targets}, the $\dot{N}_\mathrm{ion,esc} - M_\mathrm{UV}$ relation appears stronger, exhibiting a strictly monotonic trend. This is further reflected in its smaller scatter: the mean half-width of the  $16^\mathrm{th} - 84^\mathrm{th}$ percentile range (the $1\sigma$ scatter) is 0.628 dex for $\dot{N}_\mathrm{ion,esc}$ as a function of $M_\mathrm{UV}$, compared to 0.829 dex for $f_\mathrm{esc}$. Thus, the $\dot{N}\mathrm{ion,esc} - M{\mathrm{UV}}$ relation shows a tighter correlation and reduced variance relative to its scale. To verify that this difference in scatter is not driven by redshift evolution, we subdivide the galaxy sample into three redshift bins and find minimal variation in both median relations and their scatter across bins (not shown in Fig.~\ref{fig:uv_mag_vs_targets} to avoid overcrowding). This implies that the observed variance is driven primarily by intrinsic galaxy-to-galaxy stochasticity. We therefore expect $M_\mathrm{UV}$ to be a good diagnostic of both $f_\mathrm{esc}$ and $\dot{N}_\mathrm{ion,esc}$, but a decisively stronger tracer of $\dot{N}_\mathrm{ion,esc}$.

We also inspect ratios between different galactic properties in the search for potential indicators. In Fig.~\ref{fig:histograms_star_mass_vs_gas_mass}, we plot the \textsc{thesan-zoom} galaxies in 2D histograms of gas mass $M_\mathrm{gas}$ against stellar mass $M_*$, where the colour scales encode $f_\mathrm{esc}$ (left panel) and $\dot{N}_\mathrm{ion,esc}$ (right panel). The red arrows mark the direction of the steepest positive linear colour gradient on the 3D surface of each diagram, determined via a Partial Correlation Coefficient analysis \citep{bluck_are_2020}, adapted from \citet{maheson_unravelling_2024}. The corresponding angles $\theta$ are defined clockwise from the positive y-direction. In the left panel, the arrow lies at an angle of $\theta = 114.4 \pm 0.4 ^\circ > 90^\circ$. Thus, $f_\mathrm{esc}$ tends to increase as a galaxy's $M_\mathrm{gas}$ decreases and its $M_*$ increases, suggesting gas-to-stellar mass ratio is a good indicator for $f_\mathrm{esc}$. Conversely, the arrow in the right panel lies at an angle of $\theta = 86.5 \pm 0.3 ^\circ < 90^\circ$, indicating $M_\mathrm{gas}$ and $M_*$ are both positively correlated with $\dot{N}_\mathrm{ion,esc}$. The near-orthogonality of this angle further conveys that $M_\mathrm{gas}$ is only weakly correlated with $\dot{N}_\mathrm{ion,esc}$. This suggests that gas-to-stellar mass ratio is likely to be a less important diagnostic of $\dot{N}_\mathrm{ion,esc}$, highlighting that $f_\mathrm{esc}$ and $\dot{N}_\mathrm{ion,esc}$, while related, are physically distinct quantities governed by different processes and constraints. It is therefore unsurprising that they exhibit different sets of indicators.

In the search for potential indicators of $f_\mathrm{esc}$ and $\dot{N}_\mathrm{ion,esc}$, we perform the same single-variable analysis as done for $M_\mathrm{UV}$, applied to many parameters in the \textsc{thesan-zoom} catalogue suspected to have predictive power. Analogous plots for galactic properties successfully identified as indicators are shown in Appendix~\ref{sec:appendix_a} (see Figs.~\ref{fig:all_vs_f_esc} and \ref{fig:all_vs_n_esc}). We also repeat the ratio analysis using 2D histograms, as in Fig.~\ref{fig:histograms_star_mass_vs_gas_mass}, for multiple physically motivated pairs of parameters from the catalogue (see Fig.~\ref{fig:histograms_all_double} in Appendix~\ref{sec:appendix_a}). This yields a set of potential diagnostics, which, after further filtering explained in the following subsection, can be seen in Table~\ref{tab:indicators}.

\subsection{Random forests} \label{sec:random_forests}

We build our prediction models using random forest (RF) regression \citep{breiman_random_2001}. This is a machine learning method that constructs decision trees during training and outputs their mean prediction upon application. The investigation dataset, comprising features and the target variable for prediction, is split into a training set and a testing set, with the test sample used to evaluate the accuracy of the regressor. The error is typically quantified by Mean Squared Error (MSE) and Mean Absolute Error (MAE) measurements, which measure the average discrepancy between the random forest’s predictions and the dataset values. Each tree in the forest is trained on a bootstrap sample of the training data, generated by sampling with replacement, which introduces diversity among the trees. At each split, a random subset of features is considered, and node decisions are chosen to minimise the error criterion, chosen to be MSE here. The randomness of bootstrapping and feature selection enables the RF regressor to capture complex, nonlinear relationships in the data while remaining robust to noise and outliers. After training, the feature importances of the model are returned, which represent the average contribution of each feature to the decrease in the error criterion. The importances in this work, given as fractions that sum to 1, represent how dependent the RF is on each observable for the prediction of the target variables, which are $f_\mathrm{esc}$ and $\dot{N}_\mathrm{ion,esc}$ here. The random forest technique has been widely applied in recent astrophysical studies, demonstrating its versatility and effectiveness across a range of problems \citep{bluck_quenching_2022, baker_molecular_2023, maheson_unravelling_2024, kawinwanichakij_connecting_2026}.

\begin{table}
\begin{tabular}{|p{1.5cm}|p{4.75cm}|p{1.25cm}|}
\hline
Indicator                                & Definition                                                                                       & Models     \\ \hline
$\mathrm{SFR}_{10}$                      & star-formation rate over 10\,Myr                                                                 & B, D       \\
$\mathrm{SFR}_{100}$                     & star-formation rate over 100\,Myr                                                                & B, D       \\
$\Delta\mathrm{MS}_{10}$                 & Offset from the star-forming main sequence over 10\,Myr; formula in Eq.~\eqref{offset}           & A, C       \\
$\mathrm{SFR}_{10}/\mathrm{SFR}_{100}$   & 10–to–100$\,$Myr star-formation rate ratio                                                       & A, C       \\
$M_*$                                    & Stellar mass                                                                                     & A, B, C, D \\
$M_\mathrm{gas}/M_*$                     & Ratio of gas mass to stellar mass                                                                & A, B       \\
$M_*/M_\mathrm{vir}$                     & Ratio of stellar mass to halo virial mass                                                        & A, B       \\
$R_\mathrm{SFR}$                         & Half-radius of star-formation rate                                                               & A          \\
$R_\mathrm{SFR}/R_{M_*}$                 & Ratio of half-radius of star-formation rate and stellar half-mass radius                         & A          \\
$\Sigma_\mathrm{SFR_{10}}$               & star-formation rate surface density over 10\,Myr; formula given in Eq.~\eqref{SFR_density}       & A          \\
$Z$                                      & Mean stellar metallicity                                                                         & A, B       \\
$M_\mathrm{UV}$                          & Rest-frame $1500 \, \textup{\AA}$ Observed UV absolute magnitude                                 & A, B, C, D \\
$A_\mathrm{UV}$                          & Rest-frame $1500 \, \textup{\AA}$ UV Attenuation                                                 & A, B       \\
$L_\mathrm{H\alpha}$                     & $6562.8 \, \textup{\AA}$ H$\alpha$ line integrated luminosity                                    & B       \\
$L_\mathrm{UV}/L_\mathrm{H\alpha}$       & Ratio of UV ($1500 \, \textup{\AA}$) to H$\alpha$ luminosities                                   & A       \\
$R_\mathrm{UV}$                          & Half-light radius of $1500 \, \textup{\AA}$ UV luminosity                                        & A       \\
$R_\mathrm{H\alpha}$                     & Half-light radius of H$\alpha$ luminosity                                                        & A \\
\hline
\end{tabular}
\caption{The indirect diagnostics, their definitions and which of our $f_\mathrm{esc}$ and $\dot{N}_\mathrm{ion,esc}$ prediction models use them in training. All quantities are calculated based on particles within the virial radius, except for gas masses, which are calculated within the galaxy (defined as within twice the UV half-light radius) in order to trace the gas within the ISM. Model A: our best $f_\mathrm{esc}$ predictor trained with all the relevant \textsc{thesan-zoom}-derived diagnostics; Model B: similarly our best $\dot{N}_\mathrm{ion,esc}$ predictor; Model C: our $f_\mathrm{esc}$ predictor trained only with features compatible with photometric JWST surveys; Model D: our photometrically compatible $\dot{N}_\mathrm{ion,esc}$ predictor.}
\label{tab:indicators}
\end{table}

We used the RF regressor from the \textsc{python} package \textsc{scikit-learn} for this analysis \citep{pedregosa_scikit-learn_2011}. In each training sequence, the data from the \textsc{thesan-zoom} catalogue was first divided into a train and test sample, with a 75:25 split. We use subsets of the identified observable indicators as the features of the dataset, applying base-10 logarithms to each (excluding those already in magnitudes). In all models, we include $\log_{10}(1+z)$ to track cosmic evolution and a uniform random variable for benchmarking. To avoid log(0) infinities, we add a small $\epsilon_x$ to every value of feature $x$, defined as $\epsilon_x = 0.01 \, \mathrm{min}\{x\}$, where $\mathrm{min}\{x\}$ denotes the smallest non-zero instance of feature $x$. To enhance the regressors' accuracy and minimise overfitting, we fine-tune specific execution parameters within the RF function, referred to as hyperparameters; Appendix \ref{sec:appendix_b} explains this optimisation.

To finalise the indicator sets, we trained 100 preliminary models for each of the targets $f_\mathrm{esc}$ and $\dot{N}_\mathrm{ion,esc}$, using the previously identified potential indicators to obtain the mean feature importances. Regression was then applied to each of the 100 test samples to compute the mean prediction MAE. For any feature with an importance below 0.01, a further 100 models were trained without it; if the removal of that feature resulted in a lower mean test MAE, it was classified as irrelevant and removed from the potential indicator set. This iterative procedure ultimately yielded 14 indicators for $f_\mathrm{esc}$ and 9 for $\dot{N}_\mathrm{ion,esc}$ which we present in Tab.~\ref{tab:indicators}. 

For completeness, we provide below the expressions for several derived parameters listed in Tab.~\ref{tab:indicators}. The specific star-formation rate, $\mathrm{sSFR_\tau}$, for a galaxy with stellar mass $M_*$ and star-formation rate $\mathrm{SFR}_\tau$ over a lookback time of $\tau$ in Myr, is given by $\mathrm{sSFR_\tau} = \mathrm{SFR}_\tau / M_*$. The majority of star-forming galaxies follow a relatively tight relation between $\mathrm{sSFR}$ and $M_*$ known as the ``main sequence'' \citep{speagle_highly_2014, tacchella_confinement_2016, popesso_main_2023, clarke_star-forming_2024, mcclymont_thesan-zoom_2025, simmonds_bursting_2025}. Offset from the star-forming main sequence over $\tau$ Myr, $\Delta\mathrm{MS}_\tau$, is defined at $z$ for a galaxy with stellar mass $M_*$ as:
\begin{equation} \label{offset}
    \Delta\mathrm{MS}_{\tau} = \log_{10} \left(\frac{\mathrm{sSFR_\tau}}{\mathrm{sSFR}_{\rm MS}(M_*,z)}\right) \, ,
\end{equation}
where the main sequence specific star-formation rate has the functional form \citep{tacchella_confinement_2016}:
\begin{equation} \label{main_sequence}
    \mathrm{sSFR}_{\rm MS}(M_*,z) = s_b \, \left(\frac{M_*}{10^{10} \, \mathrm{M_\odot}}\right)^\beta \, (1+z)^\mu \quad \mathrm{Gyr^{-1}} \, .
\end{equation}
For a look back time of $\tau=10$\,Myr, we use the best fit parameters $s_b = 0.033 \pm 0.002$, $\beta = 0.041 \pm 0.004$ and $\mu = 2.64 \pm 0.03$, calculated by \citet{mcclymont_thesan-zoom_2025}. We also define the SFR surface density, $\Sigma_\mathrm{SFR_{10}}$, using the half-radius of SFR, $R_\mathrm{SFR}$:
\begin{equation} \label{SFR_density}
    \Sigma_\mathrm{SFR_{10}} = \frac{\mathrm{SFR_{10}} \, / \, 2}{\pi \, (R_\mathrm{SFR})^2} \, .
\end{equation}
Finally, the UV Attenuation is defined as the difference between the observed $1500 \, \textup{\AA}$ UV magnitude and the intrinsic (unobscured by absorption and scattering) $1500 \, \textup{\AA}$ UV magnitude of a galaxy: $A_\mathrm{UV} = M_\mathrm{UV,obs} - M_\mathrm{UV,int}$.

The features of our models are limited to observables obtainable through COLT post-processing. In addition to the diagnostics identified here, the UV spectral slope, $\beta$, is often considered a useful tracer of $f_\mathrm{esc}$ and is incorporated in the models of \citet{chisholm_far-ultraviolet_2022}, \citet{choustikov_physics_2024}, and \citet{stoffers_challenge_2025}. As $\beta$ is strongly correlated with UV dust attenuation \citep{shen_high-redshift_2020, jecmen_glimpse_2026}, which we already include, its information content is partially captured by our framework. Additional diagnostics have also been proposed in other recent studies \citep{katz_nature_2022, xu_tracing_2022, rosdahl_lyc_2022, katz_two_2023, mascia_closing_2023, mascia_new_2024, choustikov_great_2024, jaskot_multivariate_2024}.

\begin{table}
\centering
\begin{tabular}{cccccc}
\hline
Model & Target                       & MAE   & MSE   & RMSE  & PCC   \\ \hline
A     & $f_\mathrm{esc}$             & 0.420 & 0.305 & 0.552 & 0.809 \\
B     & $\dot{N}_\mathrm{ion,esc}$   & 0.407 & 0.289 & 0.537 & 0.919 \\
C     & $f_\mathrm{esc}$             & 0.476 & 0.386 & 0.621 & 0.739 \\
D     & $\dot{N}_\mathrm{ion,esc}$   & 0.446 & 0.351 & 0.592 & 0.899 \\ \hline
\end{tabular}
\caption{The mean absolute error (MAE; in dex), mean squared error (MSE; in dex$^2$), root mean squared error (RMSE; in dex), and Pearson Correlation Coefficient (PCC; ranging from $-1$ to 1), for our models calculated from their respective test samples. Model A: our best $f_\mathrm{esc}$ predictor trained with all the relevant \textsc{thesan-zoom}-derived diagnostics. Model B: similarly our best $\dot{N}_\mathrm{ion,esc}$ predictor. Model C: our $f_\mathrm{esc}$ predictor trained only with features compatible with the JWST observational sample from \citet{simmonds_bursting_2025}. Model D: our observationally compatible $\dot{N}_\mathrm{ion,esc}$ predictor.}
\label{tab:model_results}
\end{table}

\begin{figure*}
	\includegraphics[width=\textwidth]{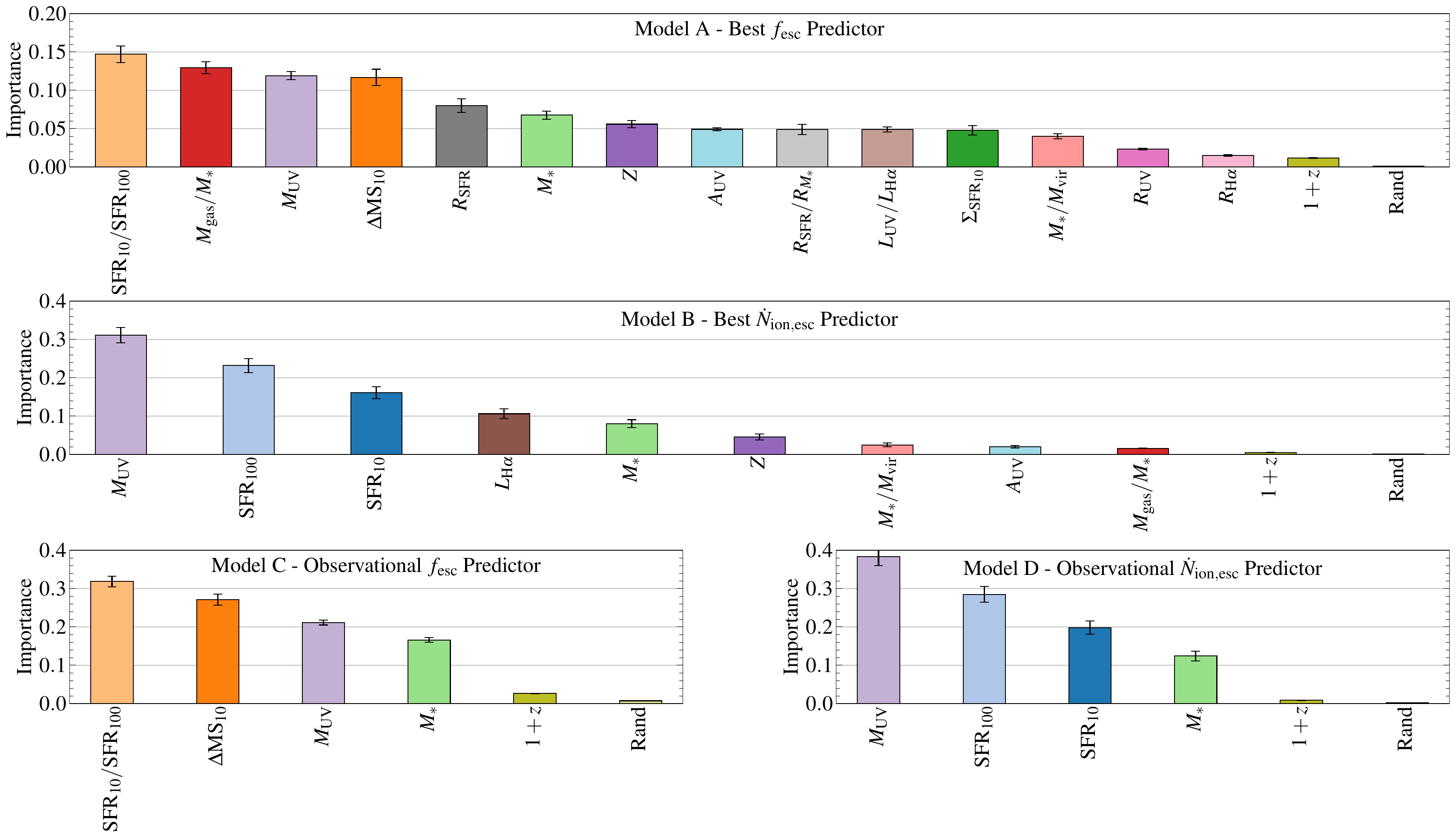}
    \caption{Mean feature importances for the four random forest models, averaged over 1000 train--test realisations. Importances represent the fractional contribution of each feature to reducing the mean squared error. The error bars on each importance denote the respective standard deviation across the 1000 runs. \textit{Top panel}: Model A, our best $f_\mathrm{esc}$ predictor trained with all the identified \textsc{thesan-zoom} $f_\mathrm{esc}$ indicators. \textit{Middle panel}: Model B, similarly our best $\dot{N}_\mathrm{ion,esc}$ predictor. \textit{Bottom left panel}: Model C, our $f_\mathrm{esc}$ predictor trained using only features compatible with photometric JWST observations. \textit{Bottom right panel}:  Model D, our photometrically compatible $\dot{N}_\mathrm{ion,esc}$ predictor. The indicators are defined in Tab.~\ref{tab:indicators}. The uniform random variable is reassuringly the least important predictive feature for each model relative to the chosen model indicators. This ranking reveals that the ratios $\mathrm{SFR_{10}/SFR_{100}}$ and $M_\mathrm{gas}/M_*$ are the strongest diagnostics of $f_\mathrm{esc}$, whereas the rest-frame UV absolute magnitude, $M_\mathrm{UV}$, clearly dominates predictions of $\dot{N}_\mathrm{ion,esc}$.}
    \label{fig:importances}
\end{figure*}

\subsection{Model results} \label{sec:results}

We present four models trained using the \textsc{thesan-zoom} catalogue with the indicator sets specified in Tab.~\ref{tab:indicators}. For each indicator configuration, 1000 random forests were trained, and the regressor yielding the smallest mean absolute error (MAE; in dex) on prediction for its corresponding test sample was selected as the final model. Their MAE, mean squared error (MSE; in dex$^2$), root mean squared error (RMSE; in dex), and Pearson correlation coefficient (PCC), which quantifies the strength of the linear relationship between the true and predicted values and ranges from $-1$ to 1, are reported in Tab.~\ref{tab:model_results}. Model A is our best $f_\mathrm{esc}$ predictor, trained using all 14 of the finalised $f_\mathrm{esc}$ indicators. Model B is our best $\dot{N}_\mathrm{ion,esc}$ predictor, similarly trained using all 9 of the identified $\dot{N}_\mathrm{ion,esc}$ indicators. 

Models C and D likewise predict $f_\mathrm{esc}$ and $\dot{N}_\mathrm{ion,esc}$, respectively, but are trained using only observables that are typically available in JWST photometric catalogues. Their indicator sets are therefore subsets of those of Models A and B, reflecting the more limited variable coverage of photometric catalogues relative to \textsc{thesan-zoom}. Consequently, Models C and D perform marginally worse than A and B, respectively, across all error and correlation metrics shown in Tab.~\ref{tab:model_results}.

The feature importances of the models are presented in Fig.~\ref{fig:importances} as the mean importances averaged over their respective 1000 RF train-test realisations. We find the 10–to–100$\,$Myr star-formation rate ratio, $\mathrm{SFR_{10}/SFR_{100}}$, to be the best diagnostic of $f_\mathrm{esc}$, followed closely by the gas-to-stellar mass ratio, $M_\mathrm{gas}/M_*$, and then rest-frame UV ($1500 \, \textup{\AA}$) absolute magnitude, $M_\mathrm{UV}$. Conversely, $M_\mathrm{UV}$ is found to dominate predictions of $\dot{N}_\mathrm{ion,esc}$, with significant importance also placed on the star-formation rate observables $\mathrm{SFR}_{100}$ and $\mathrm{SFR}_{10}$. The uniform random variable, included only for benchmarking, is reassuringly the least important predictive feature for each model.

\begin{figure*}
	\includegraphics[width=\textwidth]{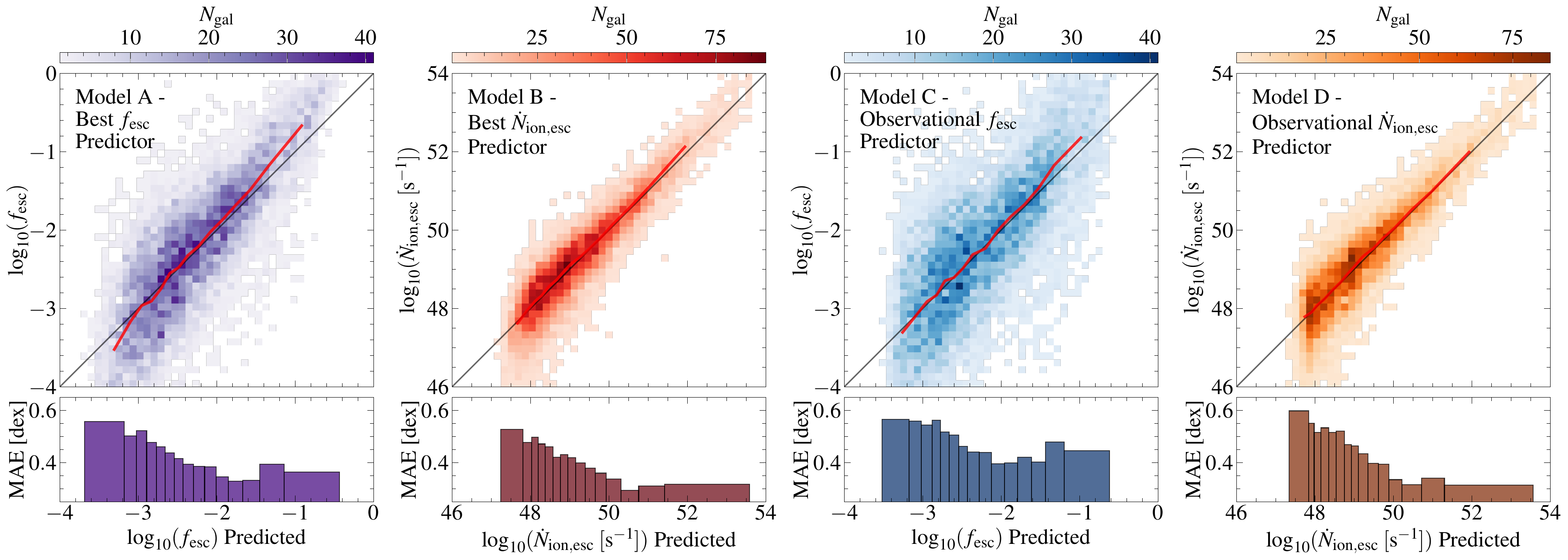}
    \caption{True versus predicted values for the four random forest models presented. From left to right they are: Model A, our best $f_\mathrm{esc}$ predictor trained with all the identified \textsc{thesan-zoom} $f_\mathrm{esc}$ indicators; Model B, similarly our best $\dot{N}_\mathrm{ion,esc}$ predictor; Model C, our $f_\mathrm{esc}$ predictor trained using only features compatible with photometric JWST surveys; Model D, our photometrically compatible $\dot{N}_\mathrm{ion,esc}$ predictor. Top panels show predicted against true $\log_{10}(f_\mathrm{esc})$ or $\log_{10}(\dot{N}_\mathrm{ion,esc})$, with the colour scale indicating the number of galaxies per bin. Red curves trace the median relation. Bottom panels show the mean absolute error (MAE; in dex) as a function of predicted value. All models achieve higher accuracy when predicting larger values of $f_\mathrm{esc}$ and $\dot{N}_\mathrm{ion,esc}$. Moreover, predictions of $\dot{N}_\mathrm{ion,esc}$ exhibit systematically reduced scatter and higher accuracy, highlighting their suitability for reionization modelling.}
    \label{fig:test_predictions_and_errors}
\end{figure*}

The predictions of the models are plotted against the true $f_\mathrm{esc}$ or $\dot{N}_\mathrm{ion,esc}$ values of their respective test samples in Fig.~\ref{fig:test_predictions_and_errors}. The prediction errors for all these models tend to decrease for higher predicted values, as demonstrated in the lower panels of Fig.~\ref{fig:test_predictions_and_errors}, which is encouraging given that high-escape galaxies are the most relevant for reionization. We see a larger scatter around the $1:1$ line for the $f_\mathrm{esc}$ models when compared to the $\dot{N}_\mathrm{ion,esc}$ models, which results in higher values of the PCC for the $\dot{N}_\mathrm{ion,esc}$ predictors, as seen in Tab.~\ref{tab:model_results}. This implies our models constrain $\dot{N}_\mathrm{ion,esc}$ more effectively, suggesting that $\dot{N}_\mathrm{ion,esc}$ may therefore be a more reliable quantity for reionization modelling. 

A further limitation of our $f_\mathrm{esc}$ models is their inability to produce predictions at the highest escape fractions, particularly in the range $\log_{10}(f_\mathrm{esc}) > -0.5$ (see the left two panels of Fig.~\ref{fig:test_predictions_and_errors}). This reflects the inherent nature of random forests, which average over decision tree outputs, thereby smoothing extreme values and restricting predictions to the range spanned by the \textsc{thesan-zoom} training data. However, this limitation affects only a small subset of galaxies, and the resulting upper bound on predicted values lies approximately within one MAE of the physical maximum $\log_{10}(f_\mathrm{esc}) = 0$, limiting its practical impact. While our $\dot{N}_\mathrm{ion,esc}$ predictors exhibit a similar, albeit softer, capping behaviour, this quantity has no intrinsic upper limit, rendering the effect largely inconsequential.

\begin{figure*}
	\includegraphics[width=\textwidth]{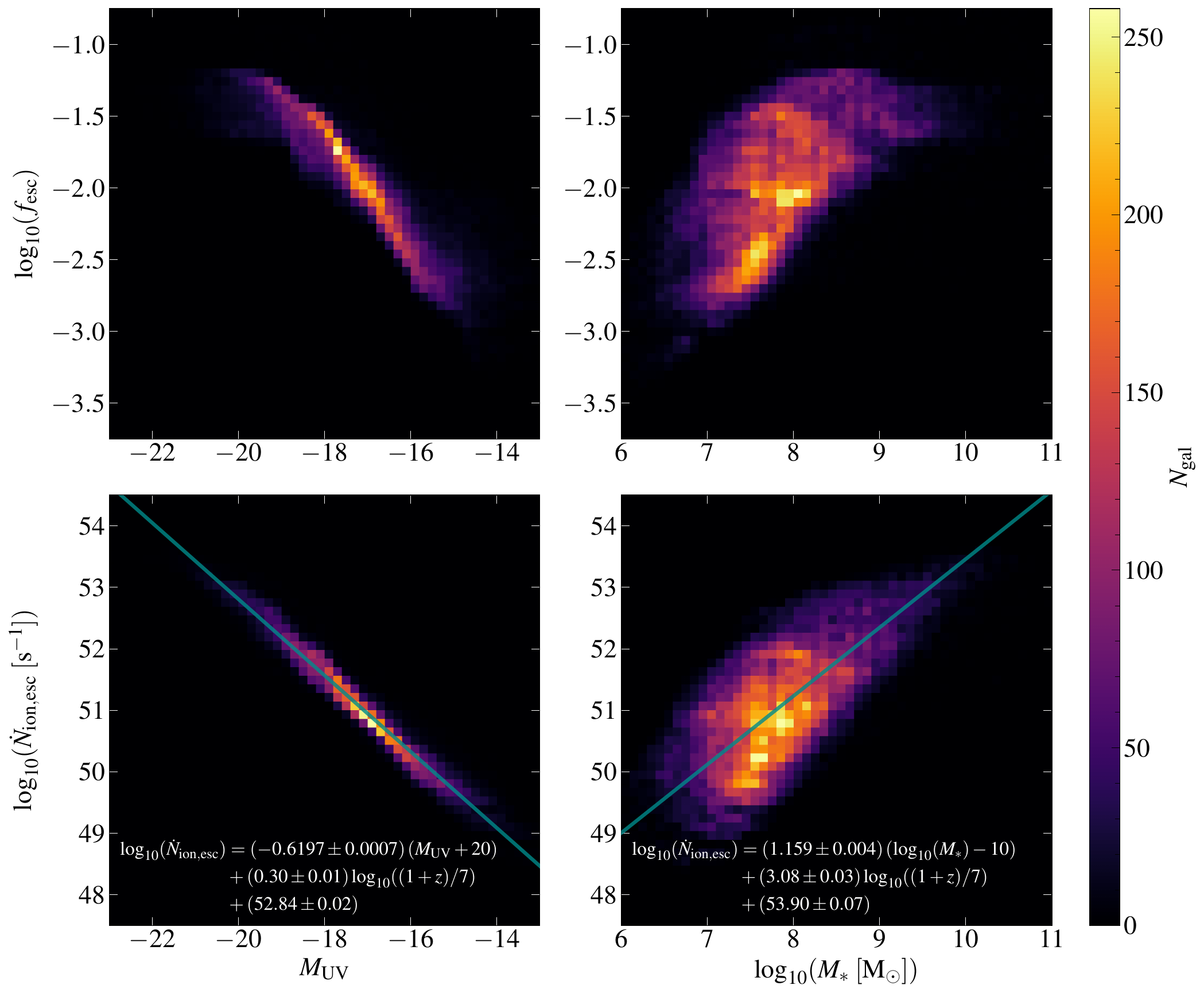}
    \caption{Model predictions of $f_\mathrm{esc}$ and $\dot{N}_\mathrm{ion,esc}$ for galaxies in the JWST photometric catalogue from \citet{simmonds_bursting_2025}. \textit{Top left:} Predicted $f_\mathrm{esc}$ as a function of $M_\mathrm{UV}$. \textit{Top right:} predicted $f_\mathrm{esc}$ as a function of $M_*$. \textit{Bottom left:} predicted $\dot{N}_\mathrm{ion,esc}$ as a function of $M_\mathrm{UV}$. \textit{Bottom right:} predicted $\dot{N}_\mathrm{ion,esc}$ as a function of $M_*$. The colour scale indicates the number of galaxies per bin. Both $f_\mathrm{esc}$ and $\dot{N}_\mathrm{ion,esc}$ show clear correlations with $M_\mathrm{UV}$ and $M_*$, consistent with expectations from \textsc{thesan-zoom}. In particular, $\log_{10}(\dot{N}_\mathrm{ion,esc})$ exhibits approximately linear behaviour with both $M_\mathrm{UV}$ and $\log_{10}(M_*)$. Therefore, in both lower panels, we present the orthogonal distance regression linear relations for $\log_{10}(\dot{N}_\mathrm{ion,esc})$, in which we also include a redshift dependence, and plot them in cyan. The relationship between predicted $\dot{N}_\mathrm{ion,esc}$ and $M_\mathrm{UV}$ exhibits very little scatter, implying that the model relies most strongly on $M_\mathrm{UV}$ in prediction, in agreement with the importance analysis presented in Fig.~\ref{fig:importances}.}
    \label{fig:observational_fits}
\end{figure*}

We test Models C and D using the photometric catalogue presented in \citet{simmonds_bursting_2025}, which is based on JWST NIRCam imaging of the JWST Advanced Deep Extragalactic Survey (JADES) in the GOODS-South and GOODS-North fields \citep{eisenstein_overview_2023, bunker_jades_2024}. The catalogue includes stellar masses, star-formation rate, and rest-frame $1500 \, \textup{\AA}$ UV magnitudes inferred via SED fitting with \textsc{\textsc{PROSPECTOR}} \citep{johnson_stellar_2021}, adopting a Chabrier initial mass function (IMF) \citep{chabrier_galactic_2003}, a two-component dust model \citep{charlot_simple_2000, conroy_propagation_2009}, and non-parametric star-formation histories (SFHs) \citep{leja_how_2019}, where SFH is described with eight different SFR bins. Photometric redshifts were obtained using the fitting code \textsc{EAzY} \citep{brammer_eazy_2008} and used as priors in the SED fitting. The following selection criteria were imposed on the JADES photometry dataset: sources must have a signal-to-noise ratio SNR$\geq3$ in the F444W NIRCam band, photometric redshift in the range $3 \leq z \leq 9$; and a reduced $\chi^2 \leq 10$. The resulting catalogue contains 41,430 galaxies. 99.7\% of these galaxies fall within $-21.7 < M_\mathrm{UV} < -13.1$ and $6.2 < \log_{10}(M_* / \mathrm{M_\odot}) < 10.5$ (the 3$\sigma$ observed UV magnitude and stellar mass ranges).

Our predictions of $f_\mathrm{esc}$ and $\dot{N}_\mathrm{ion,esc}$, using models C and D respectively for the JWST photometric sample from \citet{simmonds_bursting_2025}, are plotted in Fig.~\ref{fig:observational_fits} against $M_\mathrm{UV}$ and $M_*$. In both lower panels, $\log_{10}(\dot{N}_\mathrm{ion,esc})$ shows an approximately linear trend with $M_\mathrm{UV}$ and $\log_{10}(M_*)$. We therefore use orthogonal distance regression to fit linear relationships to each \citep{hughes_measurements_2010}, with the resulting fit parameters presented in their respective panels. We include redshift dependence in both to account for the cosmic evolution of $\dot{N}_\mathrm{ion,esc}$. For comparison, the analogous $f_\mathrm{esc}$ and $\dot{N}_\mathrm{ion,esc}$ plots and fits for the intrinsic \textsc{thesan-zoom} values are shown in Fig.~\ref{fig:thesan_fits_histogram} in Appendix~\ref{sec:appendix_a}. The $\dot{N}_\mathrm{ion,esc}-M_\mathrm{UV}$ relation obtained from this JWST analysis closely matches that found in \textsc{thesan-zoom}, albeit with an even weaker redshift dependence. A similarly weak redshift dependence of $\dot{n}_\mathrm{ion}$ inferred from JWST/NIRCam photometry has also been reported by \citet{choustikov_inferring_2024}. Our JWST-based relation also exhibits very little scatter, implying that the model relies most strongly on $M_\mathrm{UV}$ in prediction, in agreement with the feature importances shown in Fig.~\ref{fig:importances}.

\subsection{Indirect indicator physics} \label{sec:indicator_physics}

\subsubsection{$f_\mathrm{esc}$ indicator physics} \label{sec:fesc_indicator_physics}

Massive O-type and B-type stars, with lifetimes of less than 10\,Myr \citep{sternberg_ionizing_2003}, are responsible for the majority of LyC photon emission. Furthermore, their winds and supernovae create the low-density channels through which LyC photons can escape \citep{sharma_winds_2017}. Consequently, large populations of these short-lived stars are linked to high $f_\mathrm{esc}$ and $\dot{N}_\mathrm{ion,esc}$ \citep{choustikov_physics_2024, bhagwat_spice_2024}, in agreement with the high importance our models place on the star-formation rate observables $\Delta \mathrm{MS_{10}}$, $\mathrm{SFR_{10}}$, and $\mathrm{SFR_{100}}$ (see Fig.~\ref{fig:importances}). 

Many studies argue that periods of rapid, bursty star formation generate the intense stellar feedback that drives large LyC escape \citep{heckman_extreme_2011, katz_two_2023, bhagwat_spice_2024, jaskot_ionizing_2025}. In particular, \citet{bhagwat_spice_2024} find that galaxies with high $\mathrm{SFR}_{10}/\mathrm{SFR}_{100}$, indicative of ongoing starbursts, tend to exhibit high $f_\mathrm{esc}$. Our results likewise demonstrate that this ratio is the most important parameter governing $f_\mathrm{esc}$ (see Fig.~\ref{fig:importances}); however, in \textsc{thesan-zoom} we find the opposite behaviour: $\mathrm{SFR}_{10}/\mathrm{SFR}_{100}$ has a strict negative correlation with $f_\mathrm{esc}$ (see Fig.~\ref{fig:all_vs_f_esc}), where $f_\mathrm{esc}$ is highest when $\mathrm{SFR}_{10} / \mathrm{SFR}_{100} \lesssim 0.1$, corresponding to a post-burst or temporarily quenched phase. Physically, this can be understood as dense gas surrounding young massive stars during the burst absorbing most ionizing photons, whereas subsequent feedback clears the ISM and opens the low-density channels that enable delayed LyC escape. This interpretation is consistent with \citet{katz_two_2023}, who identify a population of ``Remnant Leakers'' with low $\mathrm{SFR}_{10} / \mathrm{SFR}_{100}$ and high $f_\mathrm{esc}$ in the \textsc{SPHINX}$^{20}$ simulation \citep{rosdahl_lyc_2022}, where supernova-driven feedback from previous bursts has disrupted the ISM and facilitated leakage. 

\citet{rosdahl_lyc_2022} claim that smaller mass galaxies have less sustained starbursts, meaning their feedback is less disruptive and weaker in facilitating ionizing photon escape, supporting $M_*$ as a key indicator of $f_\mathrm{esc}$, consistent with our \textsc{thesan-zoom} findings. $M_\mathrm{UV}$ is similarly important to our $f_\mathrm{esc}$ models due to its strong association with $M_*$ and star-formation rates, reflecting its dependence on the number of young, massive stars.  Likewise, the well-established correlation between $M_*$ and UV dust attenuation supports the relevance we find for $A_\mathrm{UV}$ in predicting $f_\mathrm{esc}$ \citep{pannella_goods-herschel_2015, shen_high-redshift_2020}.

Morphological variables seem to play only a limited role in $f_\mathrm{esc}$ regression and have negligible importance in $\dot{N}_\mathrm{ion,esc}$ prediction (see Fig.~\ref{fig:importances}). We note that these results highlight the relative strength of different predictors. Additionally, where there are variables with degenerate predictive power, this method generally assigns importance to the ``better'', or more fundamental, predictor. This means that even if there is a correlation between a variable and $f_\mathrm{esc}$, that variable may still be assigned a low importance. This may explain why $R_\mathrm{UV}$ is assigned a relatively weak importance in our models relative to observational works \citep{mascia_closing_2023, jaskot_multivariate_2024}. $R_\mathrm{UV}$ has been shown to heavily trace a galaxy's star-formation history in \textsc{thesan-zoom} \citep{mcclymont_thesan-zoom_2025-1}, and so most of its predictive power may be simply in tracing star formation which is better encapsulated by $\mathrm{SFR_{10}/SFR_{100}}$ and $\Delta\mathrm{MS}_{10}$. Our inclusion of $R_\mathrm{SFR}$, which shows much greater importance than $R_\mathrm{UV}$, suggests that it is the more fundamental morphological property related to $f_\mathrm{esc}$. Therefore, much of the observed correlation between  $R_\mathrm{UV}$ and $f_\mathrm{esc}$ may be due to both $R_\mathrm{UV}$ and $f_\mathrm{esc}$ being more fundamentally affected by a galaxy's star-formation history.

Although stellar metallicity $Z$ is not a dominant diagnostic in the models, it remains positively correlated with $f_\mathrm{esc}$ and $\dot{N}_\mathrm{ion,esc}$, consistent with \citet{kostyuk_ionizing_2023}, who attribute this to longer-lived, metal-rich stars whose LyC emission is more likely to persist after their birth clouds have dispersed. The gas-to-stellar mass ratio is a proxy for both LyC absorption (via gas content $M_\mathrm{gas}$) and LyC production (through its $M_*$ dependence), explaining why it is one of the strongest indicators in our best $f_\mathrm{esc}$ predictor Model A (see Fig.~\ref{fig:importances}). A similar reasoning applies to $M_*/M_\mathrm{vir}$, though its lack of direct sensitivity to LyC absorption makes it a weaker diagnostic. Likewise, we find $L_\mathrm{UV} / L_\mathrm{H\alpha}$ to be a relevant indicator of $f_\mathrm{esc}$ since H$\alpha$ traces ionizing photon absorption, while $L_\mathrm{UV}$ is correlated with their production. We see that many of the features identified as important by the random forests are correlated, indicating some level of redundancy among the selected features. This may explain why models C and D exhibit only slightly worse performance than models A and B, respectively, across the metrics reported in Tab.~\ref{tab:model_results}, despite relying on more restrictive feature sets.

We find that redshift plays a relatively minor role in determining $f_\mathrm{esc}$ which is in agreement with the findings of \citet{papovich_galaxies_2025}; we also find an even weaker redshift dependence in $\dot{N}_\mathrm{ion,esc}$. However, this likely reflects that any redshift evolution in properties, such as the merger rate or galaxy burstiness \citep{mcclymont_thesan-zoom_2025}, is already captured by other predictors (e.g., $\mathrm{SFR}_{10} / \mathrm{SFR}_{100}$) that more directly measure these physical drivers of escape. The relative unimportance of redshift therefore suggests that there is little redshift evolution relevant for ionizing escape in variables not accounted for here.

\subsubsection{$\dot{N}_\mathrm{ion,esc}$ indicator physics} \label{sec:nion_indicator_physics}

\begin{figure*}
	\includegraphics[width=\textwidth]{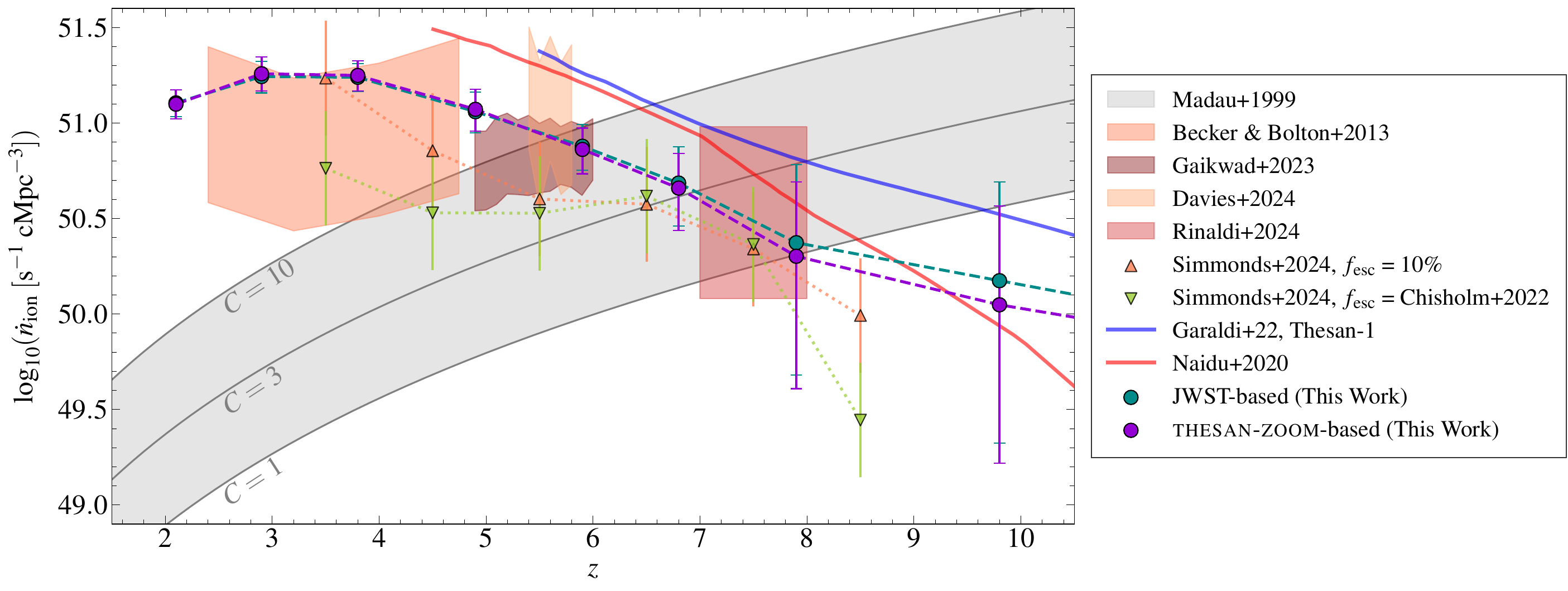}
    \caption{Comoving cosmic ionizing photon emissivity, $\dot{n}_{\mathrm{ion}}$, as a function of redshift, $z$, derived from the \textsc{thesan-zoom}-based (violet points) and JWST-based (cyan points) $\dot{N}_\mathrm{ion,esc}$ relations. These emissivities have been integrated using the UV luminosity functions from \citet{bouwens_new_2021} for $z \in [1.5, 8.5]$ and from \citet{whitler_z_2025} for $z \in [8.5, 16]$. The grey bands indicate the critical ionizing emissivity required to maintain hydrogen ionization in the IGM, determined using the recombination model of \citet{madau_radiative_1999}, for possible clumping factors $C \in [1, 10]$. For comparison, we include the $\dot{n}_{\mathrm{ion}}$ estimates of \citet{simmonds_ionizing_2024}, derived from a related JWST/NIRCam-detected JADES sample assuming either a constant $f_\mathrm{esc} = 10\%$ or the $f_\mathrm{esc}$ model of \citet{chisholm_far-ultraviolet_2022} (shown by the orange and green points respectively). We also include the curve derived from the parent \textsc{thesan-1} simulation run, as presented by \citet{garaldi_thesan_2022} (blue), and that of \citet{naidu_rapid_2020}, derived from an empirical galaxy formation model with a fitted escape fraction (red). Observational constraints obtained from Ly$\alpha$ forest measurements and JWST observations of H$\alpha$ emitters are overplotted  \citep{becker_new_2013, gaikwad_measuring_2023, davies_constraints_2024, rinaldi_midis_2024}. Both of our $\dot{n}_{\mathrm{ion}}$ histories are mutually consistent across all relevant redshifts and agree with the observational constraints. These results therefore avoid any crisis in the ionizing photon budget as they predict the critical emissivity is unlikely to have been reached until after $z \sim 8$.}
    \label{fig:reionisation_timeline}
\end{figure*}

A galaxy's $\dot{N}_\mathrm{ion,esc}$ is defined as its total UV luminosity multiplied by $f_\mathrm{esc}$ and its ionizing photon production efficiency $\xi_{\mathrm{ion}}$. Observations indicate $\xi_{\mathrm{ion}}$ varies relatively weakly across the high-redshift population \citep{simmonds_low-mass_2024, simmonds_ionizing_2024, papovich_galaxies_2025, llerena_ionizing_2025, austin_resolving_2025}, with \citet{llerena_ionizing_2025} finding a mean value of $\log_{10}(\xi_\mathrm{ion} \mathrm{[Hz \, erg^{-1}]}) = 25.22$ with an observed scatter of 0.42 dex in a JWST/NIRSpec spectroscopic sample of 761 galaxies at $4 \leq z \leq 10$. We therefore expect extensive observables highly correlated with luminosity to dominate predictions of $\dot{N}_\mathrm{ion,esc}$ (such as $M_*$ and stellar formation rates). This explains why $M_\mathrm{UV}$, a logarithmic reparametrization of the UV luminosity, is the most important feature in both the relevant Models (B and D). It also explains why ratio-based features, which remove the leading-order multiplicative dependence on luminosity by probing relative rather than absolute quantities, as well as morphological parameters, contribute little additional predictive power once the dominant luminosity scaling is accounted for.

\section{Implications for Cosmic Reionization} \label{sec:implications}

Recent JWST studies have reported a high abundance of UV bright early-Universe galaxies \citep{carniani_spectroscopic_2024, donnan_jwst_2024, robertson_earliest_2024, tacchella_star_2024, whitler_z_2025}, raising concerns about a potential ``crisis in the ionizing photon budget'' in which reionization appears to reach completion too early, in conflict with existing observations \citep{chakraborty_modelling_2024, munoz_reionization_2024, simmonds_ionizing_2024, bera_towards_2025}. Here we discuss the implications for the evolution of reionization inferred from both the \textsc{thesan-zoom} simulations and our predictions for the JWST/NIRCam-detected JADES sample of \citet{simmonds_bursting_2025}, within the context of current observational constraints.

\subsection{Ionizing emissivity} \label{sec:ionising_emissivity}

The relation obtained from application of our observational $\dot{N}_\mathrm{ion,esc}$ predictor (model D) to the JWST sample is shown in the lower-left panel of Fig.~\ref{fig:observational_fits} and is restated here:
\begin{align} \label{JWST_linear_fit}
    \log_{10}(\dot{N}_\mathrm{ion,esc})
    &= (-0.6197 \pm 0.0007) \, (M_\mathrm{UV} + 20) \nonumber \\
    &+ (0.30 \pm 0.01) \, \log_{10} \! \left( \frac{1 + z}{7} \right) \nonumber \\
    &+ (52.84 \pm 0.02) \, .
\end{align}
We also state the corresponding relation derived from the \textsc{thesan-zoom} simulations (which is analogously presented in Fig.~\ref{fig:thesan_fits_histogram} in Appendix~\ref{sec:appendix_a}):
\begin{align} \label{thesan_linear_fit}
    \log_{10}(\dot{N}_\mathrm{ion,esc})
    &= (-0.670 \pm 0.002) \, (M_\mathrm{UV} + 20) \nonumber \\
    &+ (0.28 \pm 0.03) \, \log_{10} \! \left( \frac{1 + z}{7} \right) \nonumber \\
    &+ (52.84 \pm 0.07) \, .
\end{align}
We note that we have included all galaxies in our catalogue when carrying out this fit, including those with $\mathrm{SFR}_{50} = 0$, which were neglected in the previous sections. This is because, despite typically having low luminosities, these galaxies may still be observable and included in the UVLF, thereby influencing the fitted relation. We additionally restrict the sample to galaxies with $M_\mathrm{UV} < -13$. Our \textsc{thesan-zoom} catalogue includes only systems resolved with at least 100 stellar particles, imposing a stellar-mass completeness limit of $M_* \approx 9 \times 10^5 \, \mathrm{M_\odot}$; with this mass scale roughly corresponding to a UV magnitude of $M_\mathrm{UV} \sim -13$. The sample is therefore incomplete at fainter magnitudes, motivating the aforementioned restriction.

We use the UVLFs from \citet{bouwens_new_2021} for $z \in [1.5, 8.5]$ and from \citet{whitler_z_2025} for $z \in [8.5, 16]$, which we parametrize in the Schechter form \citep{schechter_analytic_1976}. This choice is motivated in Appendix~\ref{sec:appendix_c}, where we compare the ionizing emissivities resulting from the Schechter and Double Power Law (DPL) parametrizations of the UVLF \citep{bowler_lack_2020, whitler_z_2025}, as well as the Schechter parametrization of the stellar mass function \citep{weibel_galaxy_2024}, and demonstrate that the UVLF in Schechter form is most suitable for this work. Uncertainties in our $\dot{n}_{\mathrm{ion}}$ values are primarily driven by the reported uncertainties in these Schechter parameters.

In Fig.~\ref{fig:uv_mag_vs_targets} we see that there is an approximately Gaussian scatter in $\log_{10}(\dot{N}_\mathrm{ion,esc})$ about the median $\dot{N}_\mathrm{ion,esc} - M_\mathrm{UV}$ relation for the \textsc{thesan-zoom} galaxies. Since the logarithm is nonlinear, $\langle \log_{10}\dot{N}_\mathrm{ion,esc} \rangle \neq \log_{10} \langle \dot{N}_\mathrm{ion,esc} \rangle$; consequently, Gaussian scatter in $\log_{10}(\dot{N}_\mathrm{ion,esc})$ boosts the linear-space mean escape rate, as upward fluctuations are weighted more strongly than downward fluctuations. As the relations stated in Eqs.~\eqref{JWST_linear_fit} and \eqref{thesan_linear_fit} trace the mean of $\log_{10}(\dot{N}_\mathrm{ion,esc})$, we therefore multiply them by the standard lognormal correction factor to recover $\langle \dot{N}_\mathrm{ion,esc} \rangle$ for use in Eq.~\eqref{integral_magnitude} \citep{shen_impact_2023, jun_scatter_2025}:
\begin{equation} \label{correction}
    \langle \dot{N}_\mathrm{ion,esc} \rangle = 10^{\frac{1}{2} \ln(10) \sigma_{\dot{N}}^2} \, \dot{N}_\mathrm{ion,esc}.
\end{equation}
Here $\sigma_{\dot{N}}$ is taken to be half the $16^\mathrm{th} - 84^\mathrm{th}$ percentile range of the residuals in the \textsc{thesan-zoom} $\dot{N}_\mathrm{ion,esc} - M_\mathrm{UV}$ relation, which gives $\sigma_{\dot{N}} = 0.725$ dex. We note that this differs from the value of $\sigma_{\dot{N}}$ quoted in Section~\ref{sec:identifying_indicators}, due to the aforementioned exclusion in that section of  \textsc{thesan-zoom} galaxies with $\mathrm{SFR}_{50} = 0$. The validity of assuming a constant scatter in this relation is discussed in Appendix~\ref{sec:appendix_c}. As shown in Fig.~\ref{fig:observational_fits}, our JWST sample predictions of $\dot{N}_\mathrm{ion,esc}$ are very strongly constrained by $M_\mathrm{UV}$, artificially suppressing the intrinsic scatter. We therefore adopt the \textsc{thesan-zoom} value of $\sigma_{\dot{N}}$ for both the \textsc{thesan-zoom}- and JWST-based calculations of $\dot{n}_{\mathrm{ion}}$. We discuss the consequences of different $\sigma_{\dot{N}}$ on reionization histories in Section~\ref{sec:discussion}. 

\begin{figure*}
	\includegraphics[width=\textwidth]{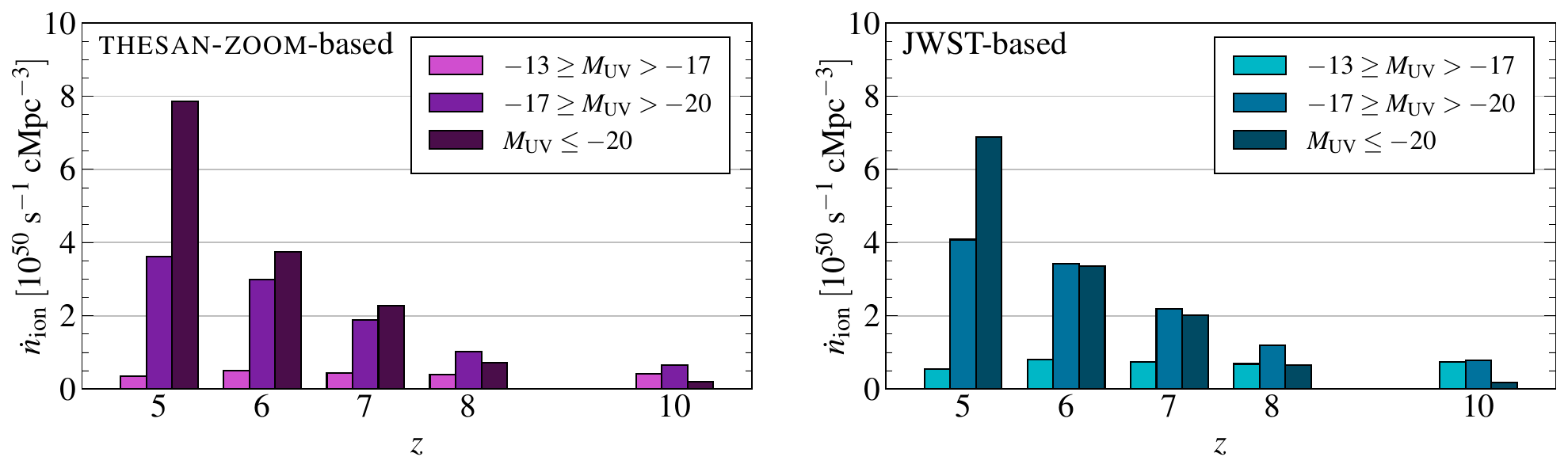}
    \caption{Contributions to the cosmic ionizing emissivity, $\dot{n}_\mathrm{ion}$, from galaxies in three separate UV magnitude bins: $-13 \geq M_\mathrm{UV} > -17$, $-17 \geq M_\mathrm{UV} > -20$ and $M_\mathrm{UV} \leq -20$. \textit{Left panel:} $\dot{n}_\mathrm{ion}$ derived using the \textsc{thesan-zoom}-based $\dot{N}_\mathrm{ion,esc}$ relation (Eq.~\eqref{thesan_linear_fit}), as shown in Fig.~\ref{fig:reionisation_timeline}, decomposed into contributions from the UV magnitude bins. \textit{right panel:}  $\dot{n}_\mathrm{ion}$ derived using the \textsc{JWST}-based $\dot{N}_\mathrm{ion,esc}$ relation (Eq.~\eqref{JWST_linear_fit}), as shown in Fig.~\ref{fig:reionisation_timeline}, decomposed into contributions from the UV magnitude bins. For $z < 8.5$ we adopt the UVLF of \citet{bouwens_new_2021}, which is reported in redshift bins of width $\Delta_z = 1$. For $z > 8.5$ we use a UVLF from \citet{whitler_z_2025}, measured over the broad interval $z \in [8.5, 12]$ and reported at its bin-averaged redshift $\langle z \rangle \approx 10$, hence no emissivity values are shown at $z = 9$ here. At high redshifts, both analyses show an approximately egalitarian picture of reionization, with neither the faintest galaxies nor the brightest galaxies dominating the $\dot{n}_\mathrm{ion}$ budget. However, both histories find that brighter galaxies contribute increasingly more toward lower redshifts, with the $M_\mathrm{UV} \leq -20$ population providing the greatest contribution to the ionizing photon budget by the completion of reionization.}
    \label{fig:reionisation_contributions}
\end{figure*}

Eq.~\eqref{integral_magnitude} converges at the high luminosity limit and is therefore insensitive to the choice of $M_\mathrm{UV,min}$. However, the integral diverges slowly toward faint galaxies, requiring a finite $M_\mathrm{UV,max}$ cutoff when evaluating $\dot{n}_\mathrm{ion}$. Consistent with the completeness limit of our \textsc{thesan-zoom} catalogue discussed above, we adopt $M_\mathrm{UV,max} = -13$, which is in agreement with values assumed in previous studies (e.g. \citet{robertson_cosmic_2015} and \citet{larson_searching_2022}). In Section~\ref{sec:discussion}, we examine the sensitivity of our results to the choice of $M_\mathrm{UV,max}$ and show that our inferred reionization histories depend only weakly on this assumption.

Strong constraints on $\dot{n}_\mathrm{ion}$ have been placed by the Ly$\alpha$ forest at lower redshifts (z < 6); accordingly, we include the observational results of \citet{becker_new_2013}, \citet{gaikwad_measuring_2023} and \citet{davies_constraints_2024} in Fig.~\ref{fig:reionisation_timeline}. From such measurements, $\dot{n}_\mathrm{ion}$ has been observed to flatten at the end of the EoR with a value of $\log_{10}(\dot{n}_\mathrm{ion} \; [\mathrm{s^{-1} \; cMpc^{-3}}]) \approx 51$  at $z \lesssim 5-6$ \citep{becker_new_2013, davies_constraints_2024}. Our \textsc{thesan-zoom}-based and JWST-based $\dot{n}_\mathrm{ion}$ histories agree across all redshifts. Furthermore, both our $\dot{n}_\mathrm{ion}(z)$ histories are consistent with these constraints and with the expected post-reionization flattening. Although early-Universe $\dot{n}_\mathrm{ion}$ values are highly uncertain, we also include the estimates of \citet{rinaldi_midis_2024}, inferred from JWST measurements of strong H$\alpha$ emitters for $z \in [7, 8]$, and find that our results are in good agreement with these as well.

For comparison to the two $\dot{n}_\mathrm{ion}$ curves of this work in Fig.~\ref{fig:reionisation_timeline}, we add two more from \citet{simmonds_ionizing_2024}, derived via \textsc{PROSPECTOR}-inferred $\xi_\mathrm{ion} - M_\mathrm{UV}$ relations of a related JWST sample based on NIRCam imaging of the JADES in the GOODS-South region and the UVLFs of \citet{bouwens_new_2021}. The first assumes a constant LyC escape fraction of $f_\mathrm{esc} = 10\%$, while the second employs the UV continuum slope-dependent empirical $f_\mathrm{esc}$ derived by \citep{chisholm_far-ultraviolet_2022} from a low-redshift sample. The $\dot{n}_\mathrm{ion}$ histories of this work show broad consistency with the results of \citet{simmonds_ionizing_2024} across redshifts relevant to cosmic reionization ($z \gtrsim 5$), with closer agreement to the constant $f_\mathrm{esc}$ curve than to the curve using an $f_\mathrm{esc}$ prescription derived from a low-redshift analogues. We also include the $\dot{n}_\mathrm{ion}$ curve derived by \citet{garaldi_thesan_2022} from the parent \textsc{thesan} simulations for the \textsc{thesan-1} run. While somewhat consistent with our results at high redshifts, the $\dot{n}_\mathrm{ion}$ inferred from \textsc{thesan} diverges from Ly$\alpha$ forest observational constraints at $z \lesssim 6$, particularly the Ly$\alpha$ opacity constraints of \citet{gaikwad_measuring_2023}. The improved resolution and physical modeling of the \textsc{thesan-zoom} simulations could plausibly explain why our results remain consistent with observational constraints, highlighting the advantage of the zoom-in technique for determining the ionizing emissivity. A similar behaviour is seen in the $\dot{n}_\mathrm{ion}$ curve inferred by \citet{naidu_rapid_2020}, based on the empirical galaxy formation model of \citet{tacchella_redshift-independent_2018} with SED-computed ionizing photon production efficiencies and a fitted escape fraction, which likewise departs from the observational constraints of \citet{gaikwad_measuring_2023}. \citet{austin_resolving_2025} measure a constant $\xi_\mathrm{ion}$ with UV magnitude for a JWST/NIRCam sample across $5.6 < z < 6.5$ galaxies, and construct Schechter UVLFs from the same sample. For $M_\mathrm{UV,max} = -13.5$, and adopting the same $f_\mathrm{esc}$ prescriptions as \citet{simmonds_ionizing_2024}, \citet{austin_resolving_2025} obtain $\log_{10}(\dot{n}_\mathrm{ion} \; [\mathrm{s^{-1} \; cMpc^{-3}}]) = 51.0\pm0.2$ assuming $f_\mathrm{esc} = 10\%$ and $\log_{10}(\dot{n}_\mathrm{ion} \; [\mathrm{s^{-1} \; cMpc^{-3}}]) = 50.9\pm0.2$ assuming the $f_\mathrm{esc}$ calibration of \citet{chisholm_far-ultraviolet_2022}. Their results are consistent with our $\dot{n}_\mathrm{ion}$ values at $z \approx 6$.

We include curves in Fig.~\ref{fig:reionisation_timeline} quantifying the critical ionizing emissivity, $\dot{n}_{\mathrm{ion}}^*(z)$, required to balance the IGM recombination rate and maintain the full ionization of hydrogen at $z$. It is given as the ratio of the mean comoving cosmic number density of hydrogen atoms, $\langle n_\mathrm{H} \rangle = 1.88 \cdot 10^{-7} \, \mathrm{cm^{-3}}$ \citep{ade_planck_2016}, to the effective IGM $\mathrm{H}$\textsc{ii} recombination timescale at $z$, $\bar{t}_\mathrm{rec}(z)$ \citep{madau_radiative_1999}. We compute this effective timescale $\bar{t}_\mathrm{rec}$ using the model of \citet{madau_cosmic_2024}:
\begin{equation} \label{timescale}
    \frac{1}{\bar{t}_\mathrm{rec}(z)} = (1 + \chi) \, \alpha_B(T_0) \, \langle n_\mathrm{H} \rangle \, (1 + z)^3 \, C \, ,
\end{equation}
where $\alpha_B = 2.58 \cdot 10^{-13} \, \mathrm{cm^3\,s^{-1}}$ is the recombination coefficient at a fixed temperature of $T_0 = 10^4 \, \mathrm{K}$ \citep{madau_cosmic_2024}, and $\chi \equiv Y/4X = 0.0819$ which assumes helium is only singly ionized at the same time as hydrogen \citep{ade_planck_2016}. Finally, $C$ is the clumping factor of ionized hydrogen and accounts for the density inhomogeneities in the IGM; it is poorly constrained \citep{frenk_cosmological_2000, so_fully_2014}, so we plot curves for clumping factors of 1, 3 and 10 in Fig.~\ref{fig:reionisation_timeline}. Our $\dot{n}_\mathrm{ion}$ results predict that critical emissivity is unlikely to have been reached until after $z \sim 8$, with reionization completing at a much lower redshift; this is supported by our analysis of the evolution of the hydrogen ionization fraction in Section~\ref{sec:ionisation_fraction}. Consequently, this work avoids the overabundance of ionizing photons predicted in recent JWST studies \citep{chakraborty_modelling_2024, munoz_reionization_2024, austin_resolving_2025}, without the need to assume a high clumping factor. We note that \citet{papovich_galaxies_2025} find that a lower average escape fraction than that generally predicted at low redshifts can also reconcile the ionizing photon budget with constraints.

\subsection{Galaxy Population Contributions to Reionization}

We can split our $\dot{n}_\mathrm{ion}(z)$ histories into contributions from different UV magnitude populations to determine which sources drove reionization. For both the \textsc{thesan-zoom}-based and JWST-based analyses, the integration of Eq.~\eqref{integral_magnitude} is performed separately over three $M_\mathrm{UV}$ bins: $-13 \geq M_\mathrm{UV} > -17$, $-17 \geq M_\mathrm{UV} > -20$, and $M_\mathrm{UV} \leq -20$. The resulting contributions to the ionizing photon budget are presented in the left and right panels of Fig.~\ref{fig:reionisation_contributions}, respectively. This figure paints an egalitarian picture of reionization at high redshifts, with neither the faintest nor the brightest galaxies dominating the $\dot{n}_\mathrm{ion}$ budget. However, for $z \lesssim 8$, both analyses convey that brighter galaxies contribute increasingly more, with these luminous giants driving reionization by its completion. The \textsc{thesan-zoom}-based analysis, for example, predicts that the $M_\mathrm{UV} \leq -20$ population provides more than 50\% of the ionizing photon budget by $z \sim 6$. 

These findings are in agreement with the conclusions of \citet{naidu_rapid_2020}, whose $\dot{n}_\mathrm{ion}$ curve is shown in Fig~\ref{fig:reionisation_timeline}, and others \citep{larson_searching_2022, roberts-borsani_nature_2023, yeh_thesan_2023, bera_towards_2025}, but challenges alternative recent studies that attribute reionization primarily to the hidden faint-end population \citep{dayal_reionization_2020, atek_most_2024, mascia_new_2024, simmonds_ionizing_2024}. Consistent with our findings, \citet{bera_towards_2025} argue that a substantial contribution from very bright galaxies, increasing toward lower redshifts, is required to reproduce JWST observations while avoiding a history of reionization that ends too early. Similarly, \citet{jecmen_glimpse_2026} report a subdominant contribution from dim galaxies with $M_\mathrm{UV} > -14$, driven by declining escape fractions toward the faint end, in agreement with our findings.

\citet{simmonds_ionizing_2024} use SED fitting with \textsc{PROSPECTOR} to derive $\dot{n}_\mathrm{ion} - M_\mathrm{UV}$ relations for their related JWST sample; finding slopes in the range $-0.47$ to $-0.39$, which are noticeably more gentle than the values of $-0.6197$ and $-0.670$ inferred in this work (see Eq.~\eqref{JWST_linear_fit} and \eqref{thesan_linear_fit}). The steeper relations obtained here imply a larger ionizing-photon output from brighter galaxies, resulting in a significantly greater role for bright galaxies in reionization.

\subsection{Hydrogen ionization fraction} \label{sec:ionisation_fraction}

As in \citet{wu_effect_2025}, we use our $\dot{n}_\mathrm{ion}(z)$ curves to solve for the volume-averaged hydrogen ionized fraction $Q_\mathrm{HII}$ \citep{madau_radiative_1999}, employing the identity $\mathrm{d}t = -\mathrm{d}z \, (H(z) \, (1+z))^{-1}$ to re-write $Q(t)$ as a function of redshift \citep{wu_effect_2025}:
\begin{align} \label{ionisation_fraction_ode}
    \frac{\mathrm{d}Q_\mathrm{HII}}{\mathrm{d}z} 
    &= - \frac{1}{H(z) \, (1+z)} \, \frac{\mathrm{d}Q_\mathrm{HII}}{\mathrm{d}t} \notag \\
    &= - \frac{1}{H(z) \, (1+z)} \, \left( \frac{\dot{n}_\mathrm{ion}}{\langle n_\mathrm{H} \rangle} - \frac{Q_\mathrm{HII}}{\bar{t}_\mathrm{rec}}\right) \, ,
\end{align}
where we assume a flat cosmology with the cosmological parameters of the Planck Collaboration \citep{ade_planck_2016}. The relevant recombination timescale is given in Eq.~\eqref{timescale} and is dependent on the ionized hydrogen clumping factor $C$. In this analysis, we follow the IGM clumping model used by \citet{madau_cosmic_2024} and \citet{wu_effect_2025}, which employs the parametrization of $C = 9.25 - 7.21\log_{10}(1 + z)$ for the clumping factor, calibrated from the radiation-hydrodynamic simulations of ionization inhomogeneities of \citet{finlator_new_2009} and \citet{chen_scorch_2020}.

\begin{figure}
    \centering
    \includegraphics[width=\linewidth]{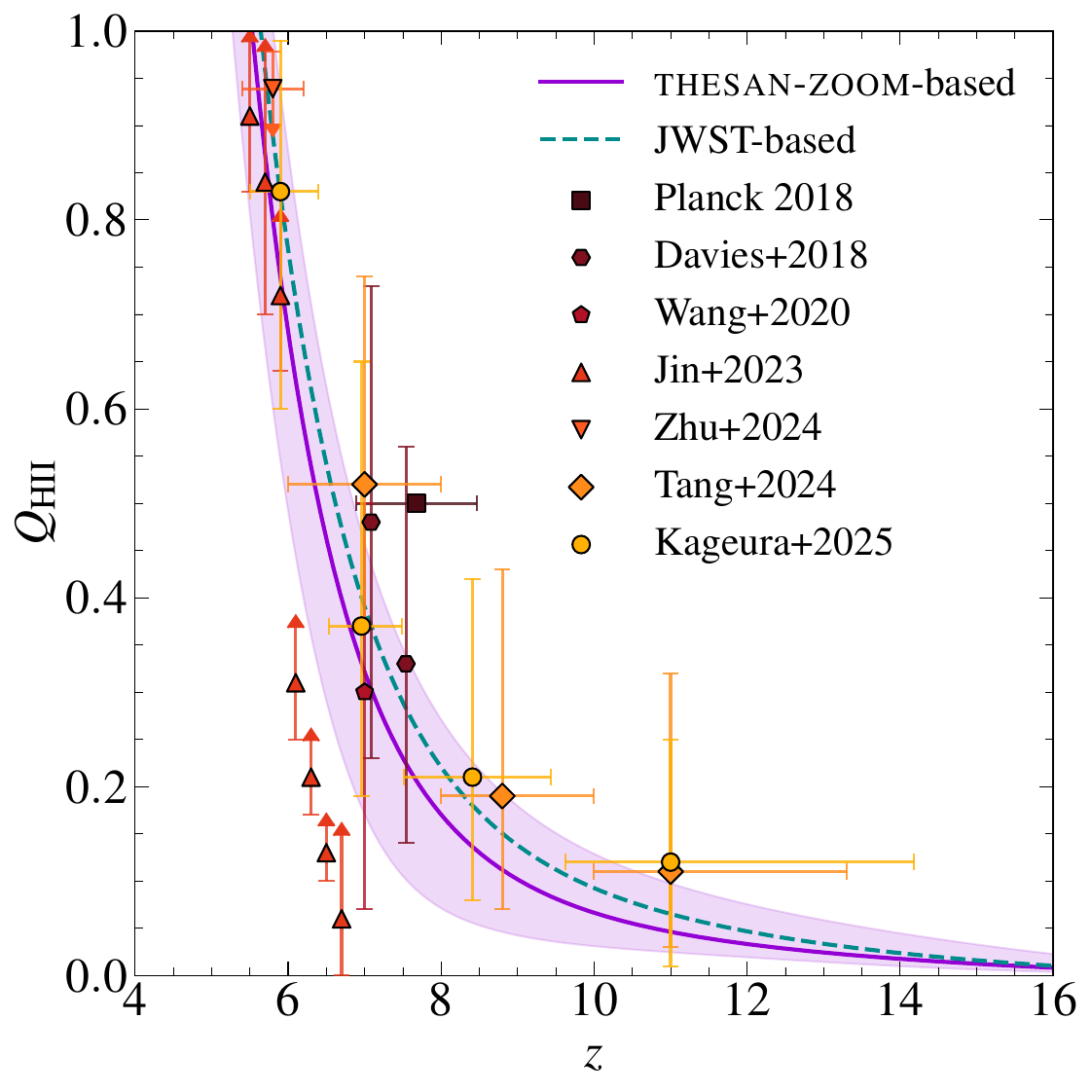}
    \caption{Hydrogen ionized fraction, $Q_\mathrm{HII}$, as a function of redshift, $z$, computed using Eq.~\eqref{ionisation_fraction_ode} together with the \textsc{thesan-zoom}-based (violet curve) and JWST-based (cyan curve) ionizing emissivities from Fig.~\ref{fig:reionisation_timeline}. The shaded violet region indicates the uncertainty in $Q_\mathrm{HII}$ for the \textsc{thesan-zoom}-based analysis (we have neglected the similar JWST-based uncertainty for figure clarity). Observational measurements and limits, represented by points and arrows, are plotted, including constraints from: the CMB \citep{aghanim_planck_2020}, quasar spectrum damping-wing analyses \citep{davies_quantitative_2018, wang_significantly_2020}, Ly$\alpha$ emissions \citep{tang_jwstnirspec_2024, kageura_census_2025} and the combined Ly$\alpha$ and Ly$\beta$ forests \citep{jin_nearly_2023, zhu_damping_2024}. Both the \textsc{thesan-zoom}-based and JWST-based ionization fraction histories are consistent with all these constraints and each other. Furthermore, they both predict a rapid, late reionization process, with $Q_\mathrm{HII}$ likely less than 0.2 at $z = 8$. }
    \label{fig:ionisation_fraction}
\end{figure}

We interpolate both our \textsc{thesan-zoom}-based and JWST-based $\dot{n}_\mathrm{ion}(z)$ histories using weighted cubic splines and integrate Eq.~\eqref{ionisation_fraction_ode} to a given $z$ from a fully neutral Universe $Q_\mathrm{HII}(z_i) = 0$ at $z_i = 20$ \citep{virtanen_scipy_2020}, with the resulting ionization fraction evolutions shown in Fig.~\ref{fig:ionisation_fraction}. We include in this figure observational constraints placed on $Q_\mathrm{HII}$ from the CMB \citep{aghanim_planck_2020}, quasar spectrum damping-wing analyses \citep{davies_quantitative_2018, wang_significantly_2020}, Ly$\alpha$ emissions \citep{tang_jwstnirspec_2024, kageura_census_2025} and the combined Ly$\alpha$ and Ly$\beta$ forests \citep{jin_nearly_2023, zhu_damping_2024}. Our \textsc{thesan-zoom}-based and JWST-based reionization histories are in good agreement with each other and these measurements. We find reionization reaches completion ($Q_\mathrm{HII}(z_f) = 1$) at $z_f = 5.5\pm0.2$ according to the \textsc{thesan-zoom}-based results, while the JWST-based yields a comparable value of $z_f = 5.6\pm0.3$. Both are consistent with current observational evidence, which indicates completion in the range $5 < z_f < 6$ \citep{becker_mean_2021, jin_nearly_2023, davies_constraints_2024, zhu_damping_2024}. We note that AGN are neglected in the \textsc{thesan-zoom} simulations, despite potentially contributing a non-negligible fraction of the ionizing photon budget, as discussed in Section~\ref{sec:introduction}. Importantly, we have again shown that the reionization results derived from our $\dot{N}_\mathrm{ion,esc} - M_\mathrm{UV}$ relations do not suffer from the ionizing photon overabundance reported in other JWST studies \citep{munoz_reionization_2024}.

In both our ionization fraction evolutions, we find $Q_\mathrm{HII}(z=8) \le 0.2$, implying the bulk of reionization occurred rapidly after $z \sim 8$. This is in agreement with other recent works \citep{kulkarni_large_2019, naidu_rapid_2020, choustikov_great_2024, jecmen_glimpse_2026}, with \citet{kulkarni_large_2019} arguing that a rapid evolution is required to reproduce Ly$\alpha$ constraints. \citet{naidu_rapid_2020}, who, as previously discussed, also find that the most luminous galaxies account for the majority of the ionizing photon budget, conclude that the increased population of such massive galaxies at lower redshifts causes this accelerated end to reionization. 

\subsection{Thomson optical depth} \label{sec:optical_depth}

\begin{figure}
    \centering
    \includegraphics[width=\linewidth]{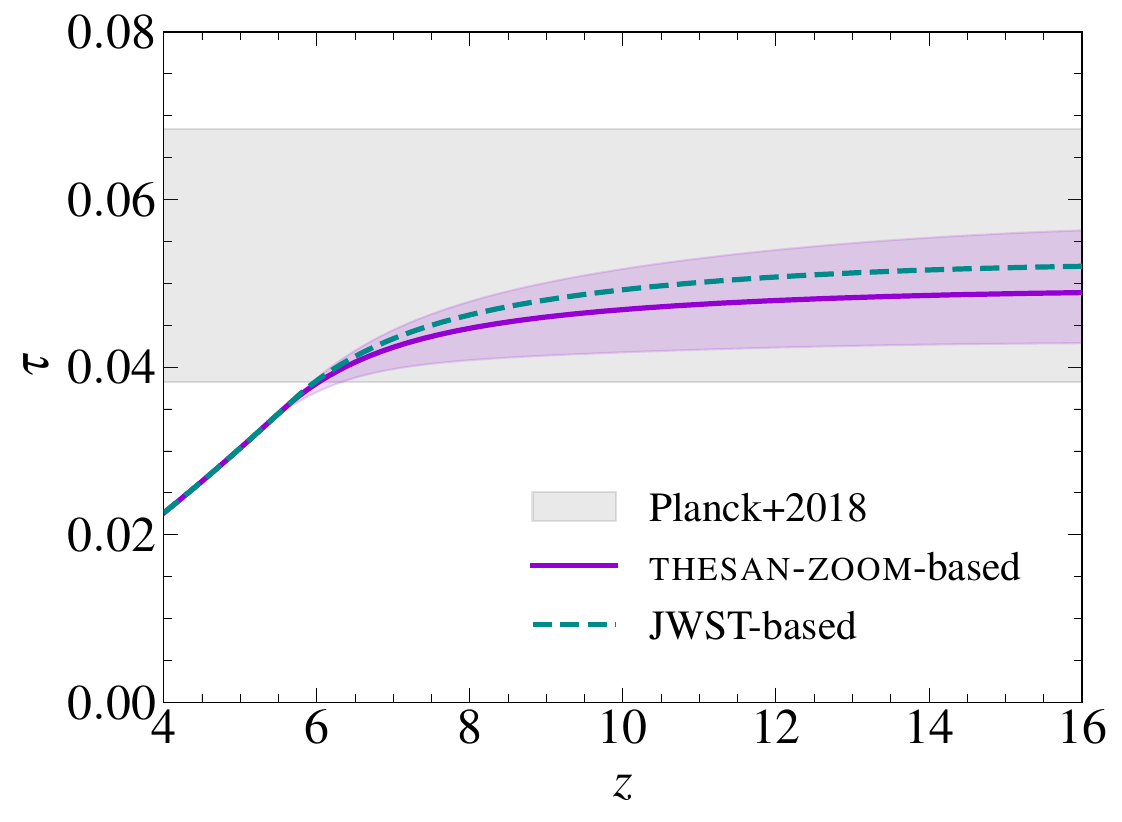}
    \caption{Thomson optical depth, $\tau$, as a function of redshift, $z$, computed using Eq.~\eqref{tau} together with the \textsc{thesan-zoom}-based (violet curve) and JWST-based (cyan based) $Q_\mathrm{HII}$ histories from Fig.~\ref{fig:ionisation_fraction}. The shaded violet region indicates the uncertainty in $\tau$ for the \textsc{thesan-zoom}-based analysis (we have neglected the similar JWST-based uncertainty for figure clarity). The grey band shows the 2$\sigma$ range of the opacity constraints measured by The 2018 Planck collaboration \citep{aghanim_planck_2020}. Both the \textsc{thesan-zoom}-based and JWST-based $\tau(z)$ asymptote well within this range. Thus, the optical depths inferred from our $\dot{N}_\mathrm{ion,esc} - M_\mathrm{UV}$ relations are consistent with constraints from the CMB.}
    \label{fig:optical_depth}
\end{figure}

The Thomson optical depth, $\tau$, quantifies the scattering of CMB photons by free electrons produced during the EoR; it therefore provides a useful constraint on reionization. Using the hydrogen ionized fractions $Q_\mathrm{HII}(z)$ shown in Fig.~\ref{fig:ionisation_fraction}, $\tau(z)$ can be computed as \citep{wu_effect_2025}:

\begin{equation} \label{tau}
    \tau(z) = c \, \sigma_T \, \langle n_\mathrm{H} \rangle \, \int^{z}_{0} (1 + \eta \, \chi) \, \frac{(1 + z')^2}{H(z')} \, Q_\mathrm{HII}(z') \, \mathrm{d}z' \, ,
\end{equation}
where $c$ is the speed of light, $\sigma_T$ is the Thomson scattering cross-section, and the prefactor $\eta$ accounts for the free electron contribution from helium, assuming helium is singly ionized for $z > 4$ ($\eta = 1)$ and fully ionized at lower redshifts ($\eta = 2$). 

Fig.~\ref{fig:optical_depth} shows $\tau(z)$ derived from both our \textsc{thesan-zoom}-based and JWST-based $Q_\mathrm{HII}(z)$ curves. We compare the high-redshift asymptotes of our predicted optical depths with the measurement $\tau_\mathrm{CMB} = 0.054^{\,+0.014}_{-0.016}$ ($2\sigma$ uncertainties) from the Planck Collaboration \citep{aghanim_planck_2020}. Some recent studies have noted a potential tension between the baryon acoustic oscillation and the Planck CMB measurements within the $\Lambda$CDM framework that could be alleviated if $\tau_\mathrm{CMB}$ were underpredicted \citep{jhaveri_turning_2025, liu_phantom_2025, sailer_dispuable_2025}. However, it has been shown that the previously discussed constraints on the hydrogen neutral fraction from the Ly$\alpha$ forest and damping-wing absorption, together with limits on the patchy kinematic Sunyaev–Zel’dovich signal, which is a CMB probe of inhomogeneous reionization \citep{chen_patchy_2023}, require reionization histories that yield optical depths consistent with the Planck value \citep{cain_cosmic_2025, kageura_new_2026}. 

Both optical depths derived in this work (from the $\dot{N}_\mathrm{ion,esc} - M_\mathrm{UV}$ relations found in \textsc{thesan-zoom} and our JWST sample $\dot{N}_\mathrm{ion,esc}$ predictions) are consistent with this CMB constraint, with uncertainty ranges that closely overlap those of the Planck result. This agreement further supports both the validity of the \textsc{thesan-zoom} simulations and the predictive models we presented in Section~\ref{sec:prediction}.

\subsection{Uncertainties in the Reionization Histories} \label{sec:discussion}

\begin{figure}
    \centering
    \includegraphics[width=\linewidth]{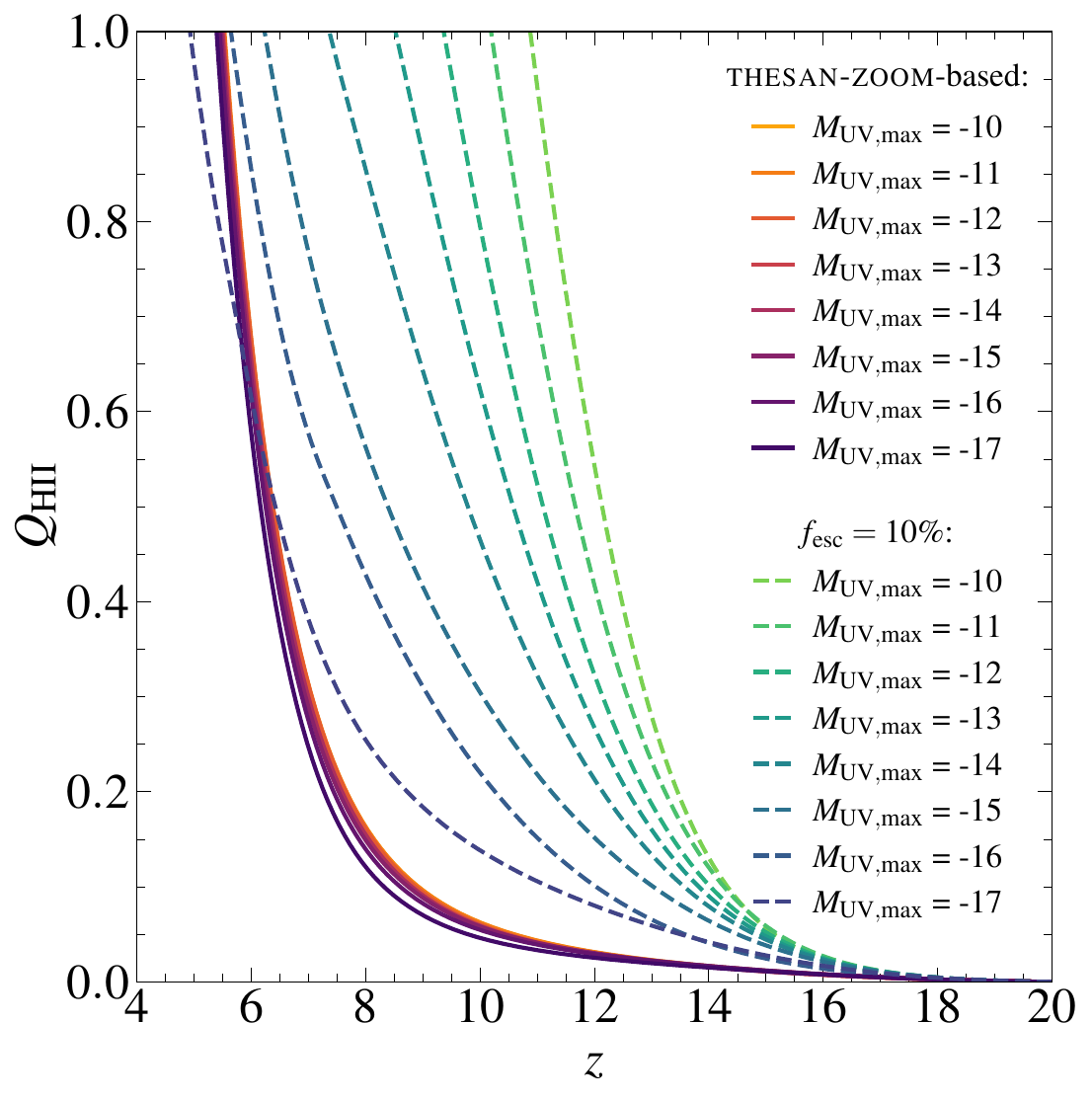}
    \caption{Hydrogen ionized fraction, $Q_\mathrm{HII}$, as a function of redshift, $z$, calculated from ionizing emissivities using different choices of the faint-end cutoff in the integral of Eq.~\eqref{integral_magnitude}, with $M_\mathrm{UV,max}$ spanning $[-10, -17]$. Solid lines are derived from the \textsc{thesan-zoom}-based $\dot{N}_\mathrm{ion,esc} - M_\mathrm{UV}$ relation given in Eq.~\ref{thesan_linear_fit}, while dashed lines are obtained assuming a constant ionizing photon production efficiency of $\log_{10}(\xi_\mathrm{ion} \mathrm{[Hz \, erg^{-1}]}) = 25.05$ \citep{austin_resolving_2025}, and a constant ionizing photon escape fraction of $f_\mathrm{esc} = 10\%$. We find that the evolution of reionization in this work is largely insensitive to the choice of the faint-end cutoff. However, this choice has a substantial impact on both the evolution and completion of reionization in models that assume a constant ionizing photon production efficiency and a constant ionizing photon escape fraction.}
    \label{fig:cut_off_comparison}
\end{figure}

In computing the ionizing emissivities of Fig.~\ref{fig:reionisation_timeline}, we adopt $M_\mathrm{UV,max} = -13$ as the faint-end cutoff of Eq.~\eqref{integral_magnitude}. This choice was well motivated as the \textsc{thesan-zoom} catalogue is approximately complete down to this luminosity. However, the selection of faint-end cutoffs varies across the literature \citep{naidu_rapid_2020, larson_searching_2022, simmonds_ionizing_2024, papovich_galaxies_2025, wu_effect_2025, austin_resolving_2025}. To investigate the impact of the choice of $M_\mathrm{UV,max}$ on our reionization histories, we repeat the \textsc{thesan-zoom}-based computation of the hydrogen ionized fraction, $Q_\mathrm{HII}$, shown in Fig.~\ref{fig:ionisation_fraction}, using a range of $\dot{n}_\mathrm{ion}$ histories calculated with faint-end cutoffs $M_\mathrm{UV,max}$ spanning $[-10, -17]$. We find that the evolution of reionization is largely insensitive to the choice of this integration limit, with changes in $M_\mathrm{UV,max}$ of $\pm 3$ magnitudes about the adopted value of $-13$ resulting in a shift in the redshift of reionization completion of less than $0.1$. This behaviour likely reflects the relatively minor contribution of the more numerous populations of faint galaxies to the overall ionizing photon budget inferred in this work. For comparison, we plot the evolution of $Q_\mathrm{HII}$ across the same range of $M_\mathrm{UV,max}$, for a model that assumes a constant ionizing photon production efficiency of $\log_{10}(\xi_\mathrm{ion} \mathrm{[Hz \, erg^{-1}]}) = 25.05)$ \citep{austin_resolving_2025}, and a constant ionizing photon escape fraction of $f_\mathrm{esc} = 10\%$, such that $\dot{N}_\mathrm{ion,esc}$ depends solely on galactic UV luminosity. We see that the evolution of reionization for such a model is highly dependent on the choice of faint-end cutoff. Such assumptions of a constant escape fraction are not uncommon in recent studies \citep{simmonds_low-mass_2024, simmonds_ionizing_2024, wu_effect_2025, austin_resolving_2025}; consequently, ionizing emissivities inferred under assumptions of a constant escape fraction and a relatively bright faint-end cutoff should be interpreted as lower limits, as the contribution from the faintest galaxies is neglected. By contrast, the weak sensitivity to $M_\mathrm{UV,max}$ in our framework gives us confidence that our inferred $\dot{n}_\mathrm{ion}$ values are not systematically underpredicted.

\begin{figure}
    \centering
    \includegraphics[width=\linewidth]{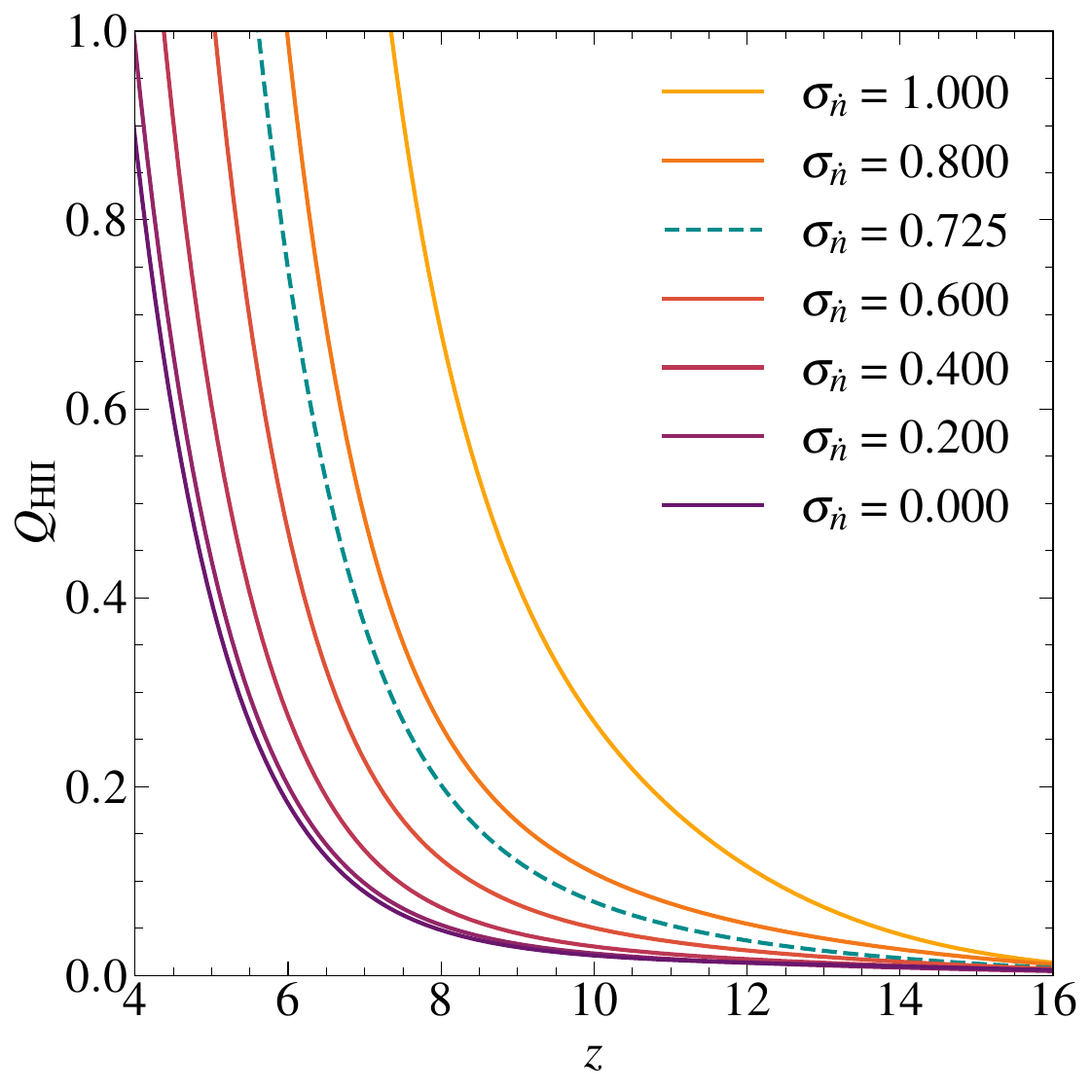}
    \caption{Hydrogen ionized fraction, $Q_\mathrm{HII}$, as a function of redshift, $z$, for JWST-based $\dot{n}_\mathrm{ion}$ histories. The unbroken curves show results obtained by assuming $\sigma_{\dot{N}} \in [0.000, 1.000]$ dex, corresponding to a range of possible scatter in the JWST-based $\dot{N}_\mathrm{ion,esc} - M_\mathrm{UV}$ relation of Eq.~\eqref{JWST_linear_fit}, when applying the lognormal correction factor described in Section~\ref{sec:ionising_emissivity}. The fiducial curve, with $\sigma_{\dot{N}} = 0.725$ corresponding to the scatter in the \textsc{thesan-zoom} relation, is also included in cyan. Even modest increases in stochasticity produce significant shifts in the redshift at which reionization is predicted to complete.}
    \label{fig:sigma_comparison}
\end{figure}

In Section~\ref{sec:ionising_emissivity} we discussed how scatter in the $\dot{N}_\mathrm{ion,esc} - M_\mathrm{UV}$ relation introduces a systematic offset between the derived mean $\log_{10}(\dot{N}_\mathrm{ion,esc})$ relation and the mean ionizing photon escape rate, $\langle \dot{N}_\mathrm{ion,esc} \rangle$, requiring a correction factor when inferring $\dot{n}_\mathrm{ion}$. We do not have direct access to the intrinsic scatter of the $\dot{N}_\mathrm{ion,esc} - M_\mathrm{UV}$ relation of the JWST sample, since the $\dot{N}_\mathrm{ion,esc}$ values have been predicted using our feature-compatible  $\dot{N}_\mathrm{ion,esc}$ model (Model D) presented in Section~\ref{sec:prediction}. Accordingly, our JWST-based reionization analysis assumed a scatter of $\sigma_{\dot{N}} = 0.725$ dex, equal to that of the \textsc{thesan-zoom} relation. We investigate the significance of $\dot{N}_\mathrm{ion,esc} - M_\mathrm{UV}$ intrinsic scatter to the volume-averaged hydrogen ionization fraction $Q_\mathrm{HII}$ in Fig.~\ref{fig:sigma_comparison}. The $Q_\mathrm{HII}(z)$ curves shown in this figure are derived using the same JWST-based $\dot{n}_\mathrm{ion}(z)$ framework as in Fig.~\ref{fig:ionisation_fraction}, but with the assumed scatter of the $\dot{N}_\mathrm{ion,esc} - M_\mathrm{UV}$ JWST sample relation varied from 0.000 to 1.000 dex in steps of 0.200 dex. Fig.~\ref{fig:sigma_comparison} reveals that even modest increases in relation variability can noticeably change the evolution of the EoR and significantly shift the redshift at which reionization completes, with $z_f$ increasing from 3.8 to 7.3 over the $\sigma_{\dot{N}}$ range. In this work, we find the $\dot{N}_\mathrm{ion,esc} - M_\mathrm{UV}$ relation to be steep, and that reionization is driven primarily by UV-bright galaxies. In this regime, upward fluctuations toward higher $log_{10}(\dot{N}_\mathrm{ion,esc})$ values can substantially increase a galaxy's ionizing photon emission, while the steep decline in galaxy number density toward the bright end of the UVLF provides many more intermediate-luminosity galaxies that can scatter into high $\dot{N}_\mathrm{ion,esc}$ values than vice versa \citep{ren_brightest_2019}. Consequently, increasing the intrinsic scatter strongly boosts the abundance of strong LyC emitters, leading to the large changes in $\dot{n}_\mathrm{ion}$ seen in Fig.~\ref{fig:sigma_comparison}. This reinforces the importance of accounting for intrinsic scatter when calculating $\dot{n}_\mathrm{ion}$ in reionization studies.

For the hydrogen recombination timescale in our $Q_\mathrm{HII}$ analysis, we have followed \citet{wu_effect_2025} in adopting the redshift-dependent clumping factor $ C$ parametrization calibrated by \citet{finlator_new_2009} and \citet{chen_scorch_2020}. However, the clumping factor is known to be poorly constrained \citep{frenk_cosmological_2000, so_fully_2014} and this parametrization should be interpreted with appropriate caution. To investigate how $C$ affects reionization, we again integrate Eq.~\eqref{ionisation_fraction_ode} using the \textsc{thesan-zoom}-based $\dot{n}_\mathrm{ion}(z)$ curve of Fig.~\ref{fig:reionisation_timeline} (as done in Fig.~\ref{fig:ionisation_fraction}), but now assume a constant IGM clumping factor of $C =$ 1, 3, or 10. This yields reionization completion at $z_f =$ $5.8\pm0.3$, $5.5\pm0.2$, and $5.0\pm0.2$, respectively. As expected, a higher clumping factor delays reionization, as a larger number of ionizing photons is required to reionize the IGM \citep{wu_effect_2025}. A perfectly uniform and homogeneous IGM ($C = 1$) is unrealistic, and \citet{wu_effect_2025} suggests that an upper bound of $C \leq 10$ for all redshifts is likely an overestimate. Consequently, while the clumping factor is clearly relevant, its effect on reionization is less significant than that associated with variability in the $\dot{N}_\mathrm{ion,esc} - M_\mathrm{UV}$ relation.

Most recent studies of the EoR rely on a combination of models for both $f_\mathrm{esc}$ and either $\dot{N}_\mathrm{ion}$ or $\xi_\mathrm{ion}$ for predictions of the ionizing emissivity, $\dot{n}_\mathrm{ion}$, the hydrogen ionization fraction, $Q_\mathrm{HII}$, and the Thomson optical depth, $\tau$. Many find that their predictions are highly sensitive to assumptions regarding the dependence of $f_\mathrm{esc}$ and $\xi_\mathrm{ion}$ on galaxy properties \citep{price_reconstructing_2016, naidu_rapid_2020, chisholm_far-ultraviolet_2022, kostyuk_ionizing_2023, simmonds_low-mass_2024, simmonds_ionizing_2024, chakraborty_modelling_2024, wu_effect_2025, austin_resolving_2025}. Moreover, the challenges associated with reliably estimating $f_\mathrm{esc}$ have been discussed throughout this paper, and the predictive performance of our $f_\mathrm{esc}$ models is lower than that of our $\dot{N}_\mathrm{ion,esc}$ models. In contrast to previous works, we derive our reionization histories directly from linear fits to $\log_{10}(\dot{N}_\mathrm{ion,esc})$, thereby avoiding the need to separately model the  poorly constrained quantities $f_\mathrm{esc}$ and $\xi_\mathrm{ion}$. This results in a simpler framework that nonetheless remains consistent with all the observational constraints on $\dot{n}_\mathrm{ion}$, $Q_\mathrm{HII}$, and $\tau$ shown in Figs.~\ref{fig:reionisation_timeline}, \ref{fig:ionisation_fraction}, and \ref{fig:optical_depth}.

\section{Conclusions} \label{sec:conculsions}

We have combined high-resolution cosmological radiation-hydrodynamic simulations with the random forest machine learning technique to investigate the history of cosmic reionization predicted by state-of-the-art theoretical models in the context of the latest observational constraints. We have trained random forest regressor models on \textsc{colt}-forward-modelled observables derived from over 35\,000 galaxy realisations spanning $3 < z < 16$ in the \textsc{thesan-zoom} simulations, to predict both $f_\mathrm{esc}$ (Model A) and $\dot{N}_\mathrm{ion,esc}$ (Model B). Our main findings are as follows:

\begin{itemize}
    \item \textbf{Ionizing escape governed by bursty star formation.} We find the 10–to–100$\,$Myr star-formation rate ratio ($\mathrm{SFR}_{10} / \mathrm{SFR}_{100}$) and the gas-to-stellar mass ratio ($M_\mathrm{gas} / M_*$) to be the most important predictors of $f_\mathrm{esc}$. We interpret the leading diagnostic as a signature of bursty star-formation cycles in which feedback-driven gas outflows clear the ISM and facilitate LyC escape. Interestingly, $M_\mathrm{UV}$ is the third most important variable, likely because it traces total stellar mass, recent star-formation rate, and dust content, all of which are relevant to ionizing photon escape.
    \item \textbf{Distinct predictors for $f_\mathrm{esc}$ and $\dot{N}_\mathrm{ion,esc}$.} In contrast to $f_\mathrm{esc}$, we find that $M_\mathrm{UV}$ dominates $\dot{N}_\mathrm{ion,esc}$ predictions, with star-formation rates (both $\mathrm{SFR}_{10}$ and $\mathrm{SFR}_{100}$) also relevant. This reflects the important role of the scaling of $\dot{N}_\mathrm{ion,esc}$ with stellar photon production as a galaxy grows and contains more stars. $M_\mathrm{UV}$ is therefore particularly powerful as it encapsulates not just information about this scaling, but also about the ionizing escape as discussed above.
    \item \textbf{Probing the sources of reionization seen by  JWST.} We have created two further random forest models to predict $f_\mathrm{esc}$ (Model C) and $\dot{N}_\mathrm{ion,esc}$ (Model D) with exclusively photometrically-accessible galaxy properties. We have applied these models to a catalogue of JWST/NIRCam-detected galaxies to predict their ionizing properties, enabling us to derive the JWST-based $M_\mathrm{UV}$--$\dot{N}_\mathrm{ion,esc}$ relation for this observed sample. We find a steeper slope than in previous works, implying a larger contribution from bright galaxies to the cosmic ionizing photon budget.
    \item \textbf{Reionization models consistent with observations.} We use both the  JWST-based $M_\mathrm{UV}$--$\dot{N}_\mathrm{ion,esc}$ relation and the relation directly measured from the simulations (\textsc{thesan-zoom}-based) combined with observed UVLFs to obtain the cosmic ionizing emissivity, $\dot{n}_{\mathrm{ion}}$. When used to construct reionization histories, we find that both the  JWST-based and \textsc{thesan-zoom}-based models are consistent with observational constraints, with reionization completing at $z = 5.6\pm0.3$ and $5.5\pm0.2$, respectively, and avoid the ``crisis in the ionizing photon budget'' that has plagued some recent JWST studies.
    \item \textbf{Late reionization driven by bright sources.} Our results indicate that the bulk of reionization occurs rapidly after $z \sim 8$, driven by the most luminous galaxies, which dominate the cosmic reionization budget for $z \leq 8$.
    \item \textbf{Robustness to the poorly constrained UVLF faint-end.} Due to the dominance of bright galaxies, our model is remarkably robust to variations in the UVLF integration limits, in contrast to models of reionization dominated by faint dwarfs which are sensitive to the shape of the poorly-constrained faint end of the UVLF. However, we find that the intrinsic scatter of the $M_\mathrm{UV}$--$\dot{N}_\mathrm{ion,esc}$ relation has a strong impact on the timing of reionization.
\end{itemize}

Our work highlights the use of machine learning techniques to apply physical insights from smaller, high-resolution simulations to large statistical samples of galaxies. Furthermore, we have demonstrated the utility of directly modelling $\dot{N}_\mathrm{ion,esc}$ for statistical samples of galaxies, thereby avoiding the need to separately model the poorly constrained quantities $f_\mathrm{esc}$ and $\xi_\mathrm{ion}$.

\section*{Acknowledgements}

This work was initially undertaken as part of a Part III Project at the Cavendish Laboratory. 
The authors gratefully acknowledge the Gauss Centre for Supercomputing e.V. (\url{www.gauss-centre.eu}) for funding this project by providing computing time on the GCS Supercomputer SuperMUC-NG at Leibniz Supercomputing Centre (\url{www.lrz.de}), under project pn29we. WM thanks the Science and Technology Facilities Council (STFC) Centre for Doctoral Training (CDT) in Data Intensive Science at the University of Cambridge (STFC grant number 2742968) for a PhD studentship. WM and ST acknowledge support by the Royal Society Research Grant G125142. AS acknowledges support through JWST AR-08709. JB is supported the National Science Foundation (Award number: 2513426). RK acknowledges support of the Natural Sciences and Engineering Research Council of Canada (NSERC) through a Discovery Grant and a Discovery Launch Supplement (funding reference numbers RGPIN-2024-06222 and DGECR-2024-00144) and York University's Global Research Excellence Initiative. ALD thanks the University of Cambridge Harding Distinguished Postgraduate Scholars Programme and the Science and Technology Facilities Council (STFC) Centre for Doctoral Training (CDT) in Data intensive science at the University of Cambridge (STFC grant number 2742605) for a PhD studentship. ALD acknowledges support by the Royal Society Research Grant G125142. EG is supported by the JSPS KAKENHI grant ILR 23K20035. XS acknowledges the support from the National Aeronautics and Space Administration (NASA) theory grant JWST-AR-04814. LK acknowledges the support of a Royal Society University Research Fellowship (grant number URF$\backslash$R1$\backslash$251793).
Support for OZ was provided by Harvard University through
the Institute for Theory and Computation Fellowship.

\section*{Data Availability}

All simulation data, including snapshots, group, and subhalo catalogs and merger trees will be made publicly available in the near future. Data will be distributed via \url{www.thesan-project.com}. Before the public data release, data underlying this article will be shared on reasonable request to the corresponding authors.



\bibliographystyle{mnras}
\bibliography{reionisation_bibliography}

@misc{choustikov_inferring_2024,
	title = {Inferring the {Ionizing} {Photon} {Contributions} of {High}-{Redshift} {Galaxies} to {Reionization} with {JWST} {NIRCam} {Photometry}},
	url = {http://arxiv.org/abs/2405.09720},
	doi = {10.48550/arXiv.2405.09720},
	urldate = {2024-10-31},
	publisher = {arXiv},
	author = {Choustikov, Nicholas and Stiskalek, Richard and Saxena, Aayush and Katz, Harley and Devrient, Julien and Slyz, Adrianne},
	month = may,
	year = {2024},
	note = {arXiv:2405.09720},
}

@article{choustikov_physics_2024,
	title = {The {Physics} of {Indirect} {Estimators} of {Lyman} {Continuum} {Escape} and their {Application} to {High}-{Redshift} {JWST} {Galaxies}},
	volume = {529},
	issn = {0035-8711},
	url = {https://ui.adsabs.harvard.edu/abs/2024MNRAS.529.3751C},
	doi = {10.1093/mnras/stae776},
	urldate = {2024-10-31},
	journal = {Monthly Notices of the Royal Astronomical Society},
	publisher = {OUP},
	author = {Choustikov, Nicholas and Katz, Harley and Saxena, Aayush and Cameron, Alex J. and Devriendt, Julien and Slyz, Adrianne and Rosdahl, Joki and Blaizot, Jeremy and Michel-Dansac, Leo},
	month = apr,
	year = {2024},
	note = {ADS Bibcode: 2024MNRAS.529.3751C},
	pages = {3751--3767},
}

@article{flury_low-redshift_2022,
	title = {The {Low}-redshift {Lyman} {Continuum} {Survey}. {I}. {New}, {Diverse} {Local} {Lyman} {Continuum} {Emitters}},
	volume = {260},
	issn = {0067-0049},
	url = {https://ui.adsabs.harvard.edu/abs/2022ApJS..260....1F},
	doi = {10.3847/1538-4365/ac5331},
	urldate = {2024-11-13},
	journal = {The Astrophysical Journal Supplement Series},
	publisher = {IOP},
	author = {Flury, Sophia R. and Jaskot, Anne E. and Ferguson, Harry C. and Worseck, Gábor and Makan, Kirill and Chisholm, John and Saldana-Lopez, Alberto and Schaerer, Daniel and McCandliss, Stephan and Wang, Bingjie and Ford, N. M. and Heckman, Timothy and Ji, Zhiyuan and Giavalisco, Mauro and Amorin, Ricardo and Atek, Hakim and Blaizot, Jeremy and Borthakur, Sanchayeeta and Carr, Cody and Castellano, Marco and Cristiani, Stefano and De Barros, Stephane and Dickinson, Mark and Finkelstein, Steven L. and Fleming, Brian and Fontanot, Fabio and Garel, Thibault and Grazian, Andrea and Hayes, Matthew and Henry, Alaina and Mauerhofer, Valentin and Micheva, Genoveva and Oey, M. S. and Ostlin, Goran and Papovich, Casey and Pentericci, Laura and Ravindranath, Swara and Rosdahl, Joakim and Rutkowski, Michael and Santini, Paola and Scarlata, Claudia and Teplitz, Harry and Thuan, Trinh and Trebitsch, Maxime and Vanzella, Eros and Verhamme, Anne and Xu, Xinfeng},
	month = may,
	year = {2022},
	note = {ADS Bibcode: 2022ApJS..260....1F},
	pages = {1},
}

@article{choustikov_great_2024,
	title = {The great escape: understanding the connection between {Ly} α emission and {LyC} escape in simulated {JWST} analogues},
	volume = {532},
	issn = {0035-8711},
	shorttitle = {The great escape},
	url = {https://ui.adsabs.harvard.edu/abs/2024MNRAS.532.2463C},
	doi = {10.1093/mnras/stae1586},
	urldate = {2024-11-13},
	journal = {Monthly Notices of the Royal Astronomical Society},
	publisher = {OUP},
	author = {Choustikov, Nicholas and Katz, Harley and Saxena, Aayush and Garel, Thibault and Devriendt, Julien and Slyz, Adrianne and Kimm, Taysun and Blaizot, Jeremy and Rosdahl, Joki},
	month = aug,
	year = {2024},
	note = {ADS Bibcode: 2024MNRAS.532.2463C},
	pages = {2463--2484},
}

@misc{kannan_introducing_2022,
	title = {Introducing the {THESAN} project: radiation-magnetohydrodynamic simulations of the epoch of reionization},
	shorttitle = {Introducing the {THESAN} project},
	url = {http://arxiv.org/abs/2110.00584},
	doi = {10.48550/arXiv.2110.00584},
	urldate = {2024-11-19},
	publisher = {arXiv},
	author = {Kannan, R. and Garaldi, E. and Smith, A. and Pakmor, R. and Springel, V. and Vogelsberger, M. and Hernquist, L.},
	month = apr,
	year = {2022},
	note = {arXiv:2110.00584},
}

@article{vogelsberger_cosmological_2020,
	title = {Cosmological simulations of galaxy formation},
	volume = {2},
	url = {https://ui.adsabs.harvard.edu/abs/2020NatRP...2...42V},
	doi = {10.1038/s42254-019-0127-2},
	urldate = {2024-11-19},
	journal = {Nature Reviews Physics},
	author = {Vogelsberger, Mark and Marinacci, Federico and Torrey, Paul and Puchwein, Ewald},
	month = jan,
	year = {2020},
	note = {ADS Bibcode: 2020NatRP...2...42V},
	pages = {42--66},
}

@misc{yeh_thesan_2023,
	title = {The {THESAN} project: ionizing escape fractions of reionization-era galaxies},
	shorttitle = {The {THESAN} project},
	url = {http://arxiv.org/abs/2205.02238},
	doi = {10.48550/arXiv.2205.02238},
	urldate = {2024-11-28},
	publisher = {arXiv},
	author = {Yeh, Jessica Y.-C. and Smith, Aaron and Kannan, Rahul and Garaldi, Enrico and Vogelsberger, Mark and Borrow, Josh and Pakmor, Rüdiger and Springel, Volker and Hernquist, Lars},
	month = jan,
	year = {2023},
	note = {arXiv:2205.02238},
}

@article{smith_physics_2022,
	title = {The physics of {Lyman}-α escape from disc-like galaxies},
	volume = {517},
	issn = {0035-8711},
	url = {https://ui.adsabs.harvard.edu/abs/2022MNRAS.517....1S},
	doi = {10.1093/mnras/stac2641},
	urldate = {2024-12-04},
	journal = {Monthly Notices of the Royal Astronomical Society},
	publisher = {OUP},
	author = {Smith, Aaron and Kannan, Rahul and Tacchella, Sandro and Vogelsberger, Mark and Hernquist, Lars and Marinacci, Federico and Sales, Laura V. and Torrey, Paul and Li, Hui and Yeh, Jessica Y. -C. and Qi, Jia},
	month = nov,
	year = {2022},
	note = {ADS Bibcode: 2022MNRAS.517....1S},
	pages = {1--27},
}

@article{tacchella_h_2022,
	title = {H α emission in local galaxies: star formation, time variability, and the diffuse ionized gas},
	volume = {513},
	issn = {0035-8711},
	shorttitle = {H α emission in local galaxies},
	url = {https://ui.adsabs.harvard.edu/abs/2022MNRAS.513.2904T},
	doi = {10.1093/mnras/stac818},
	urldate = {2024-12-04},
	journal = {Monthly Notices of the Royal Astronomical Society},
	publisher = {OUP},
	author = {Tacchella, Sandro and Smith, Aaron and Kannan, Rahul and Marinacci, Federico and Hernquist, Lars and Vogelsberger, Mark and Torrey, Paul and Sales, Laura and Li, Hui},
	month = jun,
	year = {2022},
	note = {ADS Bibcode: 2022MNRAS.513.2904T},
	pages = {2904--2929},
}

@article{maheson_unravelling_2024,
	title = {Unravelling the dust attenuation scaling relations and their evolution},
	volume = {527},
	issn = {0035-8711},
	url = {https://doi.org/10.1093/mnras/stad3685},
	doi = {10.1093/mnras/stad3685},
	number = {3},
	urldate = {2025-02-04},
	journal = {Monthly Notices of the Royal Astronomical Society},
	author = {Maheson, Gabriel and Maiolino, Roberto and Curti, Mirko and Sanders, Ryan and Tacchella, Sandro and Sandles, Lester},
	month = jan,
	year = {2024},
	pages = {8213--8233},
}

@article{heinrich_reionization_2021,
	title = {Reionization effective likelihood from {Planck} 2018 data},
	volume = {104},
	url = {https://link.aps.org/doi/10.1103/PhysRevD.104.063505},
	doi = {10.1103/PhysRevD.104.063505},
	number = {6},
	urldate = {2025-04-12},
	journal = {Physical Review D},
	publisher = {American Physical Society},
	author = {Heinrich, Chen and Hu, Wayne},
	month = sep,
	year = {2021},
	pages = {063505},
}

@article{yoo_origin_2020,
	title = {On the origin of low escape fractions of ionizing radiation from massive star-forming galaxies at high redshift},
	volume = {499},
	issn = {0035-8711},
	url = {https://doi.org/10.1093/mnras/staa3187},
	doi = {10.1093/mnras/staa3187},
	number = {4},
	urldate = {2025-04-11},
	journal = {Monthly Notices of the Royal Astronomical Society},
	author = {Yoo, Taehwa and Kimm, Taysun and Rosdahl, Joakim},
	month = nov,
	year = {2020},
	pages = {5175--5193},
}

@misc{price_reconstructing_2016,
	title = {Reconstructing the redshift evolution of escaped ionizing flux from early galaxies with {Planck} and {HST} observations},
	url = {http://arxiv.org/abs/1605.03970},
	doi = {10.48550/arXiv.1605.03970},
	urldate = {2025-04-11},
	publisher = {arXiv},
	author = {Price, Layne C. and Trac, Hy and Cen, Renyue},
	month = may,
	year = {2016},
	note = {arXiv:1605.03970 [astro-ph]},
}

@article{inoue_monte_2008,
	title = {A {Monte} {Carlo} simulation of the intergalactic absorption and the detectability of the {Lyman} continuum from distant galaxies},
	volume = {387},
	issn = {0035-8711},
	url = {https://doi.org/10.1111/j.1365-2966.2008.13350.x},
	doi = {10.1111/j.1365-2966.2008.13350.x},
	number = {4},
	urldate = {2025-04-11},
	journal = {Monthly Notices of the Royal Astronomical Society},
	author = {Inoue, Akio K. and Iwata, Ikuru},
	month = jul,
	year = {2008},
	pages = {1681--1692},
}

@article{katz_nature_2022,
	title = {The nature of high [{O} iii]88 μ m/[{C} ii]158 μm galaxies in the epoch of reionization: {Low} carbon abundance and a top-heavy {IMF}?},
	volume = {510},
	issn = {0035-8711},
	shorttitle = {The nature of high [{O} iii]88 μ m/[{C} ii]158 μm galaxies in the epoch of reionization},
	url = {https://doi.org/10.1093/mnras/stac028},
	doi = {10.1093/mnras/stac028},
	number = {4},
	urldate = {2025-04-11},
	journal = {Monthly Notices of the Royal Astronomical Society},
	author = {Katz, Harley and Rosdahl, Joakim and Kimm, Taysun and Garel, Thibault and Blaizot, Jérémy and Haehnelt, Martin G and Michel-Dansac, Léo and Martin-Alvarez, Sergio and Devriendt, Julien and Slyz, Adrianne and Teyssier, Romain and Ocvirk, Pierre and Laporte, Nicolas and Ellis, Richard},
	month = mar,
	year = {2022},
	pages = {5603--5622},
}

@article{becker_mean_2021,
	title = {The mean free path of ionizing photons at 5 < z < 6: evidence for rapid evolution near reionization},
	volume = {508},
	issn = {0035-8711},
	shorttitle = {The mean free path of ionizing photons at 5 < z < 6},
	url = {https://doi.org/10.1093/mnras/stab2696},
	doi = {10.1093/mnras/stab2696},
	number = {2},
	urldate = {2025-04-11},
	journal = {Monthly Notices of the Royal Astronomical Society},
	author = {Becker, George D and D’Aloisio, Anson and Christenson, Holly M and Zhu, Yongda and Worseck, Gábor and Bolton, James S},
	month = dec,
	year = {2021},
	pages = {1853--1869},
}

@misc{mcclymont_thesan-zoom_2025,
	title = {The {THESAN}-{ZOOM} project: {Burst}, quench, repeat -- unveiling the evolution of high-redshift galaxies along the star-forming main sequence},
	shorttitle = {The {THESAN}-{ZOOM} project},
	url = {https://ui.adsabs.harvard.edu/abs/2025arXiv250300106M},
	doi = {10.48550/arXiv.2503.00106},
	urldate = {2025-04-11},
	publisher = {arXiv},
	author = {McClymont, William and Tacchella, Sandro and Smith, Aaron and Kannan, Rahul and Puchwein, Ewald and Borrow, Josh and Garaldi, Enrico and Keating, Laura and Vogelsberger, Mark and Zier, Oliver and Shen, Xuejian and Popovic, Filip and Simmonds, Charlotte},
	month = feb,
	year = {2025},
	note = {ADS Bibcode: 2025arXiv250300106M},
}

@misc{mcclymont_thesan-zoom_2025-1,
	title = {The {THESAN}-{ZOOM} project: central starbursts and inside-out quenching govern galaxy sizes in the early {Universe}},
	shorttitle = {The {THESAN}-{ZOOM} project},
	url = {https://ui.adsabs.harvard.edu/abs/2025arXiv250304894M},
	doi = {10.48550/arXiv.2503.04894},
	urldate = {2025-04-11},
	publisher = {arXiv},
	author = {McClymont, William and Tacchella, Sandro and Smith, Aaron and Kannan, Rahul and Puchwein, Ewald and Borrow, Josh and Garaldi, Enrico and Keating, Laura and Vogelsberger, Mark and Zier, Oliver and Shen, Xuejian and Popovic, Filip},
	month = mar,
	year = {2025},
	note = {ADS Bibcode: 2025arXiv250304894M},
}

@article{rosdahl_lyc_2022,
	title = {{LyC} escape from sphinx galaxies in the {Epoch} of {Reionization}},
	volume = {515},
	issn = {0035-8711},
	url = {https://doi.org/10.1093/mnras/stac1942},
	doi = {10.1093/mnras/stac1942},
	number = {2},
	urldate = {2025-04-11},
	journal = {Monthly Notices of the Royal Astronomical Society},
	author = {Rosdahl, Joakim and Blaizot, Jérémy and Katz, Harley and Kimm, Taysun and Garel, Thibault and Haehnelt, Martin and Keating, Laura C and Martin-Alvarez, Sergio and Michel-Dansac, Léo and Ocvirk, Pierre},
	month = sep,
	year = {2022},
	pages = {2386--2414},
}

@misc{kannan_introducing_2025,
	title = {Introducing the {THESAN}-{ZOOM} project: radiation-hydrodynamic simulations of high-redshift galaxies with a multi-phase interstellar medium},
	shorttitle = {Introducing the {THESAN}-{ZOOM} project},
	url = {https://ui.adsabs.harvard.edu/abs/2025arXiv250220437K},
	doi = {10.48550/arXiv.2502.20437},
	urldate = {2025-04-11},
	publisher = {arXiv},
	author = {Kannan, Rahul and Puchwein, Ewald and Smith, Aaron and Borrow, Josh and Garaldi, Enrico and Keating, Laura and Vogelsberger, Mark and Zier, Oliver and McClymont, William and Shen, Xuejian and Popovic, Filip and Tacchella, Sandro and Hernquist, Lars and Springel, Volker},
	month = feb,
	year = {2025},
	note = {ADS Bibcode: 2025arXiv250220437K},
}

@article{dayal_reionization_2020,
	title = {Reionization with galaxies and active galactic nuclei},
	volume = {495},
	issn = {0035-8711},
	url = {https://doi.org/10.1093/mnras/staa1138},
	doi = {10.1093/mnras/staa1138},
	number = {3},
	urldate = {2025-05-05},
	journal = {Monthly Notices of the Royal Astronomical Society},
	author = {Dayal, Pratika and Volonteri, Marta and Choudhury, Tirthankar Roy and Schneider, Raffaella and Trebitsch, Maxime and Gnedin, Nickolay Y and Atek, Hakim and Hirschmann, Michaela and Reines, Amy},
	month = jul,
	year = {2020},
	pages = {3065--3078},
}

@article{parsa_no_2018,
	title = {No evidence for a significant {AGN} contribution to cosmic hydrogen reionization},
	volume = {474},
	issn = {0035-8711},
	url = {https://doi.org/10.1093/mnras/stx2887},
	doi = {10.1093/mnras/stx2887},
	number = {3},
	urldate = {2025-05-05},
	journal = {Monthly Notices of the Royal Astronomical Society},
	author = {Parsa, Shaghayegh and Dunlop, James S and McLure, Ross J},
	month = mar,
	year = {2018},
	pages = {2904--2923},
}

@article{atek_most_2024,
	title = {Most of the photons that reionized the {Universe} came from dwarf galaxies},
	volume = {626},
	copyright = {2024 The Author(s), under exclusive licence to Springer Nature Limited},
	issn = {1476-4687},
	url = {https://www.nature.com/articles/s41586-024-07043-6},
	doi = {10.1038/s41586-024-07043-6},
	language = {en},
	number = {8001},
	urldate = {2025-05-05},
	journal = {Nature},
	publisher = {Nature Publishing Group},
	author = {Atek, Hakim and Labbé, Ivo and Furtak, Lukas J. and Chemerynska, Iryna and Fujimoto, Seiji and Setton, David J. and Miller, Tim B. and Oesch, Pascal and Bezanson, Rachel and Price, Sedona H. and Dayal, Pratika and Zitrin, Adi and Kokorev, Vasily and Weaver, John R. and Brammer, Gabriel and Dokkum, Pieter van and Williams, Christina C. and Cutler, Sam E. and Feldmann, Robert and Fudamoto, Yoshinobu and Greene, Jenny E. and Leja, Joel and Maseda, Michael V. and Muzzin, Adam and Pan, Richard and Papovich, Casey and Nelson, Erica J. and Nanayakkara, Themiya and Stark, Daniel P. and Stefanon, Mauro and Suess, Katherine A. and Wang, Bingjie and Whitaker, Katherine E.},
	month = feb,
	year = {2024},
	pages = {975--978},
}

@article{naidu_rapid_2020,
	title = {Rapid {Reionization} by the {Oligarchs}: {The} {Case} for {Massive}, {UV}-bright, {Star}-forming {Galaxies} with {High} {Escape} {Fractions}},
	volume = {892},
	issn = {0004-637X},
	shorttitle = {Rapid {Reionization} by the {Oligarchs}},
	url = {https://dx.doi.org/10.3847/1538-4357/ab7cc9},
	doi = {10.3847/1538-4357/ab7cc9},
	language = {en},
	number = {2},
	urldate = {2025-05-05},
	journal = {The Astrophysical Journal},
	publisher = {The American Astronomical Society},
	author = {Naidu, Rohan P. and Tacchella, Sandro and Mason, Charlotte A. and Bose, Sownak and Oesch, Pascal A. and Conroy, Charlie},
	month = apr,
	year = {2020},
	pages = {109},
}

@article{ocvirk_cosmic_2020,
	title = {Cosmic {Dawn} {II} ({CoDa} {II}): a new radiation-hydrodynamics simulation of the self-consistent coupling of galaxy formation and reionization},
	volume = {496},
	issn = {0035-8711},
	shorttitle = {Cosmic {Dawn} {II} ({CoDa} {II})},
	url = {https://doi.org/10.1093/mnras/staa1266},
	doi = {10.1093/mnras/staa1266},
	number = {4},
	urldate = {2025-05-06},
	journal = {Monthly Notices of the Royal Astronomical Society},
	author = {Ocvirk, Pierre and Aubert, Dominique and Sorce, Jenny G and Shapiro, Paul R and Deparis, Nicolas and Dawoodbhoy, Taha and Lewis, Joseph and Teyssier, Romain and Yepes, Gustavo and Gottlöber, Stefan and Ahn, Kyungjin and Iliev, Ilian T and Hoffman, Yehuda},
	month = aug,
	year = {2020},
	pages = {4087--4107},
}

@article{bhagwat_spice_2024,
	title = {{SPICE}: the connection between cosmic reionization and stellar feedback in the first galaxies},
	volume = {531},
	issn = {0035-8711},
	shorttitle = {{SPICE}},
	url = {https://doi.org/10.1093/mnras/stae1125},
	doi = {10.1093/mnras/stae1125},
	number = {3},
	urldate = {2025-05-06},
	journal = {Monthly Notices of the Royal Astronomical Society},
	author = {Bhagwat, Aniket and Costa, Tiago and Ciardi, Benedetta and Pakmor, Rüdiger and Garaldi, Enrico},
	month = jul,
	year = {2024},
	pages = {3406--3430},
}

@misc{tacchella_star_2024,
	title = {Star formation in cosmic-dawn galaxies},
	url = {http://arxiv.org/abs/2410.04227},
	doi = {10.48550/arXiv.2410.04227},
	urldate = {2025-05-07},
	publisher = {arXiv},
	author = {Tacchella, Sandro},
	month = oct,
	year = {2024},
	note = {arXiv:2410.04227 [astro-ph]},
}

@article{kannan_arepo-rt_2019,
	title = {{AREPO}-{RT}: {Radiation} hydrodynamics on a moving mesh},
	volume = {485},
	issn = {0035-8711, 1365-2966},
	shorttitle = {{AREPO}-{RT}},
	url = {http://arxiv.org/abs/1804.01987},
	doi = {10.1093/mnras/stz287},
	number = {1},
	urldate = {2025-05-07},
	journal = {Monthly Notices of the Royal Astronomical Society},
	author = {Kannan, Rahul and Vogelsberger, Mark and Marinacci, Federico and McKinnon, Ryan and Pakmor, Rüdiger and Springel, Volker},
	month = may,
	year = {2019},
	note = {arXiv:1804.01987 [astro-ph]},
	pages = {117--149},
}

@article{springel_first_2018,
	title = {First results from the {IllustrisTNG} simulations: matter and galaxy clustering},
	volume = {475},
	issn = {0035-8711},
	shorttitle = {First results from the {IllustrisTNG} simulations},
	url = {https://doi.org/10.1093/mnras/stx3304},
	doi = {10.1093/mnras/stx3304},
	number = {1},
	urldate = {2025-05-07},
	journal = {Monthly Notices of the Royal Astronomical Society},
	author = {Springel, Volker and Pakmor, Rüdiger and Pillepich, Annalisa and Weinberger, Rainer and Nelson, Dylan and Hernquist, Lars and Vogelsberger, Mark and Genel, Shy and Torrey, Paul and Marinacci, Federico and Naiman, Jill},
	month = mar,
	year = {2018},
	pages = {676--698},
}

@article{marinacci_first_2018,
	title = {First results from the {IllustrisTNG} simulations: radio haloes and magnetic fields},
	volume = {480},
	issn = {0035-8711},
	shorttitle = {First results from the {IllustrisTNG} simulations},
	url = {https://doi.org/10.1093/mnras/sty2206},
	doi = {10.1093/mnras/sty2206},
	number = {4},
	urldate = {2025-05-07},
	journal = {Monthly Notices of the Royal Astronomical Society},
	author = {Marinacci, Federico and Vogelsberger, Mark and Pakmor, Rüdiger and Torrey, Paul and Springel, Volker and Hernquist, Lars and Nelson, Dylan and Weinberger, Rainer and Pillepich, Annalisa and Naiman, Jill and Genel, Shy},
	month = nov,
	year = {2018},
	pages = {5113--5139},
}

@misc{nelson_illustristng_2021,
	title = {The {IllustrisTNG} {Simulations}: {Public} {Data} {Release}},
	shorttitle = {The {IllustrisTNG} {Simulations}},
	url = {http://arxiv.org/abs/1812.05609},
	doi = {10.48550/arXiv.1812.05609},
	urldate = {2025-05-07},
	publisher = {arXiv},
	author = {Nelson, Dylan and Springel, Volker and Pillepich, Annalisa and Rodriguez-Gomez, Vicente and Torrey, Paul and Genel, Shy and Vogelsberger, Mark and Pakmor, Ruediger and Marinacci, Federico and Weinberger, Rainer and Kelley, Luke and Lovell, Mark and Diemer, Benedikt and Hernquist, Lars},
	month = jan,
	year = {2021},
	note = {arXiv:1812.05609 [astro-ph]},
}

@article{mckinnon_simulating_2017,
	title = {Simulating the dust content of galaxies: successes and failures},
	volume = {468},
	issn = {0035-8711},
	shorttitle = {Simulating the dust content of galaxies},
	url = {https://doi.org/10.1093/mnras/stx467},
	doi = {10.1093/mnras/stx467},
	number = {2},
	urldate = {2025-05-07},
	journal = {Monthly Notices of the Royal Astronomical Society},
	author = {McKinnon, Ryan and Torrey, Paul and Vogelsberger, Mark and Hayward, Christopher C. and Marinacci, Federico},
	month = jun,
	year = {2017},
	pages = {1505--1521},
}

@article{springel_e_2010,
	title = {E pur si muove: {Galilean}-invariant cosmological hydrodynamical simulations on a moving mesh},
	volume = {401},
	issn = {0035-8711},
	shorttitle = {E pur si muove},
	url = {https://doi.org/10.1111/j.1365-2966.2009.15715.x},
	doi = {10.1111/j.1365-2966.2009.15715.x},
	number = {2},
	urldate = {2025-05-07},
	journal = {Monthly Notices of the Royal Astronomical Society},
	author = {Springel, Volker},
	month = jan,
	year = {2010},
	pages = {791--851},
}

@article{ade_planck_2016,
	title = {Planck 2015 results - {XIII}. {Cosmological} parameters},
	volume = {594},
	copyright = {© ESO, 2016},
	issn = {0004-6361, 1432-0746},
	url = {https://www.aanda.org/articles/aa/abs/2016/10/aa25830-15/aa25830-15.html},
	doi = {10.1051/0004-6361/201525830},
	language = {en},
	urldate = {2025-05-07},
	journal = {Astronomy \& Astrophysics},
	publisher = {EDP Sciences},
	author = {Ade, P. a. R. and Aghanim, N. and Arnaud, M. and Ashdown, M. and Aumont, J. and Baccigalupi, C. and Banday, A. J. and Barreiro, R. B. and Bartlett, J. G. and Bartolo, N. and Battaner, E. and Battye, R. and Benabed, K. and Benoît, A. and Benoit-Lévy, A. and Bernard, J.-P. and Bersanelli, M. and Bielewicz, P. and Bock, J. J. and Bonaldi, A. and Bonavera, L. and Bond, J. R. and Borrill, J. and Bouchet, F. R. and Boulanger, F. and Bucher, M. and Burigana, C. and Butler, R. C. and Calabrese, E. and Cardoso, J.-F. and Catalano, A. and Challinor, A. and Chamballu, A. and Chary, R.-R. and Chiang, H. C. and Chluba, J. and Christensen, P. R. and Church, S. and Clements, D. L. and Colombi, S. and Colombo, L. P. L. and Combet, C. and Coulais, A. and Crill, B. P. and Curto, A. and Cuttaia, F. and Danese, L. and Davies, R. D. and Davis, R. J. and Bernardis, P. de and Rosa, A. de and Zotti, G. de and Delabrouille, J. and Désert, F.-X. and Valentino, E. Di and Dickinson, C. and Diego, J. M. and Dolag, K. and Dole, H. and Donzelli, S. and Doré, O. and Douspis, M. and Ducout, A. and Dunkley, J. and Dupac, X. and Efstathiou, G. and Elsner, F. and Enßlin, T. A. and Eriksen, H. K. and Farhang, M. and Fergusson, J. and Finelli, F. and Forni, O. and Frailis, M. and Fraisse, A. A. and Franceschi, E. and Frejsel, A. and Galeotta, S. and Galli, S. and Ganga, K. and Gauthier, C. and Gerbino, M. and Ghosh, T. and Giard, M. and Giraud-Héraud, Y. and Giusarma, E. and Gjerløw, E. and González-Nuevo, J. and Górski, K. M. and Gratton, S. and Gregorio, A. and Gruppuso, A. and Gudmundsson, J. E. and Hamann, J. and Hansen, F. K. and Hanson, D. and Harrison, D. L. and Helou, G. and Henrot-Versillé, S. and Hernández-Monteagudo, C. and Herranz, D. and Hildebrandt, S. R. and Hivon, E. and Hobson, M. and Holmes, W. A. and Hornstrup, A. and Hovest, W. and Huang, Z. and Huffenberger, K. M. and Hurier, G. and Jaffe, A. H. and Jaffe, T. R. and Jones, W. C. and Juvela, M. and Keihänen, E. and Keskitalo, R. and Kisner, T. S. and Kneissl, R. and Knoche, J. and Knox, L. and Kunz, M. and Kurki-Suonio, H. and Lagache, G. and Lähteenmäki, A. and Lamarre, J.-M. and Lasenby, A. and Lattanzi, M. and Lawrence, C. R. and Leahy, J. P. and Leonardi, R. and Lesgourgues, J. and Levrier, F. and Lewis, A. and Liguori, M. and Lilje, P. B. and Linden-Vørnle, M. and López-Caniego, M. and Lubin, P. M. and Macías-Pérez, J. F. and Maggio, G. and Maino, D. and Mandolesi, N. and Mangilli, A. and Marchini, A. and Maris, M. and Martin, P. G. and Martinelli, M. and Martínez-González, E. and Masi, S. and Matarrese, S. and McGehee, P. and Meinhold, P. R. and Melchiorri, A. and Melin, J.-B. and Mendes, L. and Mennella, A. and Migliaccio, M. and Millea, M. and Mitra, S. and Miville-Deschênes, M.-A. and Moneti, A. and Montier, L. and Morgante, G. and Mortlock, D. and Moss, A. and Munshi, D. and Murphy, J. A. and Naselsky, P. and Nati, F. and Natoli, P. and Netterfield, C. B. and Nørgaard-Nielsen, H. U. and Noviello, F. and Novikov, D. and Novikov, I. and Oxborrow, C. A. and Paci, F. and Pagano, L. and Pajot, F. and Paladini, R. and Paoletti, D. and Partridge, B. and Pasian, F. and Patanchon, G. and Pearson, T. J. and Perdereau, O. and Perotto, L. and Perrotta, F. and Pettorino, V. and Piacentini, F. and Piat, M. and Pierpaoli, E. and Pietrobon, D. and Plaszczynski, S. and Pointecouteau, E. and Polenta, G. and Popa, L. and Pratt, G. W. and Prézeau, G. and Prunet, S. and Puget, J.-L. and Rachen, J. P. and Reach, W. T. and Rebolo, R. and Reinecke, M. and Remazeilles, M. and Renault, C. and Renzi, A. and Ristorcelli, I. and Rocha, G. and Rosset, C. and Rossetti, M. and Roudier, G. and d’Orfeuil, B. Rouillé and Rowan-Robinson, M. and Rubiño-Martín, J. A. and Rusholme, B. and Said, N. and Salvatelli, V. and Salvati, L. and Sandri, M. and Santos, D. and Savelainen, M. and Savini, G. and Scott, D. and Seiffert, M. D. and Serra, P. and Shellard, E. P. S. and Spencer, L. D. and Spinelli, M. and Stolyarov, V. and Stompor, R. and Sudiwala, R. and Sunyaev, R. and Sutton, D. and Suur-Uski, A.-S. and Sygnet, J.-F. and Tauber, J. A. and Terenzi, L. and Toffolatti, L. and Tomasi, M. and Tristram, M. and Trombetti, T. and Tucci, M. and Tuovinen, J. and Türler, M. and Umana, G. and Valenziano, L. and Valiviita, J. and Tent, F. Van and Vielva, P. and Villa, F. and Wade, L. A. and Wandelt, B. D. and Wehus, I. K. and White, M. and White, S. D. M. and Wilkinson, A. and Yvon, D. and Zacchei, A. and Zonca, A.},
	month = oct,
	year = {2016},
	pages = {A13},
}

@article{marinacci_simulating_2019,
	title = {Simulating the interstellar medium and stellar feedback on a moving mesh: implementation and isolated galaxies},
	volume = {489},
	issn = {0035-8711},
	shorttitle = {Simulating the interstellar medium and stellar feedback on a moving mesh},
	url = {https://doi.org/10.1093/mnras/stz2391},
	doi = {10.1093/mnras/stz2391},
	number = {3},
	urldate = {2025-05-07},
	journal = {Monthly Notices of the Royal Astronomical Society},
	author = {Marinacci, Federico and Sales, Laura V and Vogelsberger, Mark and Torrey, Paul and Springel, Volker},
	month = nov,
	year = {2019},
	pages = {4233--4260},
}

@article{smith_lyman_2015,
	title = {The {Lyman} α signature of the first galaxies},
	volume = {449},
	issn = {0035-8711},
	url = {https://doi.org/10.1093/mnras/stv565},
	doi = {10.1093/mnras/stv565},
	number = {4},
	urldate = {2025-05-08},
	journal = {Monthly Notices of the Royal Astronomical Society},
	author = {Smith, Aaron and Safranek-Shrader, Chalence and Bromm, Volker and Milosavljević, Miloš},
	month = jun,
	year = {2015},
	pages = {4336--4362},
}

@article{sternberg_ionizing_2003,
	title = {Ionizing {Photon} {Emission} {Rates} from {O}- and {Early} {B}-{Type} {Stars} and {Clusters}},
	volume = {599},
	issn = {0004-637X},
	url = {https://iopscience.iop.org/article/10.1086/379506/meta},
	doi = {10.1086/379506},
	language = {en},
	number = {2},
	urldate = {2025-05-13},
	journal = {The Astrophysical Journal},
	publisher = {IOP Publishing},
	author = {Sternberg, Amiel and Hoffmann, Tadziu L. and Pauldrach, A. W. A.},
	month = dec,
	year = {2003},
	pages = {1333},
}

@article{bluck_are_2020,
	title = {Are galactic star formation and quenching governed by local, global, or environmental phenomena?},
	volume = {492},
	issn = {0035-8711},
	url = {https://doi.org/10.1093/mnras/stz3264},
	doi = {10.1093/mnras/stz3264},
	number = {1},
	urldate = {2025-05-14},
	journal = {Monthly Notices of the Royal Astronomical Society},
	author = {Bluck, Asa F L and Maiolino, Roberto and Sánchez, Sebastian F and Ellison, Sara L and Thorp, Mallory D and Piotrowska, Joanna M and Teimoorinia, Hossen and Bundy, Kevin A},
	month = feb,
	year = {2020},
	pages = {96--139},
}

@article{mascia_new_2024,
	title = {New insight on the nature of cosmic reionizers from the {CEERS} survey},
	volume = {685},
	copyright = {© The Authors 2024},
	issn = {0004-6361, 1432-0746},
	url = {https://www.aanda.org/articles/aa/abs/2024/05/aa47884-23/aa47884-23.html},
	doi = {10.1051/0004-6361/202347884},
	language = {en},
	urldate = {2025-05-15},
	journal = {Astronomy \& Astrophysics},
	publisher = {EDP Sciences},
	author = {Mascia, S. and Pentericci, L. and Calabrò, A. and Santini, P. and Napolitano, L. and Haro, P. Arrabal and Castellano, M. and Dickinson, M. and Ocvirk, P. and Lewis, J. S. W. and Amorín, R. and Bagley, M. and Bhatawdekar, R. and Cleri, N. J. and Costantin, L. and Dekel, A. and Finkelstein, S. L. and Fontana, A. and Giavalisco, M. and Grogin, N. A. and Hathi, N. P. and Hirschmann, M. and Holwerda, B. W. and Jung, I. and Kartaltepe, J. S. and Koekemoer, A. M. and Lucas, R. A. and Papovich, C. and Pérez-González, P. G. and Pirzkal, N. and Trump, J. R. and Wilkins, S. M. and Yung, L. Y. A.},
	month = may,
	year = {2024},
	pages = {A3},
}

@misc{jaskot_multivariate_2024,
	title = {Multivariate {Predictors} of {LyC} {Escape} {II}: {Predicting} {LyC} {Escape} {Fractions} for {High}-{Redshift} {Galaxies}},
	shorttitle = {Multivariate {Predictors} of {LyC} {Escape} {II}},
	url = {http://arxiv.org/abs/2406.10179},
	doi = {10.48550/arXiv.2406.10179},
	urldate = {2025-05-15},
	publisher = {arXiv},
	author = {Jaskot, Anne E. and Silveyra, Anneliese C. and Plantinga, Anna and Flury, Sophia R. and Hayes, Matthew and Chisholm, John and Heckman, Timothy and Pentericci, Laura and Schaerer, Daniel and Trebitsch, Maxime and Verhamme, Anne and Carr, Cody and Ferguson, Henry C. and Ji, Zhiyuan and Giavalisco, Mauro and Henry, Alaina and Marques-Chaves, Rui and Östlin, Göran and Saldana-Lopez, Alberto and Scarlata, Claudia and Worseck, Gábor and Xu, Xinfeng},
	month = sep,
	year = {2024},
	note = {arXiv:2406.10179 [astro-ph]},
}

@article{jensen_machine-learning_2016,
	title = {A {MACHINE}-{LEARNING} {APPROACH} {TO} {MEASURING} {THE} {ESCAPE} {OF} {IONIZING} {RADIATION} {FROM} {GALAXIES} {IN} {THE} {REIONIZATION} {EPOCH}},
	volume = {827},
	issn = {0004-637X},
	url = {https://dx.doi.org/10.3847/0004-637X/827/1/5},
	doi = {10.3847/0004-637X/827/1/5},
	language = {en},
	number = {1},
	urldate = {2025-05-15},
	journal = {The Astrophysical Journal},
	publisher = {The American Astronomical Society},
	author = {Jensen, Hannes and Zackrisson, Erik and Pelckmans, Kristiaan and Binggeli, Christian and Ausmees, Kristiina and Lundholm, Ulrika},
	month = aug,
	year = {2016},
	pages = {5},
}

@article{xu_tracing_2022,
	title = {Tracing {Lyman}-alpha and {Lyman} {Continuum} {Escape} in {Galaxies} with {Mg} {II} {Emission}},
	volume = {933},
	issn = {0004-637X, 1538-4357},
	url = {http://arxiv.org/abs/2205.11317},
	doi = {10.3847/1538-4357/ac7225},
	number = {2},
	urldate = {2025-05-15},
	journal = {The Astrophysical Journal},
	author = {Xu, Xinfeng and Henry, Alaina and Heckman, Timothy and Chisholm, John and Worseck, Gábor and Gronke, Max and Jaskot, Anne and McCandliss, Stephan R. and Flury, Sophia R. and Giavalisco, Mauro and Ji, Zhiyuan and Amorín, Ricardo O. and Berg, Danielle A. and Borthakur, Sanchayeeta and Bouche, Nicolas and Carr, Cody and Erb, Dawn K. and Ferguson, Harry and Garel, Thibault and Hayes, Matthew and Makan, Kirill and Marques-Chaves, Rui and Rutkowski, Michael and Östlin, Göran and Rafelski, Marc and Saldana-Lopez, Alberto and Scarlata, Claudia and Schaerer, Daniel and Trebitsch, Maxime and Tremonti, Christy and Verhamme, Anne and Wang, Bingjie},
	month = jul,
	year = {2022},
	note = {arXiv:2205.11317 [astro-ph]},
	pages = {202},
}

@article{pedregosa_scikit-learn_2011,
	title = {Scikit-learn: {Machine} {Learning} in {Python}},
	volume = {12},
	issn = {1533-7928},
	shorttitle = {Scikit-learn},
	url = {http://jmlr.org/papers/v12/pedregosa11a.html},
	number = {85},
	urldate = {2025-05-15},
	journal = {Journal of Machine Learning Research},
	author = {Pedregosa, Fabian and Varoquaux, Gaël and Gramfort, Alexandre and Michel, Vincent and Thirion, Bertrand and Grisel, Olivier and Blondel, Mathieu and Prettenhofer, Peter and Weiss, Ron and Dubourg, Vincent and Vanderplas, Jake and Passos, Alexandre and Cournapeau, David and Brucher, Matthieu and Perrot, Matthieu and Duchesnay, Édouard},
	year = {2011},
	pages = {2825--2830},
}

@article{simmonds_low-mass_2024,
	title = {Low-mass bursty galaxies in {JADES} efficiently produce ionizing photons and could represent the main drivers of reionization},
	volume = {527},
	issn = {0035-8711},
	url = {https://ui.adsabs.harvard.edu/abs/2024MNRAS.527.6139S},
	doi = {10.1093/mnras/stad3605},
	urldate = {2025-05-16},
	journal = {Monthly Notices of the Royal Astronomical Society},
	publisher = {OUP},
	author = {Simmonds, C. and Tacchella, S. and Hainline, K. and Johnson, B. D. and McClymont, W. and Robertson, B. and Saxena, A. and Sun, F. and Witten, C. and Baker, W. M. and Bhatawdekar, R. and Boyett, K. and Bunker, A. J. and Charlot, S. and Curtis-Lake, E. and Egami, E. and Eisenstein, D. J. and Hausen, R. and Maiolino, R. and Maseda, M. V. and Scholtz, J. and Williams, C. C. and Willott, C. and Witstok, J.},
	month = jan,
	year = {2024},
	note = {ADS Bibcode: 2024MNRAS.527.6139S},
	pages = {6139--6157},
}

@article{simmonds_ionizing_2024,
	title = {Ionizing properties of galaxies in {JADES} for a stellar mass complete sample: resolving the cosmic ionizing photon budget crisis at the {Epoch} of {Reionization}},
	volume = {535},
	issn = {0035-8711},
	shorttitle = {Ionizing properties of galaxies in {JADES} for a stellar mass complete sample},
	url = {https://ui.adsabs.harvard.edu/abs/2024MNRAS.535.2998S},
	doi = {10.1093/mnras/stae2537},
	urldate = {2025-05-16},
	journal = {Monthly Notices of the Royal Astronomical Society},
	publisher = {OUP},
	author = {Simmonds, C. and Tacchella, S. and Hainline, K. and Johnson, B. D. and Puskás, D. and Robertson, B. and Baker, W. M. and Bhatawdekar, R. and Boyett, K. and Bunker, A. J. and Cargile, P. A. and Carniani, S. and Chevallard, J. and Curti, M. and Curtis-Lake, E. and Ji, Z. and Jones, G. C. and Kumari, N. and Laseter, I. and Maiolino, R. and Maseda, M. V. and Rinaldi, P. and Stoffers, A. and Übler, H. and Villanueva, N. C. and Williams, C. C. and Willott, C. and Witstok, J. and Zhu, Y.},
	month = dec,
	year = {2024},
	note = {ADS Bibcode: 2024MNRAS.535.2998S},
	pages = {2998--3019},
}

@article{frenk_cosmological_2000,
	title = {Cosmological reionization},
	volume = {358},
	url = {https://royalsocietypublishing.org/doi/abs/10.1098/rsta.2000.0627},
	doi = {10.1098/rsta.2000.0627},
	number = {1772},
	urldate = {2025-05-16},
	journal = {Philosophical Transactions of the Royal Society of London. Series A: Mathematical, Physical and Engineering Sciences},
	publisher = {Royal Society},
	author = {Frenk, C. S. and White, S. D. M. and Madau, Piero},
	month = jul,
	year = {2000},
	pages = {2021--2033},
}

@article{zhu_damping_2024,
	title = {Damping wing-like features in the stacked {Ly} α forest: {Potential} neutral hydrogen islands at z < 6},
	volume = {533},
	issn = {1745-3925},
	shorttitle = {Damping wing-like features in the stacked {Ly} α forest},
	url = {https://doi.org/10.1093/mnrasl/slae061},
	doi = {10.1093/mnrasl/slae061},
	number = {1},
	urldate = {2025-05-16},
	journal = {Monthly Notices of the Royal Astronomical Society: Letters},
	author = {Zhu, Yongda and Becker, George D and Bosman, Sarah E I and Cain, Christopher and Keating, Laura C and Nasir, Fahad and D’Odorico, Valentina and Bañados, Eduardo and Bian, Fuyan and Bischetti, Manuela and Bolton, James S and Chen, Huanqing and D’Aloisio, Anson and Davies, Frederick B and Davies, Rebecca L and Eilers, Anna-Christina and Fan, Xiaohui and Gaikwad, Prakash and Greig, Bradley and Haehnelt, Martin G and Kulkarni, Girish and Lai, Samuel and Puchwein, Ewald and Qin, Yuxiang and Ryan-Weber, Emma V and Satyavolu, Sindhu and Spina, Benedetta and Walter, Fabian and Wang, Feige and Wolfson, Molly and Yang, Jinyi},
	month = sep,
	year = {2024},
	pages = {L49--L56},
}

@article{tacchella_confinement_2016,
	title = {The confinement of star-forming galaxies into a main sequence through episodes of gas compaction, depletion and replenishment},
	volume = {457},
	issn = {0035-8711},
	url = {https://doi.org/10.1093/mnras/stw131},
	doi = {10.1093/mnras/stw131},
	number = {3},
	urldate = {2025-05-17},
	journal = {Monthly Notices of the Royal Astronomical Society},
	author = {Tacchella, Sandro and Dekel, Avishai and Carollo, C. Marcella and Ceverino, Daniel and DeGraf, Colin and Lapiner, Sharon and Mandelker, Nir and Primack Joel, R.},
	month = apr,
	year = {2016},
	pages = {2790--2813},
}

@article{schechter_analytic_1976,
	title = {An analytic expression for the luminosity function for galaxies.},
	volume = {203},
	issn = {0004-637X},
	url = {https://ui.adsabs.harvard.edu/abs/1976ApJ...203..297S},
	doi = {10.1086/154079},
	urldate = {2025-05-20},
	journal = {The Astrophysical Journal},
	publisher = {IOP},
	author = {Schechter, P.},
	month = jan,
	year = {1976},
	note = {ADS Bibcode: 1976ApJ...203..297S},
	pages = {297--306},
}

@article{weibel_galaxy_2024,
	title = {Galaxy build-up in the first 1.5 {Gyr} of cosmic history: insights from the stellar mass function at z {\textasciitilde} 4–9 from {JWST} {NIRCam} observations},
	volume = {533},
	issn = {0035-8711},
	shorttitle = {Galaxy build-up in the first 1.5 {Gyr} of cosmic history},
	url = {https://doi.org/10.1093/mnras/stae1891},
	doi = {10.1093/mnras/stae1891},
	number = {2},
	urldate = {2025-05-20},
	journal = {Monthly Notices of the Royal Astronomical Society},
	author = {Weibel, Andrea and Oesch, Pascal A and Barrufet, Laia and Gottumukkala, Rashmi and Ellis, Richard S and Santini, Paola and Weaver, John R and Allen, Natalie and Bouwens, Rychard and Bowler, Rebecca A A and Brammer, Gabe and Carnall, Adam C and Cullen, Fergus and Dayal, Pratika and Dickinson, Mark and Donnan, Callum T and Dunlop, James S and Giavalisco, Mauro and Grogin, Norman A and Illingworth, Garth D and Koekemoer, Anton M and Labbe, Ivo and Marchesini, Danilo and McLeod, Derek J and McLure, Ross J and Naidu, Rohan P and Pérez-González, Pablo G and Shuntov, Marko and Stefanon, Mauro and Toft, Sune and Xiao, Mengyuan},
	month = sep,
	year = {2024},
	pages = {1808--1838},
}

@article{bouwens_new_2021,
	title = {New {Determinations} of the {UV} {Luminosity} {Functions} from z   9 to 2 {Show} a {Remarkable} {Consistency} with {Halo} {Growth} and a {Constant} {Star} {Formation} {Efficiency}},
	volume = {162},
	issn = {0004-6256},
	url = {https://ui.adsabs.harvard.edu/abs/2021AJ....162...47B},
	doi = {10.3847/1538-3881/abf83e},
	urldate = {2025-05-20},
	journal = {The Astronomical Journal},
	publisher = {IOP},
	author = {Bouwens, R. J. and Oesch, P. A. and Stefanon, M. and Illingworth, G. and Labbé, I. and Reddy, N. and Atek, H. and Montes, M. and Naidu, R. and Nanayakkara, T. and Nelson, E. and Wilkins, S.},
	month = aug,
	year = {2021},
	note = {ADS Bibcode: 2021AJ....162...47B},
	pages = {47},
}

@article{munoz_reionization_2024,
	title = {Reionization after {JWST}: a photon budget crisis?},
	volume = {535},
	issn = {1745-3925},
	shorttitle = {Reionization after {JWST}},
	url = {https://doi.org/10.1093/mnrasl/slae086},
	doi = {10.1093/mnrasl/slae086},
	number = {1},
	urldate = {2025-05-20},
	journal = {Monthly Notices of the Royal Astronomical Society: Letters},
	author = {Muñoz, Julian B and Mirocha, Jordan and Chisholm, John and Furlanetto, Steven R and Mason, Charlotte},
	month = nov,
	year = {2024},
	pages = {L37--L43},
}

@book{hughes_measurements_2010,
	title = {Measurements and their uncertainties : a practical guide to modern error analysis},
	isbn = {978-0-19-956632-7 978-0-19-956633-4},
	shorttitle = {Measurements and their uncertainties},
	url = {http://archive.org/details/measurementsthei0000hugh},
	language = {eng},
	urldate = {2025-05-21},
	publisher = {Oxford : New York : Oxford University Press},
	author = {Hughes, Ifan},
	collaborator = {{Internet Archive}},
	year = {2010},
}

@misc{mascia_little_2025,
	title = {Little impact of mergers and galaxy morphology on the production and escape of ionizing photons in the early {Universe}},
	url = {http://arxiv.org/abs/2501.08268},
	doi = {10.48550/arXiv.2501.08268},
	urldate = {2025-05-22},
	publisher = {arXiv},
	author = {Mascia, S. and Pentericci, L. and Llerena, M. and Calabrò, A. and Matthee, J. and Flury, S. and Pacucci, F. and Jaskot, A. and Amorín, R. O. and Bhatawdekar, R. and Castellano, M. and Cleri, N. and Costantin, L. and Davis, K. and Cesare, C. Di and Dickinson, M. and Fontana, A. and Guo, Y. and Giavalisco, M. and Holwerda, B. W. and Hu, W. and Huertas-Company, M. and Jung, Intae and Kartaltepe, J. and Kashino, D. and Koekemoer, Anton M. and Lucas, R. A. and Lotz, J. and Napolitano, L. and Jogee, S. and Wilkins, S.},
	month = jan,
	year = {2025},
	note = {arXiv:2501.08268 [astro-ph]},
}

@article{heckman_extreme_2011,
	title = {Extreme {Feedback} and the {Epoch} of {Reionization}: {Clues} in the {Local} {Universe}},
	volume = {730},
	issn = {0004-637X},
	shorttitle = {Extreme {Feedback} and the {Epoch} of {Reionization}},
	url = {https://ui.adsabs.harvard.edu/abs/2011ApJ...730....5H},
	doi = {10.1088/0004-637X/730/1/5},
	urldate = {2025-05-22},
	journal = {The Astrophysical Journal},
	publisher = {IOP},
	author = {Heckman, Timothy M. and Borthakur, Sanchayeeta and Overzier, Roderik and Kauffmann, Guinevere and Basu-Zych, Antara and Leitherer, Claus and Sembach, Ken and Martin, D. Chris and Rich, R. Michael and Schiminovich, David and Seibert, Mark},
	month = mar,
	year = {2011},
	note = {ADS Bibcode: 2011ApJ...730....5H},
	pages = {5},
}

@article{katz_two_2023,
	title = {Two modes of {LyC} escape from bursty star formation: implications for [{C} ii] deficits and the sources of reionization},
	volume = {518},
	issn = {0035-8711},
	shorttitle = {Two modes of {LyC} escape from bursty star formation},
	url = {https://doi.org/10.1093/mnras/stac3019},
	doi = {10.1093/mnras/stac3019},
	number = {1},
	urldate = {2025-05-22},
	journal = {Monthly Notices of the Royal Astronomical Society},
	author = {Katz, Harley and Saxena, Aayush and Rosdahl, Joki and Kimm, Taysun and Blaizot, Jeremy and Garel, Thibault and Michel-Dansac, Leo and Haehnelt, Martin and Ellis, Richard S and Penterrici, Laura and Devriendt, Julien and Slyz, Adrianne},
	month = jan,
	year = {2023},
	pages = {270--285},
}

@article{sharma_winds_2017,
	title = {Winds of change: reionization by starburst galaxies},
	volume = {468},
	issn = {0035-8711, 1365-2966},
	shorttitle = {Winds of change},
	url = {https://durham-repository.worktribe.com/output/1387906/winds-of-change-reionization-by-starburst-galaxies},
	doi = {10.1093/mnras/stx578},
	language = {en},
	number = {2},
	urldate = {2025-05-22},
	journal = {Monthly Notices of the Royal Astronomical Society},
	publisher = {Royal Astronomical Society},
	author = {Sharma, M. and Theuns, T. and Frenk, C. and Bower, R. G. and Crain, R. A. and Schaller, M. and Schaye, J.},
	month = mar,
	year = {2017},
}

@article{madau_radiative_1999,
	title = {Radiative {Transfer} in a {Clumpy} {Universe}. {III}. {The} {Nature} of {Cosmological} {Ionizing} {Sources}},
	volume = {514},
	issn = {0004-637X},
	url = {https://ui.adsabs.harvard.edu/abs/1999ApJ...514..648M},
	doi = {10.1086/306975},
	urldate = {2025-05-23},
	journal = {The Astrophysical Journal},
	publisher = {IOP},
	author = {Madau, Piero and Haardt, Francesco and Rees, Martin J.},
	month = apr,
	year = {1999},
	note = {ADS Bibcode: 1999ApJ...514..648M},
	pages = {648--659},
}

@article{so_fully_2014,
	title = {{FULLY} {COUPLED} {SIMULATION} {OF} {COSMIC} {REIONIZATION}. {II}. {RECOMBINATIONS}, {CLUMPING} {FACTORS}, {AND} {THE} {PHOTON} {BUDGET} {FOR} {REIONIZATION}},
	volume = {789},
	issn = {0004-637X},
	url = {https://dx.doi.org/10.1088/0004-637X/789/2/149},
	doi = {10.1088/0004-637X/789/2/149},
	language = {en},
	number = {2},
	urldate = {2025-05-23},
	journal = {The Astrophysical Journal},
	publisher = {The American Astronomical Society},
	author = {So, Geoffrey C. and Norman, Michael L. and Reynolds, Daniel R. and Wise, John H.},
	month = jun,
	year = {2014},
	pages = {149},
}

@article{davies_constraints_2024,
	title = {Constraints on the {Evolution} of the {Ionizing} {Background} and {Ionizing} {Photon} {Mean} {Free} {Path} at the {End} of {Reionization}},
	volume = {965},
	issn = {0004-637X},
	url = {https://dx.doi.org/10.3847/1538-4357/ad1d5d},
	doi = {10.3847/1538-4357/ad1d5d},
	language = {en},
	number = {2},
	urldate = {2025-05-23},
	journal = {The Astrophysical Journal},
	publisher = {The American Astronomical Society},
	author = {Davies, Frederick B. and Bosman, Sarah E. I. and Gaikwad, Prakash and Nasir, Fahad and Hennawi, Joseph F. and Becker, George D. and Haehnelt, Martin G. and D’Odorico, Valentina and Bischetti, Manuela and Eilers, Anna-Christina and Keating, Laura C. and Kulkarni, Girish and Lai, Samuel and Mazzucchelli, Chiara and Qin, Yuxiang and Satyavolu, Sindhu and Wang, Feige and Yang, Jinyi and Zhu, Yongda},
	month = apr,
	year = {2024},
	pages = {134},
}

@article{rosdahl_sphinx_2018,
	title = {The {SPHINX} cosmological simulations of the first billion years: the impact of binary stars on reionization},
	volume = {479},
	issn = {0035-8711},
	shorttitle = {The {SPHINX} cosmological simulations of the first billion years},
	url = {https://ui.adsabs.harvard.edu/abs/2018MNRAS.479..994R},
	doi = {10.1093/mnras/sty1655},
	urldate = {2025-05-24},
	journal = {Monthly Notices of the Royal Astronomical Society},
	publisher = {OUP},
	author = {Rosdahl, Joakim and Katz, Harley and Blaizot, Jérémy and Kimm, Taysun and Michel-Dansac, Léo and Garel, Thibault and Haehnelt, Martin and Ocvirk, Pierre and Teyssier, Romain},
	month = sep,
	year = {2018},
	note = {ADS Bibcode: 2018MNRAS.479..994R},
	pages = {994--1016},
}

@article{kannan_simulating_2020,
	title = {Simulating the interstellar medium of galaxies with radiative transfer, non-equilibrium thermochemistry, and dust},
	volume = {499},
	issn = {0035-8711},
	url = {https://doi.org/10.1093/mnras/staa3249},
	doi = {10.1093/mnras/staa3249},
	number = {4},
	urldate = {2025-05-31},
	journal = {Monthly Notices of the Royal Astronomical Society},
	author = {Kannan, Rahul and Marinacci, Federico and Vogelsberger, Mark and Sales, Laura V and Torrey, Paul and Springel, Volker and Hernquist, Lars},
	month = nov,
	year = {2020},
	pages = {5732--5748},
}

@article{breiman_random_2001,
	title = {Random {Forests}},
	volume = {45},
	issn = {1573-0565},
	url = {https://doi.org/10.1023/A:1010933404324},
	doi = {10.1023/A:1010933404324},
	language = {en},
	number = {1},
	urldate = {2025-05-31},
	journal = {Machine Learning},
	author = {Breiman, Leo},
	month = oct,
	year = {2001},
	pages = {5--32},
}

@article{becker_new_2013,
	title = {New measurements of the ionizing ultraviolet background over 2 < z < 5 and implications for hydrogen reionization},
	volume = {436},
	issn = {0035-8711},
	url = {https://doi.org/10.1093/mnras/stt1610},
	doi = {10.1093/mnras/stt1610},
	number = {2},
	urldate = {2025-10-29},
	journal = {Monthly Notices of the Royal Astronomical Society},
	author = {Becker, George D. and Bolton, James S.},
	month = dec,
	year = {2013},
	pages = {1023--1039},
}

@article{johnson_stellar_2021,
	title = {Stellar {Population} {Inference} with {Prospector}},
	volume = {254},
	issn = {0067-0049},
	url = {https://ui.adsabs.harvard.edu/abs/2021ApJS..254...22J},
	doi = {10.3847/1538-4365/abef67},
	urldate = {2025-10-31},
	journal = {The Astrophysical Journal Supplement Series},
	publisher = {IOP},
	author = {Johnson, Benjamin D. and Leja, Joel and Conroy, Charlie and Speagle, Joshua S.},
	month = jun,
	year = {2021},
	note = {ADS Bibcode: 2021ApJS..254...22J},
	pages = {22},
}

@article{brammer_eazy_2008,
	title = {{EAZY}: {A} {Fast}, {Public} {Photometric} {Redshift} {Code}},
	volume = {686},
	issn = {0004-637X},
	shorttitle = {{EAZY}},
	url = {https://ui.adsabs.harvard.edu/abs/2008ApJ...686.1503B},
	doi = {10.1086/591786},
	urldate = {2025-10-31},
	journal = {The Astrophysical Journal},
	publisher = {IOP},
	author = {Brammer, Gabriel B. and van Dokkum, Pieter G. and Coppi, Paolo},
	month = oct,
	year = {2008},
	note = {ADS Bibcode: 2008ApJ...686.1503B},
	pages = {1503--1513},
}

@article{gaikwad_measuring_2023,
	title = {Measuring the photoionization rate, neutral fraction, and mean free path of {H} i ionizing photons at 4.9 ≤ z ≤ 6.0 from a large sample of {XShooter} and {ESI} spectra},
	volume = {525},
	issn = {0035-8711},
	url = {https://doi.org/10.1093/mnras/stad2566},
	doi = {10.1093/mnras/stad2566},
	number = {3},
	urldate = {2025-11-05},
	journal = {Monthly Notices of the Royal Astronomical Society},
	author = {Gaikwad, Prakash and Haehnelt, Martin G and Davies, Fredrick B and Bosman, Sarah E I and Molaro, Margherita and Kulkarni, Girish and D’Odorico, Valentina and Becker, George D and Davies, Rebecca L and Nasir, Fahad and Bolton, James S and Keating, Laura C and Iršič, Vid and Puchwein, Ewald and Zhu, Yongda and Asthana, Shikhar and Yang, Jinyi and Lai, Samuel and Eilers, Anna-Christina},
	month = nov,
	year = {2023},
	pages = {4093--4120},
}

@misc{eisenstein_overview_2023,
	title = {Overview of the {JWST} {Advanced} {Deep} {Extragalactic} {Survey} ({JADES})},
	url = {http://arxiv.org/abs/2306.02465},
	doi = {10.48550/arXiv.2306.02465},
	urldate = {2025-11-06},
	publisher = {arXiv},
	author = {Eisenstein, Daniel J. and Willott, Chris and Alberts, Stacey and Arribas, Santiago and Bonaventura, Nina and Bunker, Andrew J. and Cameron, Alex J. and Carniani, Stefano and Charlot, Stephane and Curtis-Lake, Emma and D'Eugenio, Francesco and Endsley, Ryan and Ferruit, Pierre and Giardino, Giovanna and Hainline, Kevin and Hausen, Ryan and Jakobsen, Peter and Johnson, Benjamin D. and Maiolino, Roberto and Rieke, Marcia and Rieke, George and Rix, Hans-Walter and Robertson, Brant and Stark, Daniel P. and Tacchella, Sandro and Williams, Christina C. and Willmer, Christopher N. A. and Baker, William M. and Baum, Stefi and Bhatawdekar, Rachana and Boyett, Kristan and Chen, Zuyi and Chevallard, Jacopo and Circosta, Chiara and Curti, Mirko and Danhaive, A. Lola and DeCoursey, Christa and Graaff, Anna de and Dressler, Alan and Egami, Eiichi and Helton, Jakob M. and Hviding, Raphael E. and Ji, Zhiyuan and Jones, Gareth C. and Kumari, Nimisha and Lützgendorf, Nora and Laseter, Isaac and Looser, Tobias J. and Lyu, Jianwei and Maseda, Michael V. and Nelson, Erica and Parlanti, Eleonora and Perna, Michele and Puskás, Dávid and Rawle, Tim and Pino, Bruno Rodríguez Del and Sandles, Lester and Saxena, Aayush and Scholtz, Jan and Sharpe, Katherine and Shivaei, Irene and Silcock, Maddie S. and Simmonds, Charlotte and Skarbinski, Maya and Smit, Renske and Stone, Meredith and Suess, Katherine A. and Sun, Fengwu and Tang, Mengtao and Topping, Michael W. and Übler, Hannah and Villanueva, Natalia C. and Wallace, Imaan E. B. and Whitler, Lily and Witstok, Joris and Woodrum, Charity},
	month = jun,
	year = {2023},
	note = {arXiv:2306.02465 [astro-ph]},
}

@misc{simmonds_bursting_2025,
	title = {Bursting at the seams: the star-forming main sequence and its scatter at z=3-9 using {NIRCam} photometry from {JADES}},
	shorttitle = {Bursting at the seams},
	url = {http://arxiv.org/abs/2508.04410},
	doi = {10.48550/arXiv.2508.04410},
	urldate = {2025-11-06},
	publisher = {arXiv},
	author = {Simmonds, C. and Tacchella, S. and McClymont, W. and Curtis-Lake, E. and D'Eugenio, F. and Hainline, K. and Johnson, B. D. and Kravtsov, A. and Puskás, D. and Robertson, B. and Stoffers, A. and Willott, C. and Baker, W. M. and Belokurov, V. A. and Bhatawdekar, R. and Bunker, A. J. and Carniani, S. and Chevallard, J. and Curti, M. and Duan, Q. and Helton, J. M. and Ji, Z. and Looser, T. J. and Maiolino, R. and Maseda, M. V. and Shivaei, I. and Williams, C. C.},
	month = nov,
	year = {2025},
	note = {arXiv:2508.04410 [astro-ph]},
}

@article{bunker_jades_2024,
	title = {{JADES} {NIRSpec} initial data release for the {Hubble} {Ultra} {Deep} {Field}: {Redshifts} and line fluxes of distant galaxies from the deepest {JWST} {Cycle} 1 {NIRSpec} multi-object spectroscopy},
	volume = {690},
	issn = {0004-6361},
	shorttitle = {{JADES} {NIRSpec} initial data release for the {Hubble} {Ultra} {Deep} {Field}},
	url = {https://ui.adsabs.harvard.edu/abs/2024A&A...690A.288B},
	doi = {10.1051/0004-6361/202347094},
	urldate = {2025-11-06},
	journal = {Astronomy and Astrophysics},
	publisher = {EDP},
	author = {Bunker, Andrew J. and Cameron, Alex J. and Curtis-Lake, Emma and Jakobsen, Peter and Carniani, Stefano and Curti, Mirko and Witstok, Joris and Maiolino, Roberto and D'Eugenio, Francesco and Looser, Tobias J. and Willott, Chris and Bonaventura, Nina and Hainline, Kevin and Übler, Hannah and Willmer, Christopher N. A. and Saxena, Aayush and Smit, Renske and Alberts, Stacey and Arribas, Santiago and Baker, William M. and Baum, Stefi and Bhatawdekar, Rachana and Bowler, Rebecca A. A. and Boyett, Kristan and Charlot, Stephane and Chen, Zuyi and Chevallard, Jacopo and Circosta, Chiara and DeCoursey, Christa and de Graaff, Anna and Egami, Eiichi and Eisenstein, Daniel J. and Endsley, Ryan and Ferruit, Pierre and Giardino, Giovanna and Hausen, Ryan and Helton, Jakob M. and Hviding, Raphael E. and Ji, Zhiyuan and Johnson, Benjamin D. and Jones, Gareth C. and Kumari, Nimisha and Laseter, Isaac and Lützgendorf, Nora and Maseda, Michael V. and Nelson, Erica and Parlanti, Eleonora and Perna, Michele and Rauscher, Bernard J. and Rawle, Tim and Rix, Hans-Walter and Rieke, Marcia and Robertson, Brant and Rodríguez Del Pino, Bruno and Sandles, Lester and Scholtz, Jan and Sharpe, Katherine and Skarbinski, Maya and Stark, Daniel P. and Sun, Fengwu and Tacchella, Sandro and Topping, Michael W. and Villanueva, Natalia C. and Wallace, Imaan E. B. and Williams, Christina C. and Woodrum, Charity},
	month = oct,
	year = {2024},
	note = {ADS Bibcode: 2024A\&A...690A.288B},
	pages = {A288},
}

@article{chabrier_galactic_2003,
	title = {Galactic {Stellar} and {Substellar} {Initial} {Mass} {Function}},
	volume = {115},
	issn = {0004-6280},
	url = {https://ui.adsabs.harvard.edu/abs/2003PASP..115..763C},
	doi = {10.1086/376392},
	urldate = {2025-11-06},
	journal = {Publications of the Astronomical Society of the Pacific},
	publisher = {IOP},
	author = {Chabrier, Gilles},
	month = jul,
	year = {2003},
	note = {ADS Bibcode: 2003PASP..115..763C},
	pages = {763--795},
}

@article{charlot_simple_2000,
	title = {A {Simple} {Model} for the {Absorption} of {Starlight} by {Dust} in {Galaxies}},
	volume = {539},
	issn = {0004-637X},
	url = {https://ui.adsabs.harvard.edu/abs/2000ApJ...539..718C},
	doi = {10.1086/309250},
	urldate = {2025-11-06},
	journal = {The Astrophysical Journal},
	publisher = {IOP},
	author = {Charlot, Stéphane and Fall, S. Michael},
	month = aug,
	year = {2000},
	note = {ADS Bibcode: 2000ApJ...539..718C},
	pages = {718--731},
}

@article{conroy_propagation_2009,
	title = {The {Propagation} of {Uncertainties} in {Stellar} {Population} {Synthesis} {Modeling}. {I}. {The} {Relevance} of {Uncertain} {Aspects} of {Stellar} {Evolution} and the {Initial} {Mass} {Function} to the {Derived} {Physical} {Properties} of {Galaxies}},
	volume = {699},
	issn = {0004-637X},
	url = {https://ui.adsabs.harvard.edu/abs/2009ApJ...699..486C},
	doi = {10.1088/0004-637X/699/1/486},
	urldate = {2025-11-06},
	journal = {The Astrophysical Journal},
	publisher = {IOP},
	author = {Conroy, Charlie and Gunn, James E. and White, Martin},
	month = jul,
	year = {2009},
	note = {ADS Bibcode: 2009ApJ...699..486C},
	pages = {486--506},
}

@article{leja_how_2019,
	title = {How to {Measure} {Galaxy} {Star} {Formation} {Histories}. {II}. {Nonparametric} {Models}},
	volume = {876},
	issn = {0004-637X},
	url = {https://ui.adsabs.harvard.edu/abs/2019ApJ...876....3L},
	doi = {10.3847/1538-4357/ab133c},
	urldate = {2025-11-06},
	journal = {The Astrophysical Journal},
	publisher = {IOP},
	author = {Leja, Joel and Carnall, Adam C. and Johnson, Benjamin D. and Conroy, Charlie and Speagle, Joshua S.},
	month = may,
	year = {2019},
	note = {ADS Bibcode: 2019ApJ...876....3L},
	pages = {3},
}

@article{kostyuk_ionizing_2023,
	title = {Ionizing photon production and escape fractions during cosmic reionization in the {TNG50} simulation},
	volume = {521},
	issn = {0035-8711},
	url = {https://doi.org/10.1093/mnras/stad677},
	doi = {10.1093/mnras/stad677},
	number = {2},
	urldate = {2025-11-10},
	journal = {Monthly Notices of the Royal Astronomical Society},
	author = {Kostyuk, Ivan and Nelson, Dylan and Ciardi, Benedetta and Glatzle, Martin and Pillepich, Annalisa},
	month = may,
	year = {2023},
	pages = {3077--3097},
}

@misc{chakraborty_modelling_2024,
	title = {Modelling the star-formation activity and ionizing properties of high-redshift galaxies},
	url = {https://arxiv.org/abs/2404.02879v3},
	doi = {10.1088/1475-7516/2024/07/078},
	language = {en},
	urldate = {2025-11-13},
	journal = {arXiv.org},
	author = {Chakraborty, Anirban and Choudhury, Tirthankar Roy},
	month = apr,
	year = {2024},
}

@misc{wu_effect_2025,
	title = {Effect of ionizing photon escape fraction in faint galaxies on modeling reionization history of the universe},
	url = {http://arxiv.org/abs/2511.07543},
	doi = {10.48550/arXiv.2511.07543},
	urldate = {2025-11-13},
	publisher = {arXiv},
	author = {Wu, Zewei and Kravtsov, Andrey and Katz, Harley},
	month = nov,
	year = {2025},
	note = {arXiv:2511.07543 [astro-ph]},
}

@article{whitler_z_2025,
	title = {The z ≳ 9 {Galaxy} {UV} {Luminosity} {Function} from the {JWST} {Advanced} {Deep} {Extragalactic} {Survey}: {Insights} into {Early} {Galaxy} {Evolution} and {Reionization}},
	volume = {992},
	issn = {0004-637X},
	shorttitle = {The z ≳ 9 {Galaxy} {UV} {Luminosity} {Function} from the {JWST} {Advanced} {Deep} {Extragalactic} {Survey}},
	url = {https://ui.adsabs.harvard.edu/abs/2025ApJ...992...63W},
	doi = {10.3847/1538-4357/adfddc},
	urldate = {2025-11-14},
	journal = {The Astrophysical Journal},
	publisher = {IOP},
	author = {Whitler, Lily and Stark, Daniel P. and Topping, Michael W. and Robertson, Brant and Rieke, Marcia and Hainline, Kevin N. and Endsley, Ryan and Chen, Zuyi and Baker, William M. and Bhatawdekar, Rachana and Bunker, Andrew J. and Carniani, Stefano and Charlot, Stéphane and Chevallard, Jacopo and Curtis-Lake, Emma and Egami, Eiichi and Eisenstein, Daniel J. and Helton, Jakob M. and Ji, Zhiyuan and Johnson, Benjamin D. and Pérez-González, Pablo G. and Rinaldi, Pierluigi and Tacchella, Sandro and Williams, Christina C. and Willmer, Christopher N. A. and Willott, Chris and Witstok, Joris},
	month = oct,
	year = {2025},
	note = {ADS Bibcode: 2025ApJ...992...63W},
	pages = {63},
}

@article{madau_cosmic_2024,
	title = {Cosmic {Reionization} in the {JWST} {Era}: {Back} to {AGNs}?},
	volume = {971},
	issn = {0004-637X},
	shorttitle = {Cosmic {Reionization} in the {JWST} {Era}},
	url = {https://doi.org/10.3847/1538-4357/ad5ce8},
	doi = {10.3847/1538-4357/ad5ce8},
	language = {en},
	number = {1},
	urldate = {2025-11-17},
	journal = {The Astrophysical Journal},
	publisher = {The American Astronomical Society},
	author = {Madau, Piero and Giallongo, Emanuele and Grazian, Andrea and Haardt, Francesco},
	month = aug,
	year = {2024},
	pages = {75},
}

@article{shen_impact_2023,
	title = {The impact of {UV} variability on the abundance of bright galaxies at z ≥ 9},
	volume = {525},
	issn = {0035-8711},
	url = {https://ui.adsabs.harvard.edu/abs/2023MNRAS.525.3254S},
	doi = {10.1093/mnras/stad2508},
	urldate = {2025-11-20},
	journal = {Monthly Notices of the Royal Astronomical Society},
	publisher = {OUP},
	author = {Shen, Xuejian and Vogelsberger, Mark and Boylan-Kolchin, Michael and Tacchella, Sandro and Kannan, Rahul},
	month = nov,
	year = {2023},
	note = {ADS Bibcode: 2023MNRAS.525.3254S},
	pages = {3254--3261},
}

@misc{kageura_census_2025,
	title = {Census of {Ly}$\alpha$ {Emission} from $\sim 600$ {Galaxies} at $z=5-14$: {Evolution} of the {Ly}$\alpha$ {Luminosity} {Function} and a {Late} {Sharp} {Cosmic} {Reionization}},
	shorttitle = {Census of {Ly}$\alpha$ {Emission} from $\sim 600$ {Galaxies} at $z=5-14$},
	url = {http://arxiv.org/abs/2501.05834},
	doi = {10.48550/arXiv.2501.05834},
	urldate = {2025-11-24},
	publisher = {arXiv},
	author = {Kageura, Yuta and Ouchi, Masami and Nakane, Minami and Umeda, Hiroya and Harikane, Yuichi and Yoshiura, Shintaro and Nakajima, Kimihiko and Yajima, Hidenobu and Thai, Tran Thi},
	month = mar,
	year = {2025},
	note = {arXiv:2501.05834 [astro-ph]},
}

@article{davies_quantitative_2018,
	title = {Quantitative {Constraints} on the {Reionization} {History} from the {IGM} {Damping} {Wing} {Signature} in {Two} {Quasars} at z > 7},
	volume = {864},
	issn = {0004-637X},
	url = {https://doi.org/10.3847/1538-4357/aad6dc},
	doi = {10.3847/1538-4357/aad6dc},
	language = {en},
	number = {2},
	urldate = {2025-11-24},
	journal = {The Astrophysical Journal},
	publisher = {The American Astronomical Society},
	author = {Davies, Frederick B. and Hennawi, Joseph F. and Bañados, Eduardo and Lukić, Zarija and Decarli, Roberto and Fan, Xiaohui and Farina, Emanuele P. and Mazzucchelli, Chiara and Rix, Hans-Walter and Venemans, Bram P. and Walter, Fabian and Wang, Feige and Yang, Jinyi},
	month = sep,
	year = {2018},
	pages = {142},
}

@article{bowler_lack_2020,
	title = {A lack of evolution in the very bright end of the galaxy luminosity function from z ≃ 8 to 10},
	volume = {493},
	issn = {0035-8711},
	url = {https://ui.adsabs.harvard.edu/abs/2020MNRAS.493.2059B},
	doi = {10.1093/mnras/staa313},
	urldate = {2025-11-26},
	journal = {Monthly Notices of the Royal Astronomical Society},
	publisher = {OUP},
	author = {Bowler, R. A. A. and Jarvis, M. J. and Dunlop, J. S. and McLure, R. J. and McLeod, D. J. and Adams, N. J. and Milvang-Jensen, B. and McCracken, H. J.},
	month = apr,
	year = {2020},
	note = {ADS Bibcode: 2020MNRAS.493.2059B},
	pages = {2059--2084},
}

@article{aghanim_planck_2020,
	title = {Planck 2018 results - {VI}. {Cosmological} parameters},
	volume = {641},
	copyright = {© ESO 2020},
	issn = {0004-6361, 1432-0746},
	url = {https://www.aanda.org/articles/aa/abs/2020/09/aa33910-18/aa33910-18.html},
	doi = {10.1051/0004-6361/201833910},
	language = {en},
	urldate = {2025-11-26},
	journal = {Astronomy \& Astrophysics},
	publisher = {EDP Sciences},
	author = {Aghanim, N. and Akrami, Y. and Ashdown, M. and Aumont, J. and Baccigalupi, C. and Ballardini, M. and Banday, A. J. and Barreiro, R. B. and Bartolo, N. and Basak, S. and Battye, R. and Benabed, K. and Bernard, J.-P. and Bersanelli, M. and Bielewicz, P. and Bock, J. J. and Bond, J. R. and Borrill, J. and Bouchet, F. R. and Boulanger, F. and Bucher, M. and Burigana, C. and Butler, R. C. and Calabrese, E. and Cardoso, J.-F. and Carron, J. and Challinor, A. and Chiang, H. C. and Chluba, J. and Colombo, L. P. L. and Combet, C. and Contreras, D. and Crill, B. P. and Cuttaia, F. and Bernardis, P. de and Zotti, G. de and Delabrouille, J. and Delouis, J.-M. and Valentino, E. Di and Diego, J. M. and Doré, O. and Douspis, M. and Ducout, A. and Dupac, X. and Dusini, S. and Efstathiou, G. and Elsner, F. and Enßlin, T. A. and Eriksen, H. K. and Fantaye, Y. and Farhang, M. and Fergusson, J. and Fernandez-Cobos, R. and Finelli, F. and Forastieri, F. and Frailis, M. and Fraisse, A. A. and Franceschi, E. and Frolov, A. and Galeotta, S. and Galli, S. and Ganga, K. and Génova-Santos, R. T. and Gerbino, M. and Ghosh, T. and González-Nuevo, J. and Górski, K. M. and Gratton, S. and Gruppuso, A. and Gudmundsson, J. E. and Hamann, J. and Handley, W. and Hansen, F. K. and Herranz, D. and Hildebrandt, S. R. and Hivon, E. and Huang, Z. and Jaffe, A. H. and Jones, W. C. and Karakci, A. and Keihänen, E. and Keskitalo, R. and Kiiveri, K. and Kim, J. and Kisner, T. S. and Knox, L. and Krachmalnicoff, N. and Kunz, M. and Kurki-Suonio, H. and Lagache, G. and Lamarre, J.-M. and Lasenby, A. and Lattanzi, M. and Lawrence, C. R. and Jeune, M. Le and Lemos, P. and Lesgourgues, J. and Levrier, F. and Lewis, A. and Liguori, M. and Lilje, P. B. and Lilley, M. and Lindholm, V. and López-Caniego, M. and Lubin, P. M. and Ma, Y.-Z. and Macías-Pérez, J. F. and Maggio, G. and Maino, D. and Mandolesi, N. and Mangilli, A. and Marcos-Caballero, A. and Maris, M. and Martin, P. G. and Martinelli, M. and Martínez-González, E. and Matarrese, S. and Mauri, N. and McEwen, J. D. and Meinhold, P. R. and Melchiorri, A. and Mennella, A. and Migliaccio, M. and Millea, M. and Mitra, S. and Miville-Deschênes, M.-A. and Molinari, D. and Montier, L. and Morgante, G. and Moss, A. and Natoli, P. and Nørgaard-Nielsen, H. U. and Pagano, L. and Paoletti, D. and Partridge, B. and Patanchon, G. and Peiris, H. V. and Perrotta, F. and Pettorino, V. and Piacentini, F. and Polastri, L. and Polenta, G. and Puget, J.-L. and Rachen, J. P. and Reinecke, M. and Remazeilles, M. and Renzi, A. and Rocha, G. and Rosset, C. and Roudier, G. and Rubiño-Martín, J. A. and Ruiz-Granados, B. and Salvati, L. and Sandri, M. and Savelainen, M. and Scott, D. and Shellard, E. P. S. and Sirignano, C. and Sirri, G. and Spencer, L. D. and Sunyaev, R. and Suur-Uski, A.-S. and Tauber, J. A. and Tavagnacco, D. and Tenti, M. and Toffolatti, L. and Tomasi, M. and Trombetti, T. and Valenziano, L. and Valiviita, J. and Tent, B. Van and Vibert, L. and Vielva, P. and Villa, F. and Vittorio, N. and Wandelt, B. D. and Wehus, I. K. and White, M. and White, S. D. M. and Zacchei, A. and Zonca, A.},
	month = sep,
	year = {2020},
	pages = {A6},
}

@misc{tang_jwstnirspec_2024,
	title = {{JWST}/{NIRSpec} {Observations} of {Ly}$\alpha$ {Emission} in {Star} {Forming} {Galaxies} at $6.5\lesssim z\lesssim13$},
	url = {http://arxiv.org/abs/2408.01507},
	doi = {10.48550/arXiv.2408.01507},
	urldate = {2025-11-26},
	publisher = {arXiv},
	author = {Tang, Mengtao and Stark, Daniel P. and Topping, Michael W. and Mason, Charlotte and Ellis, Richard S.},
	month = sep,
	year = {2024},
	note = {arXiv:2408.01507 [astro-ph]},
}

@article{jin_nearly_2023,
	title = {({Nearly}) {Model}-independent {Constraints} on the {Neutral} {Hydrogen} {Fraction} in the {Intergalactic} {Medium} at z ∼ 5–7 {Using} {Dark} {Pixel} {Fractions} in {Lyα} and {Lyβ} {Forests}},
	volume = {942},
	issn = {0004-637X},
	url = {https://doi.org/10.3847/1538-4357/aca678},
	doi = {10.3847/1538-4357/aca678},
	language = {en},
	number = {2},
	urldate = {2025-11-26},
	journal = {The Astrophysical Journal},
	publisher = {The American Astronomical Society},
	author = {Jin, Xiangyu and Yang, Jinyi and Fan, Xiaohui and Wang, Feige and Bañados, Eduardo and Bian, Fuyan and Davies, Frederick B. and Eilers, Anna-Christina and Farina, Emanuele Paolo and Hennawi, Joseph F. and Pacucci, Fabio and Venemans, Bram and Walter, Fabian},
	month = jan,
	year = {2023},
	pages = {59},
}

@article{robertson_cosmic_2015,
	title = {{COSMIC} {REIONIZATION} {AND} {EARLY} {STAR}-{FORMING} {GALAXIES}: {A} {JOINT} {ANALYSIS} {OF} {NEW} {CONSTRAINTS} {FROM} {PLANCK} {AND} {THE} {HUBBLE} {SPACE} ℡{ESCOPE}},
	volume = {802},
	issn = {2041-8205},
	shorttitle = {{COSMIC} {REIONIZATION} {AND} {EARLY} {STAR}-{FORMING} {GALAXIES}},
	url = {https://doi.org/10.1088/2041-8205/802/2/L19},
	doi = {10.1088/2041-8205/802/2/L19},
	language = {en},
	number = {2},
	urldate = {2025-12-02},
	journal = {The Astrophysical Journal Letters},
	publisher = {The American Astronomical Society},
	author = {Robertson, Brant E. and Ellis, Richard S. and Furlanetto, Steven R. and Dunlop, James S.},
	month = apr,
	year = {2015},
	pages = {L19},
}

@article{rinaldi_midis_2024,
	title = {{MIDIS}: {Unveiling} the {Role} of {Strong} {Hα} {Emitters} {During} the {Epoch} of {Reionization} with {JWST}},
	volume = {969},
	issn = {0004-637X},
	shorttitle = {{MIDIS}},
	url = {https://doi.org/10.3847/1538-4357/ad4147},
	doi = {10.3847/1538-4357/ad4147},
	language = {en},
	number = {1},
	urldate = {2025-12-04},
	journal = {The Astrophysical Journal},
	publisher = {The American Astronomical Society},
	author = {Rinaldi, P. and Caputi, K. I. and Iani, E. and Costantin, L. and Gillman, S. and Perez Gonzalez, P. G. and Östlin, G. and Colina, L. and Greve, T. R. and Nørgard-Nielsen, H. U. and Wright, G. S. and Álvarez-Márquez, J. and Eckart, A. and García-Marín, M. and Hjorth, J. and Ilbert, O. and Kendrew, S. and Labiano, A. and Le Fèvre, O. and Pye, J. and Tikkanen, T. and Walter, F. and van der Werf, P. and Ward, M. and Annunziatella, M. and Azzollini, R. and Bik, A. and Boogaard, L. and Bosman, S. E. I. and Crespo Gómez, A. and Jermann, I. and Langeroodi, D. and Melinder, J. and Meyer, R. A. and Moutard, T. and Peissker, F. and van Dishoeck, E. and Güdel, M. and Henning, Th. and Lagage, P.-O. and Ray, T. and Vandenbussche, B. and Waelkens, C. and Dayal, Pratika},
	month = jun,
	year = {2024},
	pages = {12},
}

@article{chisholm_far-ultraviolet_2022,
	title = {The far-ultraviolet continuum slope as a {Lyman} {Continuum} escape estimator at high redshift},
	volume = {517},
	issn = {0035-8711},
	url = {https://doi.org/10.1093/mnras/stac2874},
	doi = {10.1093/mnras/stac2874},
	number = {4},
	urldate = {2025-12-04},
	journal = {Monthly Notices of the Royal Astronomical Society},
	author = {Chisholm, J and Saldana-Lopez, A and Flury, S and Schaerer, D and Jaskot, A and Amorín, R and Atek, H and Finkelstein, S L and Fleming, B and Ferguson, H and Fernández, V and Giavalisco, M and Hayes, M and Heckman, T and Henry, A and Ji, Z and Marques-Chaves, R and Mauerhofer, V and McCandliss, S and Oey, M S and Östlin, G and Rutkowski, M and Scarlata, C and Thuan, T and Trebitsch, M and Wang, B and Worseck, G and Xu, X},
	month = dec,
	year = {2022},
	pages = {5104--5120},
}

@article{wang_significantly_2020,
	title = {A {Significantly} {Neutral} {Intergalactic} {Medium} {Around} the {Luminous} z=7 {Quasar} {J0252}-0503},
	volume = {896},
	issn = {0004-637X, 1538-4357},
	url = {http://arxiv.org/abs/2004.10877},
	doi = {10.3847/1538-4357/ab8c45},
	number = {1},
	urldate = {2025-12-04},
	journal = {The Astrophysical Journal},
	author = {Wang, Feige and Davies, Frederick B. and Yang, Jinyi and Hennawi, Joseph F. and Fan, Xiaohui and Barth, Aaron J. and Jiang, Linhua and Wu, Xue-Bing and Mudd, Dale M. and Banados, Eduardo and Bian, Fuyan and Decarli, Roberto and Eilers, Anna-Christina and Farina, Emanuele Paolo and Venemans, Bram and Walter, Fabian and Yue, Minghao},
	month = jun,
	year = {2020},
	note = {arXiv:2004.10877 [astro-ph]},
	pages = {23},
}

@article{larson_searching_2022,
	title = {Searching for {Islands} of {Reionization}: {A} {Potential} {Ionized} {Bubble} {Powered} by a {Spectroscopic} {Overdensity} at z = 8.7},
	volume = {930},
	issn = {0004-637X},
	shorttitle = {Searching for {Islands} of {Reionization}},
	url = {https://ui.adsabs.harvard.edu/abs/2022ApJ...930..104L},
	doi = {10.3847/1538-4357/ac5dbd},
	urldate = {2025-12-09},
	journal = {The Astrophysical Journal},
	publisher = {IOP},
	author = {Larson, Rebecca L. and Finkelstein, Steven L. and Hutchison, Taylor A. and Papovich, Casey and Bagley, Micaela and Dickinson, Mark and Rojas-Ruiz, Sofía and Ferguson, Harry C. and Jung, Intae and Giavalisco, Mauro and Grazian, Andrea and Pentericci, Laura and Tacchella, Sandro},
	month = may,
	year = {2022},
	note = {ADS Bibcode: 2022ApJ...930..104L},
	pages = {104},
}

@article{roberts-borsani_nature_2023,
	title = {Nature and {Nurture}? {Comparing} {Lyα} {Detections} in {UV}-bright and {Fainter} [{O} iii]+{Hβ} {Emitters} at z ∼ 8 with {Keck}/{MOSFIRE}},
	volume = {948},
	issn = {0004-637X},
	shorttitle = {Nature and {Nurture}?},
	url = {https://doi.org/10.3847/1538-4357/acc798},
	doi = {10.3847/1538-4357/acc798},
	language = {en},
	number = {1},
	urldate = {2025-12-09},
	journal = {The Astrophysical Journal},
	publisher = {The American Astronomical Society},
	author = {Roberts-Borsani, Guido and Treu, Tommaso and Mason, Charlotte and Ellis, Richard S. and Laporte, Nicolas and Schmidt, Thomas and Bradac, Marusa and Fontana, Adriano and Morishita, Takahiro and Santini, Paola},
	month = may,
	year = {2023},
	pages = {54},
}

@article{ren_brightest_2019,
	title = {The {Brightest} {Galaxies} at {Cosmic} {Dawn} from {Scatter} in the {Galaxy} {Luminosity} versus {Halo} {Mass} {Relation}},
	volume = {878},
	issn = {0004-637X},
	url = {https://doi.org/10.3847/1538-4357/ab2117},
	doi = {10.3847/1538-4357/ab2117},
	language = {en},
	number = {2},
	urldate = {2025-12-11},
	journal = {The Astrophysical Journal},
	publisher = {The American Astronomical Society},
	author = {Ren, Keven and Trenti, Michele and Mason, Charlotte A.},
	month = jun,
	year = {2019},
	pages = {114},
}

@misc{austin_resolving_2025,
	title = {Resolving the ionizing photon budget crisis with {JWST}/{NIRCam} {HII} clumping constraints at z=6},
	url = {http://arxiv.org/abs/2512.10839},
	doi = {10.48550/arXiv.2512.10839},
	urldate = {2025-12-12},
	publisher = {arXiv},
	author = {Austin, Duncan and Harvey, Thomas and Conselice, Christopher J. and Adams, Nathan J. and Rusakov, Vadim and Li, Qiong and Westcott, Lewi and Goolsby, Caio and Madgwick, Kai and Arcidiacono, James and Ricotti, Massimo and Newman, Sophie L. and Seeyave, Louise T. C. and Trussler, James and Frye, Brenda and Grogin, Norman A. and Jansen, Rolf A. and Koekemoer, Anton M. and Pirzkal, Nor and Rutkowski, Michael and Windhorst, Rogier A.},
	month = dec,
	year = {2025},
	note = {arXiv:2512.10839 [astro-ph]},
}

@article{finlator_new_2009,
	title = {A new moment method for continuum radiative transfer in cosmological re-ionization},
	volume = {393},
	issn = {0035-8711},
	url = {https://doi.org/10.1111/j.1365-2966.2008.14190.x},
	doi = {10.1111/j.1365-2966.2008.14190.x},
	number = {4},
	urldate = {2025-12-15},
	journal = {Monthly Notices of the Royal Astronomical Society},
	author = {Finlator, Kristian and Özel, Feryal and Davé, Romeel},
	month = mar,
	year = {2009},
	pages = {1090--1106},
}

@article{chen_scorch_2020,
	title = {{SCORCH}. {III}. {Analytical} {Models} of {Reionization} with {Varying} {Clumping} {Factors}},
	volume = {905},
	issn = {0004-637X},
	url = {https://doi.org/10.3847/1538-4357/abc890},
	doi = {10.3847/1538-4357/abc890},
	language = {en},
	number = {2},
	urldate = {2025-12-15},
	journal = {The Astrophysical Journal},
	publisher = {The American Astronomical Society},
	author = {Chen, Nianyi and Doussot, Aristide and Trac, Hy and Cen, Renyue},
	month = dec,
	year = {2020},
	pages = {132},
}

@article{virtanen_scipy_2020,
	title = {{SciPy} 1.0: fundamental algorithms for scientific computing in {Python}},
	volume = {17},
	copyright = {2020 The Author(s)},
	issn = {1548-7105},
	shorttitle = {{SciPy} 1.0},
	url = {https://www.nature.com/articles/s41592-019-0686-2},
	doi = {10.1038/s41592-019-0686-2},
	language = {en},
	number = {3},
	urldate = {2025-12-15},
	journal = {Nature Methods},
	publisher = {Nature Publishing Group},
	author = {Virtanen, Pauli and Gommers, Ralf and Oliphant, Travis E. and Haberland, Matt and Reddy, Tyler and Cournapeau, David and Burovski, Evgeni and Peterson, Pearu and Weckesser, Warren and Bright, Jonathan and van der Walt, Stéfan J. and Brett, Matthew and Wilson, Joshua and Millman, K. Jarrod and Mayorov, Nikolay and Nelson, Andrew R. J. and Jones, Eric and Kern, Robert and Larson, Eric and Carey, C. J. and Polat, İlhan and Feng, Yu and Moore, Eric W. and VanderPlas, Jake and Laxalde, Denis and Perktold, Josef and Cimrman, Robert and Henriksen, Ian and Quintero, E. A. and Harris, Charles R. and Archibald, Anne M. and Ribeiro, Antônio H. and Pedregosa, Fabian and van Mulbregt, Paul},
	month = mar,
	year = {2020},
	pages = {261--272},
}

@article{pannella_goods-herschel_2015,
	title = {{GOODS}-{HERSCHEL}: {STAR} {FORMATION}, {DUST} {ATTENUATION}, {AND} {THE} {FIR}–{RADIO} {CORRELATION} {ON} {THE} {MAIN} {SEQUENCE} {OF} {STAR}-{FORMING} {GALAXIES} {UP} {TO} z ≃ 4*},
	volume = {807},
	issn = {0004-637X},
	shorttitle = {{GOODS}-{HERSCHEL}},
	url = {https://doi.org/10.1088/0004-637X/807/2/141},
	doi = {10.1088/0004-637X/807/2/141},
	language = {en},
	number = {2},
	urldate = {2025-12-16},
	journal = {The Astrophysical Journal},
	publisher = {The American Astronomical Society},
	author = {Pannella, M. and Elbaz, D. and Daddi, E. and Dickinson, M. and Hwang, H. S. and Schreiber, C. and Strazzullo, V. and Aussel, H. and Bethermin, M. and Buat, V. and Charmandaris, V. and Cibinel, A. and Juneau, S. and Ivison, R. J. and Borgne, D. Le and Floc’h, E. Le and Leiton, R. and Lin, L. and Magdis, G. and Morrison, G. E. and Mullaney, J. and Onodera, M. and Renzini, A. and Salim, S. and Sargent, M. T. and Scott, D. and Shu, X. and Wang, T.},
	month = jul,
	year = {2015},
	pages = {141},
}

@article{madau_cosmic_2014,
	title = {Cosmic {Star}-{Formation} {History}},
	volume = {52},
	issn = {0066-4146},
	url = {https://ui.adsabs.harvard.edu/abs/2014ARA&A..52..415M},
	doi = {10.1146/annurev-astro-081811-125615},
	urldate = {2025-12-17},
	journal = {Annual Review of Astronomy and Astrophysics},
	author = {Madau, Piero and Dickinson, Mark},
	month = aug,
	year = {2014},
	note = {ADS Bibcode: 2014ARA\&A..52..415M},
	pages = {415--486},
}

@article{davis_evolution_1985,
	title = {The evolution of large-scale structure in a universe dominated by cold dark matter},
	volume = {292},
	issn = {0004-637X},
	url = {https://ui.adsabs.harvard.edu/abs/1985ApJ...292..371D},
	doi = {10.1086/163168},
	urldate = {2025-12-17},
	journal = {The Astrophysical Journal},
	publisher = {IOP},
	author = {Davis, M. and Efstathiou, G. and Frenk, C. S. and White, S. D. M.},
	month = may,
	year = {1985},
	note = {ADS Bibcode: 1985ApJ...292..371D},
	pages = {371--394},
}

@article{springel_populating_2001,
	title = {Populating a cluster of galaxies – {I}. {Results} at z = 0},
	volume = {328},
	issn = {0035-8711},
	url = {https://doi.org/10.1046/j.1365-8711.2001.04912.x},
	doi = {10.1046/j.1365-8711.2001.04912.x},
	number = {3},
	urldate = {2025-12-17},
	journal = {Monthly Notices of the Royal Astronomical Society},
	author = {Springel, Volker and White, Simon D. M. and Tormen, Giuseppe and Kauffmann, Guinevere},
	month = dec,
	year = {2001},
	pages = {726--750},
}

@article{springel_simulating_2021,
	title = {Simulating cosmic structure formation with the gadget-4 code},
	volume = {506},
	issn = {0035-8711},
	url = {https://doi.org/10.1093/mnras/stab1855},
	doi = {10.1093/mnras/stab1855},
	number = {2},
	urldate = {2025-12-17},
	journal = {Monthly Notices of the Royal Astronomical Society},
	author = {Springel, Volker and Pakmor, Rüdiger and Zier, Oliver and Reinecke, Martin},
	month = sep,
	year = {2021},
	pages = {2871--2949},
}

@article{bosman_hydrogen_2022,
	title = {Hydrogen reionization ends by z = 5.3: {Lyman}-α optical depth measured by the {XQR}-30 sample},
	volume = {514},
	issn = {0035-8711},
	shorttitle = {Hydrogen reionization ends by z = 5.3},
	url = {https://doi.org/10.1093/mnras/stac1046},
	doi = {10.1093/mnras/stac1046},
	number = {1},
	urldate = {2025-12-18},
	journal = {Monthly Notices of the Royal Astronomical Society},
	author = {Bosman, Sarah E I and Davies, Frederick B and Becker, George D and Keating, Laura C and Davies, Rebecca L and Zhu, Yongda and Eilers, Anna-Christina and D’Odorico, Valentina and Bian, Fuyan and Bischetti, Manuela and Cristiani, Stefano V and Fan, Xiaohui and Farina, Emanuele P and Haehnelt, Martin G and Hennawi, Joseph F and Kulkarni, Girish and Mesinger, Andrei and Meyer, Romain A and Onoue, Masafusa and Pallottini, Andrea and Qin, Yuxiang and Ryan-Weber, Emma and Schindler, Jan-Torge and Walter, Fabian and Wang, Feige and Yang, Jinyi},
	month = jul,
	year = {2022},
	pages = {55--76},
}

@article{witstok_witnessing_2025,
	title = {Witnessing the onset of reionisation via {Lyman}-$\alpha$ emission at redshift 13},
	volume = {639},
	issn = {0028-0836, 1476-4687},
	url = {http://arxiv.org/abs/2408.16608},
	doi = {10.1038/s41586-025-08779-5},
	number = {8056},
	urldate = {2025-12-18},
	journal = {Nature},
	author = {Witstok, Joris and Jakobsen, Peter and Maiolino, Roberto and Helton, Jakob M. and Johnson, Benjamin D. and Robertson, Brant E. and Tacchella, Sandro and Cameron, Alex J. and Smit, Renske and Bunker, Andrew J. and Saxena, Aayush and Sun, Fengwu and Alberts, Stacey and Arribas, Santiago and Baker, William M. and Bhatawdekar, Rachana and Boyett, Kristan and Cargile, Phillip A. and Carniani, Stefano and Charlot, Stéphane and Chevallard, Jacopo and Curti, Mirko and Curtis-Lake, Emma and D'Eugenio, Francesco and Eisenstein, Daniel J. and Hainline, Kevin N. and Jones, Gareth C. and Kumari, Nimisha and Maseda, Michael V. and Pérez-González, Pablo G. and Rinaldi, Pierluigi and Scholtz, Jan and Übler, Hannah and Williams, Christina C. and Willmer, Christopher N. A. and Willott, Chris and Zhu, Yongda},
	month = mar,
	year = {2025},
	note = {arXiv:2408.16608 [astro-ph]},
	pages = {897--901},
}

@article{carniani_spectroscopic_2024,
	title = {Spectroscopic confirmation of two luminous galaxies at a redshift of 14},
	volume = {633},
	issn = {0028-0836},
	url = {https://ui.adsabs.harvard.edu/abs/2024Natur.633..318C},
	doi = {10.1038/s41586-024-07860-9},
	urldate = {2025-12-18},
	journal = {Nature},
	author = {Carniani, Stefano and Hainline, Kevin and D'Eugenio, Francesco and Eisenstein, Daniel J. and Jakobsen, Peter and Witstok, Joris and Johnson, Benjamin D. and Chevallard, Jacopo and Maiolino, Roberto and Helton, Jakob M. and Willott, Chris and Robertson, Brant and Alberts, Stacey and Arribas, Santiago and Baker, William M. and Bhatawdekar, Rachana and Boyett, Kristan and Bunker, Andrew J. and Cameron, Alex J. and Cargile, Phillip A. and Charlot, Stéphane and Curti, Mirko and Curtis-Lake, Emma and Egami, Eiichi and Giardino, Giovanna and Isaak, Kate and Ji, Zhiyuan and Jones, Gareth C. and Kumari, Nimisha and Maseda, Michael V. and Parlanti, Eleonora and Pérez-González, Pablo G. and Rawle, Tim and Rieke, George and Rieke, Marcia and Del Pino, Bruno Rodríguez and Saxena, Aayush and Scholtz, Jan and Smit, Renske and Sun, Fengwu and Tacchella, Sandro and Übler, Hannah and Venturi, Giacomo and Williams, Christina C. and Willmer, Christopher N. A.},
	month = sep,
	year = {2024},
	note = {ADS Bibcode: 2024Natur.633..318C},
	pages = {318--322},
}

@article{trebitsch_modelling_2020,
	title = {Modelling a bright z = 6 galaxy at the faint end of the {AGN} luminosity function},
	volume = {494},
	doi = {10.1093/mnras/staa1012},
	journal = {Monthly Notices of the Royal Astronomical Society},
	author = {Trebitsch, Maxime and Volonteri, Marta and Dubois, Yohan},
	month = may,
	year = {2020},
	pages = {3453--3463},
}

@article{kakiichi_role_2018,
	title = {The role of galaxies and {AGN} in reionizing the {IGM} – {I}. {Keck} spectroscopy of 5 < z < 7 galaxies in the {QSO} field {J1148}+5251},
	volume = {479},
	issn = {0035-8711},
	url = {https://doi.org/10.1093/mnras/sty1318},
	doi = {10.1093/mnras/sty1318},
	number = {1},
	urldate = {2025-12-18},
	journal = {Monthly Notices of the Royal Astronomical Society},
	author = {Kakiichi, Koki and Ellis, Richard S and Laporte, Nicolas and Zitrin, Adi and Eilers, Anna-Christina and Ryan-Weber, Emma and Meyer, Romain A and Robertson, Brant and Stark, Daniel P and Bosman, Sarah E I},
	month = sep,
	year = {2018},
	pages = {43--63},
}

@misc{madau_cosmic_2017,
	title = {Cosmic {Reionization} {After} {Planck} and {Before} {JWST}: {An} {Analytic} {Approach}},
	shorttitle = {Cosmic {Reionization} {After} {Planck} and {Before} {JWST}},
	url = {https://arxiv.org/abs/1710.07636v1},
	doi = {10.3847/1538-4357/aa9715},
	language = {en},
	urldate = {2025-12-18},
	journal = {arXiv.org},
	author = {Madau, Piero},
	month = oct,
	year = {2017},
}

@misc{stoffers_challenge_2025,
	title = {The {Challenge} in {Illuminating} the {Invisible}: {Constraining} {LyC} {Escape} with {Bayesian} {Modelling} and {Symbolic} {Regression}},
	shorttitle = {The {Challenge} in {Illuminating} the {Invisible}},
	url = {http://arxiv.org/abs/2511.02908},
	doi = {10.48550/arXiv.2511.02908},
	urldate = {2025-12-18},
	publisher = {arXiv},
	author = {Stoffers, Amanda and Tacchella, Sandro and Simmonds, Charlotte and Johnson, Benjamin D. and Maiolino, Roberto},
	month = nov,
	year = {2025},
	note = {arXiv:2511.02908 [astro-ph]},
}

@article{saxena_strong_2022,
	title = {Strong {C} iv emission from star-forming galaxies: a case for high {Lyman} continuum photon escape},
	volume = {517},
	issn = {0035-8711},
	shorttitle = {Strong {C} iv emission from star-forming galaxies},
	url = {https://doi.org/10.1093/mnras/stac2742},
	doi = {10.1093/mnras/stac2742},
	number = {1},
	urldate = {2025-12-18},
	journal = {Monthly Notices of the Royal Astronomical Society},
	author = {Saxena, A and Cryer, E and Ellis, R S and Pentericci, L and Calabrò, A and Mascia, S and Saldana-Lopez, A and Schaerer, D and Katz, H and Llerena, M and Amorín, R},
	month = nov,
	year = {2022},
	pages = {1098--1111},
}

@article{katz_first_2023,
	title = {First insights into the {ISM} at z > 8 with {JWST}: possible physical implications of a high [{O} iii] λ4363/[{O} iii] λ5007},
	volume = {518},
	issn = {0035-8711},
	shorttitle = {First insights into the {ISM} at z > 8 with {JWST}},
	url = {https://doi.org/10.1093/mnras/stac2657},
	doi = {10.1093/mnras/stac2657},
	number = {1},
	urldate = {2025-12-18},
	journal = {Monthly Notices of the Royal Astronomical Society},
	author = {Katz, Harley and Saxena, Aayush and Cameron, Alex J and Carniani, Stefano and Bunker, Andrew J and Arribas, Santiago and Bhatawdekar, Rachana and Bowler, Rebecca A A and Boyett, Kristan N K and Cresci, Giovanni and Curtis-Lake, Emma and D’Eugenio, Francesco and Kumari, Nimisha and Looser, Tobias J and Maiolino, Roberto and Übler, Hannah and Willott, Chris and Witstok, Joris},
	month = jan,
	year = {2023},
	pages = {592--603},
}

@article{donnan_jwst_2024,
	title = {{JWST} {PRIMER}: a new multifield determination of the evolving galaxy {UV} luminosity function at redshifts z ≃ 9 - 15},
	volume = {533},
	issn = {0035-8711},
	shorttitle = {{JWST} {PRIMER}},
	url = {https://ui.adsabs.harvard.edu/abs/2024MNRAS.533.3222D},
	doi = {10.1093/mnras/stae2037},
	urldate = {2025-12-18},
	journal = {Monthly Notices of the Royal Astronomical Society},
	publisher = {OUP},
	author = {Donnan, C. T. and McLure, R. J. and Dunlop, J. S. and McLeod, D. J. and Magee, D. and Arellano-Córdova, K. Z. and Barrufet, L. and Begley, R. and Bowler, R. A. A. and Carnall, A. C. and Cullen, F. and Ellis, R. S. and Fontana, A. and Illingworth, G. D. and Grogin, N. A. and Hamadouche, M. L. and Koekemoer, A. M. and Liu, F.-Y. and Mason, C. and Santini, P. and Stanton, T. M.},
	month = sep,
	year = {2024},
	note = {ADS Bibcode: 2024MNRAS.533.3222D},
	pages = {3222--3237},
}

@article{smith_thesan_2022,
	title = {The {THESAN} project: {Lyman}-α emission and transmission during the {Epoch} of {Reionization}},
	volume = {512},
	issn = {0035-8711},
	shorttitle = {The {THESAN} project},
	url = {https://ui.adsabs.harvard.edu/abs/2022MNRAS.512.3243S},
	doi = {10.1093/mnras/stac713},
	urldate = {2026-01-14},
	journal = {Monthly Notices of the Royal Astronomical Society},
	publisher = {OUP},
	author = {Smith, A. and Kannan, R. and Garaldi, E. and Vogelsberger, M. and Pakmor, R. and Springel, V. and Hernquist, L.},
	month = may,
	year = {2022},
	note = {ADS Bibcode: 2022MNRAS.512.3243S},
	pages = {3243--3265},
}

@article{garaldi_thesan_2022,
	title = {The {THESAN} project: properties of the intergalactic medium and its connection to reionization-era galaxies},
	volume = {512},
	issn = {0035-8711},
	shorttitle = {The {THESAN} project},
	url = {https://ui.adsabs.harvard.edu/abs/2022MNRAS.512.4909G},
	doi = {10.1093/mnras/stac257},
	urldate = {2026-01-14},
	journal = {Monthly Notices of the Royal Astronomical Society},
	publisher = {OUP},
	author = {Garaldi, E. and Kannan, R. and Smith, A. and Springel, V. and Pakmor, R. and Vogelsberger, M. and Hernquist, L.},
	month = jun,
	year = {2022},
	note = {ADS Bibcode: 2022MNRAS.512.4909G},
	pages = {4909--4933},
}

@article{garaldi_thesan_2024,
	title = {The {THESAN} project: public data release of radiation-hydrodynamic simulations matching reionization-era {JWST} observations},
	volume = {530},
	issn = {0035-8711},
	shorttitle = {The {THESAN} project},
	url = {https://ui.adsabs.harvard.edu/abs/2024MNRAS.530.3765G},
	doi = {10.1093/mnras/stae839},
	urldate = {2026-01-14},
	journal = {Monthly Notices of the Royal Astronomical Society},
	publisher = {OUP},
	author = {Garaldi, Enrico and Kannan, Rahul and Smith, Aaron and Borrow, Josh and Vogelsberger, Mark and Pakmor, Rüdiger and Springel, Volker and Hernquist, Lars and Galárraga-Espinosa, Daniela and Yeh, Jessica Y.-C. and Shen, Xuejian and Xu, Clara and Neyer, Meredith and Spina, Benedetta and Almualla, Mouza and Zhao, Yu},
	month = jun,
	year = {2024},
	note = {ADS Bibcode: 2024MNRAS.530.3765G},
	pages = {3765--3786},
}

@misc{mcclymont_nature_2024,
	title = {The nature of diffuse ionised gas in star-forming galaxies},
	url = {http://arxiv.org/abs/2403.03243},
	doi = {10.48550/arXiv.2403.03243},
	urldate = {2026-01-14},
	publisher = {arXiv},
	author = {McClymont, William and Tacchella, Sandro and Smith, Aaron and Kannan, Rahul and Maiolino, Roberto and Belfiore, Francesco and Hernquist, Lars and Li, Hui and Vogelsberger, Mark},
	month = mar,
	year = {2024},
	note = {arXiv:2403.03243 [astro-ph]},
}

@article{smith_physics_2019,
	title = {The physics of {Lyman} α escape from high-redshift galaxies},
	volume = {484},
	issn = {0035-8711},
	url = {https://doi.org/10.1093/mnras/sty3483},
	doi = {10.1093/mnras/sty3483},
	number = {1},
	urldate = {2026-01-14},
	journal = {Monthly Notices of the Royal Astronomical Society},
	author = {Smith, Aaron and Ma, Xiangcheng and Bromm, Volker and Finkelstein, Steven L and Hopkins, Philip F and Faucher-Giguère, Claude-André and Kereš, Dušan},
	month = mar,
	year = {2019},
	pages = {39--59},
}

@article{llerena_ionizing_2025,
	title = {The ionizing photon production efficiency of star-forming galaxies at z ∼ 4─10},
	volume = {698},
	issn = {0004-6361},
	url = {https://ui.adsabs.harvard.edu/abs/2025A&A...698A.302L},
	doi = {10.1051/0004-6361/202453251},
	urldate = {2026-01-14},
	journal = {Astronomy and Astrophysics},
	publisher = {EDP},
	author = {Llerena, M. and Pentericci, L. and Napolitano, L. and Mascia, S. and Amorín, R. and Calabrò, A. and Castellano, M. and Cleri, N. J. and Giavalisco, M. and Grogin, N. A. and Hathi, N. P. and Hirschmann, M. and Koekemoer, A. M. and Nanayakkara, T. and Pacucci, F. and Shen, L. and Wilkins, S. M. and Yoon, I. and Yung, L. Y. A. and Bhatawdekar, R. and Lucas, R. A. and Wang, X. and Arrabal Haro, P. and Bagley, M. B. and Finkelstein, S. L. and Kartaltepe, J. S. and Merlin, E. and Papovich, C. and Pirzkal, N. and Santini, P.},
	month = jun,
	year = {2025},
	note = {ADS Bibcode: 2025A\&A...698A.302L},
	pages = {A302},
}

@article{shen_high-redshift_2020,
	title = {High-redshift {JWST} predictions from {IllustrisTNG}: {II}. {Galaxy} line and continuum spectral indices and dust attenuation curves},
	volume = {495},
	issn = {0035-8711},
	shorttitle = {High-redshift {JWST} predictions from {IllustrisTNG}},
	url = {https://ui.adsabs.harvard.edu/abs/2020MNRAS.495.4747S},
	doi = {10.1093/mnras/staa1423},
	urldate = {2026-01-14},
	journal = {Monthly Notices of the Royal Astronomical Society},
	publisher = {OUP},
	author = {Shen, Xuejian and Vogelsberger, Mark and Nelson, Dylan and Pillepich, Annalisa and Tacchella, Sandro and Marinacci, Federico and Torrey, Paul and Hernquist, Lars and Springel, Volker},
	month = jul,
	year = {2020},
	note = {ADS Bibcode: 2020MNRAS.495.4747S},
	pages = {4747--4768},
}

@misc{papovich_galaxies_2025,
	title = {Galaxies in the {Epoch} of {Reionization} {Are} {All} {Bark} and {No} {Bite} -- {Plenty} of {Ionizing} {Photons}, {Low} {Escape} {Fractions}},
	url = {http://arxiv.org/abs/2505.08870},
	doi = {10.48550/arXiv.2505.08870},
	urldate = {2026-01-14},
	publisher = {arXiv},
	author = {Papovich, Casey and Cole, Justin W. and Hu, Weida and Finkelstein, Steven L. and Shen, Lu and Haro, Pablo Arrabal and Amorín, Ricardo O. and Backhaus, Bren and Bagley, Micaela B. and Bhatawdekar, Rachana and Calabró, Antonello and Carnall, Adam C. and Cleri, Nikko and Daddi, Emanuele and Dickinson, Mark and Grogin, Norman and Holwerda, Benne W. and Jaskot, Anne E. and Koekemoer, Anton M. and Llerena, Mario and Lucas, Ray A. and Mascia, Sara and Pacucci, Fabio and Pentericci, Laura and Pérez-González, Pablo G. and Pirzkal, Nor and Raghunathan, Srinivasan and Seillé, Lisa-Marie and Somerville, Rachel and Yung, L. Y. Aaron},
	month = may,
	year = {2025},
	note = {arXiv:2505.08870 [astro-ph]},
}

@article{kulkarni_large_2019,
	title = {Large {Ly} α opacity fluctuations and low {CMB} τ in models of late reionization with large islands of neutral hydrogen extending to z < 5.5},
	volume = {485},
	issn = {0035-8711},
	url = {https://ui.adsabs.harvard.edu/abs/2019MNRAS.485L..24K},
	doi = {10.1093/mnrasl/slz025},
	urldate = {2026-01-15},
	journal = {Monthly Notices of the Royal Astronomical Society},
	publisher = {OUP},
	author = {Kulkarni, Girish and Keating, Laura C. and Haehnelt, Martin G. and Bosman, Sarah E. I. and Puchwein, Ewald and Chardin, Jonathan and Aubert, Dominique},
	month = may,
	year = {2019},
	note = {ADS Bibcode: 2019MNRAS.485L..24K},
	pages = {L24--L28},
}

@article{robertson_earliest_2024,
	title = {Earliest {Galaxies} in the {JADES} {Origins} {Field}: {Luminosity} {Function} and {Cosmic} {Star} {Formation} {Rate} {Density} 300 {Myr} after the {Big} {Bang}},
	volume = {970},
	issn = {0004-637X},
	shorttitle = {Earliest {Galaxies} in the {JADES} {Origins} {Field}},
	url = {https://ui.adsabs.harvard.edu/abs/2024ApJ...970...31R},
	doi = {10.3847/1538-4357/ad463d},
	urldate = {2026-01-15},
	journal = {The Astrophysical Journal},
	publisher = {IOP},
	author = {Robertson, Brant and Johnson, Benjamin D. and Tacchella, Sandro and Eisenstein, Daniel J. and Hainline, Kevin and Arribas, Santiago and Baker, William M. and Bunker, Andrew J. and Carniani, Stefano and Cargile, Phillip A. and Carreira, Courtney and Charlot, Stephane and Chevallard, Jacopo and Curti, Mirko and Curtis-Lake, Emma and D'Eugenio, Francesco and Egami, Eiichi and Hausen, Ryan and Helton, Jakob M. and Jakobsen, Peter and Ji, Zhiyuan and Jones, Gareth C. and Maiolino, Roberto and Maseda, Michael V. and Nelson, Erica and Pérez-González, Pablo G. and Puskás, Dávid and Rieke, Marcia and Smit, Renske and Sun, Fengwu and Übler, Hannah and Whitler, Lily and Williams, Christina C. and Willmer, Christopher N. A. and Willott, Chris and Witstok, Joris},
	month = jul,
	year = {2024},
	note = {ADS Bibcode: 2024ApJ...970...31R},
	pages = {31},
}

@article{gnedin_cosmic_2014,
	title = {Cosmic {Reionization} {On} {Computers} {I}. {Design} and {Calibration} of {Simulations}},
	volume = {793},
	issn = {1538-4357},
	url = {http://arxiv.org/abs/1403.4245},
	doi = {10.1088/0004-637X/793/1/29},
	number = {1},
	urldate = {2026-01-22},
	journal = {The Astrophysical Journal},
	author = {Gnedin, Nickolay Y.},
	month = sep,
	year = {2014},
	note = {arXiv:1403.4245 [astro-ph]},
	pages = {29},
}

@article{puchwein_sherwood-relics_2023,
	title = {The {Sherwood}-{Relics} simulations: overview and impact of patchy reionization and pressure smoothing on the intergalactic medium},
	volume = {519},
	issn = {0035-8711, 1365-2966},
	shorttitle = {The {Sherwood}-{Relics} simulations},
	url = {http://arxiv.org/abs/2207.13098},
	doi = {10.1093/mnras/stac3761},
	number = {4},
	urldate = {2026-01-22},
	journal = {Monthly Notices of the Royal Astronomical Society},
	author = {Puchwein, Ewald and Bolton, James S. and Keating, Laura C. and Molaro, Margherita and Gaikwad, Prakash and Kulkarni, Girish and Haehnelt, Martin G. and Iršič, Vid and Šoltinský, Tomáš and Viel, Matteo and Aubert, Dominique and Becker, George D. and Meiksin, Avery},
	month = jan,
	year = {2023},
	note = {arXiv:2207.13098 [astro-ph]},
	pages = {6162--6183},
}

@article{ma_simulating_2018,
	title = {Simulating galaxies in the reionization era with {FIRE}-2: galaxy scaling relations, stellar mass functions, and luminosity functions},
	volume = {478},
	issn = {0035-8711},
	shorttitle = {Simulating galaxies in the reionization era with {FIRE}-2},
	url = {https://ui.adsabs.harvard.edu/abs/2018MNRAS.478.1694M},
	doi = {10.1093/mnras/sty1024},
	urldate = {2026-01-22},
	journal = {Monthly Notices of the Royal Astronomical Society},
	publisher = {OUP},
	author = {Ma, Xiangcheng and Hopkins, Philip F. and Garrison-Kimmel, Shea and Faucher-Giguère, Claude-André and Quataert, Eliot and Boylan-Kolchin, Michael and Hayward, Christopher C. and Feldmann, Robert and Kereš, Dušan},
	month = aug,
	year = {2018},
	note = {ADS Bibcode: 2018MNRAS.478.1694M},
	pages = {1694--1715},
}

@article{pallottini_deep_2019,
	title = {Deep into the structure of the first galaxies: {SERRA} views},
	volume = {487},
	issn = {0035-8711, 1365-2966},
	shorttitle = {Deep into the structure of the first galaxies},
	url = {http://arxiv.org/abs/1905.08254},
	doi = {10.1093/mnras/stz1383},
	number = {2},
	urldate = {2026-01-22},
	journal = {Monthly Notices of the Royal Astronomical Society},
	author = {Pallottini, A. and Ferrara, A. and Decataldo, D. and Gallerani, S. and Vallini, L. and Carniani, S. and Behrens, C. and Kohandel, M. and Salvadori, S.},
	month = aug,
	year = {2019},
	note = {arXiv:1905.08254 [astro-ph]},
	pages = {1689--1708},
}

@article{shen_thesan-hr_2024,
	title = {thesan-hr: galaxies in the {Epoch} of {Reionization} in warm dark matter, fuzzy dark matter, and interacting dark matter},
	volume = {527},
	issn = {0035-8711},
	shorttitle = {thesan-hr},
	url = {https://doi.org/10.1093/mnras/stad3397},
	doi = {10.1093/mnras/stad3397},
	number = {2},
	urldate = {2026-02-02},
	journal = {Monthly Notices of the Royal Astronomical Society},
	author = {Shen, Xuejian and Borrow, Josh and Vogelsberger, Mark and Garaldi, Enrico and Smith, Aaron and Kannan, Rahul and Tacchella, Sandro and Zavala, Jesús and Hernquist, Lars and Yeh, Jessica Y-C and Zheng, Chunyuan},
	month = jan,
	year = {2024},
	pages = {2835--2857},
}

@misc{borrow_thesan-hr_2023,
	title = {{THESAN}-{HR}: {How} does reionization impact early galaxy evolution?},
	shorttitle = {{THESAN}-{HR}},
	url = {http://arxiv.org/abs/2212.03255},
	doi = {10.48550/arXiv.2212.03255},
	urldate = {2026-02-02},
	publisher = {arXiv},
	author = {Borrow, Josh and Kannan, Rahul and Garaldi, Enrico and Smith, Aaron and Vogelsberger, Mark and Pakmor, Rüdiger and Springel, Volker and Hernquist, Lars},
	month = aug,
	year = {2023},
	note = {arXiv:2212.03255 [astro-ph]},
}

@misc{almualla_thesan_2025,
	title = {The {THESAN} project: {Lyman}-alpha intensity mapping of cosmic reionization},
	shorttitle = {The {THESAN} project},
	url = {http://arxiv.org/abs/2512.06085},
	doi = {10.48550/arXiv.2512.06085},
	urldate = {2026-02-02},
	publisher = {arXiv},
	author = {Almualla, Mouza and Smith, Aaron and Kannan, Rahul and Hernquist, Lars and Garaldi, Enrico and Lidz, Adam and Lorinc, Kevin and Chan, Jennifer Yik Ham and Vogelsberger, Mark},
	month = dec,
	year = {2025},
	note = {arXiv:2512.06085 [astro-ph]},
}

@misc{zhao_thesan_2025,
	title = {The {THESAN} project: environmental drivers of {Local} {Group} reionization},
	shorttitle = {The {THESAN} project},
	url = {http://arxiv.org/abs/2507.16245},
	doi = {10.48550/arXiv.2507.16245},
	urldate = {2026-02-02},
	publisher = {arXiv},
	author = {Zhao, Yu and Smith, Aaron and Kannan, Rahul and Garaldi, Enrico and Li, Hui and Vogelsberger, Mark and Benson, Andrew and Hernquist, Lars},
	month = jul,
	year = {2025},
	note = {arXiv:2507.16245 [astro-ph]},
}

@article{jamieson_thesan_2025,
	title = {The thesan project: tracking the expansion and merger histories of ionized bubbles during the {Epoch} of {Reionization}},
	volume = {541},
	issn = {0035-8711},
	shorttitle = {The thesan project},
	url = {https://doi.org/10.1093/mnras/staf996},
	doi = {10.1093/mnras/staf996},
	number = {2},
	urldate = {2026-02-02},
	journal = {Monthly Notices of the Royal Astronomical Society},
	author = {Jamieson, Nathan and Smith, Aaron and Neyer, Meredith and Kannan, Rahul and Garaldi, Enrico and Vogelsberger, Mark and Hernquist, Lars and Zier, Oliver and Shen, Xuejian and Kakiichi, Koki},
	month = aug,
	year = {2025},
	pages = {1088--1105},
}

@article{shen_thesan-zoom_2026,
	title = {The {THESAN}-{ZOOM} project: {Star}-formation efficiencies in high-redshift galaxies},
	volume = {545},
	issn = {0035-8711, 1365-2966},
	shorttitle = {The {THESAN}-{ZOOM} project},
	url = {http://arxiv.org/abs/2503.01949},
	doi = {10.1093/mnras/staf2119},
	number = {4},
	urldate = {2026-02-02},
	journal = {Monthly Notices of the Royal Astronomical Society},
	author = {Shen, Xuejian and Kannan, Rahul and Puchwein, Ewald and Smith, Aaron and Vogelsberger, Mark and Borrow, Josh and Garaldi, Enrico and Keating, Laura and Zier, Oliver and McClymont, William and Tacchella, Sandro and Wang, Zihao and Hernquist, Lars},
	month = jan,
	year = {2026},
	note = {arXiv:2503.01949 [astro-ph]},
	pages = {staf2119},
}

@misc{mcclymont_overmassive_2025,
	title = {Overmassive black holes in the early {Universe} can be explained by gas-rich, dark matter-dominated galaxies},
	url = {http://arxiv.org/abs/2506.13852},
	doi = {10.48550/arXiv.2506.13852},
	urldate = {2026-02-02},
	publisher = {arXiv},
	author = {McClymont, William and Tacchella, Sandro and Ji, Xihan and Kannan, Rahul and Maiolino, Roberto and Simmonds, Charlotte and Smith, Aaron and Puchwein, Ewald and Garaldi, Enrico and Vogelsberger, Mark and D'Eugenio, Francesco and Keating, Laura and Shen, Xuejian and Trefoloni, Bartolomeo and Zier, Oliver},
	month = jun,
	year = {2025},
	note = {arXiv:2506.13852 [astro-ph]},
}

@misc{mcclymont_thesan-zoom_2025-2,
	title = {The {THESAN}-{ZOOM} project: {Mystery} {N}/{O} more -- uncovering the origin of peculiar chemical abundances and a not-so-fundamental metallicity relation at $3<z<12$},
	shorttitle = {The {THESAN}-{ZOOM} project},
	url = {http://arxiv.org/abs/2507.08787},
	doi = {10.48550/arXiv.2507.08787},
	urldate = {2026-02-02},
	publisher = {arXiv},
	author = {McClymont, William and Tacchella, Sandro and Smith, Aaron and Kannan, Rahul and Garaldi, Enrico and Puchwein, Ewald and Isobe, Yuki and Ji, Xihan and Shen, Xuejian and Wang, Zihao and Belokurov, Vasily and Borrow, Josh and D'Eugenio, Francesco and Keating, Laura and Maiolino, Roberto and Monty, Stephanie and Vogelsberger, Mark and Zier, Oliver},
	month = jul,
	year = {2025},
	note = {arXiv:2507.08787 [astro-ph]},
}

@misc{zier_thesan-zoom_2025,
	title = {The {THESAN}-{ZOOM} project: {Long}-term imprints of external reionization on galaxy evolution},
	shorttitle = {The {THESAN}-{ZOOM} project},
	url = {http://arxiv.org/abs/2503.02927},
	doi = {10.48550/arXiv.2503.02927},
	urldate = {2026-02-02},
	publisher = {arXiv},
	author = {Zier, Oliver and Kannan, Rahul and Smith, Aaron and Puchwein, Ewald and Vogelsberger, Mark and Borrow, Josh and Garaldi, Enrico and Keating, Laura and McClymont, William and Shen, Xuejian and Hernquist, Lars},
	month = jun,
	year = {2025},
	note = {arXiv:2503.02927 [astro-ph]},
}

@article{zier_thesan-zoom_2025-1,
	title = {The {THESAN}-{ZOOM} project: {Population} {III} star formation continues until the end of reionization},
	volume = {544},
	issn = {0035-8711, 1365-2966},
	shorttitle = {The {THESAN}-{ZOOM} project},
	url = {http://arxiv.org/abs/2503.03806},
	doi = {10.1093/mnras/staf1053},
	number = {1},
	urldate = {2026-02-02},
	journal = {Monthly Notices of the Royal Astronomical Society},
	author = {Zier, Oliver and Kannan, Rahul and Smith, Aaron and Puchwein, Ewald and Vogelsberger, Mark and Borrow, Josh and Garaldi, Enrico and Keating, Laura and McClymont, William and Shen, Xuejian and Hernquist, Lars},
	month = oct,
	year = {2025},
	note = {arXiv:2503.03806 [astro-ph]},
	pages = {410--429},
}

@article{wang_thesan-zoom_2025,
	title = {The {THESAN}-{ZOOM} project: {Star} formation efficiency from giant molecular clouds to galactic scale in high-redshift starbursts},
	volume = {544},
	issn = {0035-8711, 1365-2966},
	shorttitle = {The {THESAN}-{ZOOM} project},
	url = {http://arxiv.org/abs/2505.05554},
	doi = {10.1093/mnras/staf1677},
	number = {3},
	urldate = {2026-02-02},
	journal = {Monthly Notices of the Royal Astronomical Society},
	author = {Wang, Zihao and Shen, Xuejian and Vogelsberger, Mark and Li, Hui and Kannan, Rahul and Puchwein, Ewald and Smith, Aaron and Borrow, Josh and Garaldi, Enrico and Keating, Laura and Zier, Oliver and McClymont, William and Tacchella, Sandro and Ni, Yang and Hernquist, Lars},
	month = nov,
	year = {2025},
	note = {arXiv:2505.05554 [astro-ph]},
	pages = {2675--2697},
}

@misc{pruto_thesan-zoom_2026,
	title = {The {THESAN}-{ZOOM} project: {The} {Hidden} {Neighbours} of {OI} {Absorbers} during {Reionization}},
	shorttitle = {The {THESAN}-{ZOOM} project},
	url = {http://arxiv.org/abs/2510.13977},
	doi = {10.48550/arXiv.2510.13977},
	urldate = {2026-02-02},
	publisher = {arXiv},
	author = {Pruto, Giulia and Keating, Laura and Kannan, Rahul and Puchwein, Ewald and Smith, Aaron and Borrow, Josh and Garaldi, Enrico and Vogelsberger, Mark and Zier, Oliver and McClymont, William and Shen, Xuejian and Tacchella, Sandro},
	month = jan,
	year = {2026},
	note = {arXiv:2510.13977 [astro-ph]},
}

@misc{mcclymont_modelling_2025,
	title = {Modelling the nebular emission of galaxies across cosmic time with {COLT}},
	url = {https://ui.adsabs.harvard.edu/abs/2025arXiv251013952M},
	doi = {10.48550/arXiv.2510.13952},
	urldate = {2026-02-03},
	publisher = {arXiv},
	author = {McClymont, William and Smith, Aaron and Tacchella, Sandro},
	month = oct,
	year = {2025},
	note = {ADS Bibcode: 2025arXiv251013952M},
}

@article{bluck_quenching_2022,
	title = {The quenching of galaxies, bulges, and disks since cosmic noon - {A} machine learning approach for identifying causality in astronomical data},
	volume = {659},
	copyright = {© ESO 2022},
	issn = {0004-6361, 1432-0746},
	url = {https://www.aanda.org/articles/aa/abs/2022/03/aa42643-21/aa42643-21.html},
	doi = {10.1051/0004-6361/202142643},
	language = {en},
	urldate = {2026-02-03},
	journal = {Astronomy \& Astrophysics},
	publisher = {EDP Sciences},
	author = {Bluck, Asa F. L. and Maiolino, Roberto and Brownson, Simcha and Conselice, Christopher J. and Ellison, Sara L. and Piotrowska, Joanna M. and Thorp, Mallory D.},
	month = mar,
	year = {2022},
	pages = {A160},
}

@article{baker_molecular_2023,
	title = {The molecular gas main sequence and {Schmidt}-{Kennicutt} relation are fundamental, the star-forming main sequence is a (useful) byproduct},
	volume = {518},
	issn = {0035-8711},
	url = {https://ui.adsabs.harvard.edu/abs/2023MNRAS.518.4767B},
	doi = {10.1093/mnras/stac3413},
	urldate = {2026-02-03},
	journal = {Monthly Notices of the Royal Astronomical Society},
	publisher = {OUP},
	author = {Baker, William M. and Maiolino, Roberto and Belfiore, Francesco and Bluck, Asa F. L. and Curti, Mirko and Wylezalek, Dominika and Bertemes, Caroline and Bothwell, M. S. and Lin, Lihwai and Thorp, Mallory and Pan, Hsi-An},
	month = jan,
	year = {2023},
	note = {ADS Bibcode: 2023MNRAS.518.4767B},
	pages = {4767--4781},
}

@article{kawinwanichakij_connecting_2026,
	title = {Connecting {Environment}, {Star} {Formation} {History}, and {Morphology} of {Massive} {Quiescent} {Galaxies} at 3 < z < 4 with {JWST}},
	volume = {997},
	issn = {0004-637X},
	url = {https://ui.adsabs.harvard.edu/abs/2026ApJ...997...29K},
	doi = {10.3847/1538-4357/ae0a18},
	urldate = {2026-02-03},
	journal = {The Astrophysical Journal},
	publisher = {IOP},
	author = {Kawinwanichakij, Lalitwadee and Glazebrook, Karl and Nanayakkara, Themiya and Kacprzak, Glenn G. and Chittenden, Harry George and Jacobs, Colin and Chandro-Gómez, Ángel and Lagos, Claudia and Marchesini, Danilo and Martìnez-Marìn, M. and Oesch, Pascal A. and Remus, Rhea-Silvia},
	month = jan,
	year = {2026},
	note = {ADS Bibcode: 2026ApJ...997...29K},
	pages = {29},
}

@article{popesso_main_2023,
	title = {The main sequence of star-forming galaxies across cosmic times},
	volume = {519},
	issn = {0035-8711},
	url = {https://ui.adsabs.harvard.edu/abs/2023MNRAS.519.1526P},
	doi = {10.1093/mnras/stac3214},
	urldate = {2026-02-03},
	journal = {Monthly Notices of the Royal Astronomical Society},
	publisher = {OUP},
	author = {Popesso, P. and Concas, A. and Cresci, G. and Belli, S. and Rodighiero, G. and Inami, H. and Dickinson, M. and Ilbert, O. and Pannella, M. and Elbaz, D.},
	month = feb,
	year = {2023},
	note = {ADS Bibcode: 2023MNRAS.519.1526P},
	pages = {1526--1544},
}

@article{clarke_star-forming_2024,
	title = {The {Star}-forming {Main} {Sequence} in {JADES} and {CEERS} at z > 1.4: {Investigating} the {Burstiness} of {Star} {Formation}},
	volume = {977},
	issn = {0004-637X},
	shorttitle = {The {Star}-forming {Main} {Sequence} in {JADES} and {CEERS} at z > 1.4},
	url = {https://ui.adsabs.harvard.edu/abs/2024ApJ...977..133C},
	doi = {10.3847/1538-4357/ad8ba4},
	urldate = {2026-02-03},
	journal = {The Astrophysical Journal},
	publisher = {IOP},
	author = {Clarke, Leonardo and Shapley, Alice E. and Sanders, Ryan L. and Topping, Michael W. and Brammer, Gabriel B. and Bento, Trinity and Reddy, Naveen A. and Kehoe, Emily},
	month = dec,
	year = {2024},
	note = {ADS Bibcode: 2024ApJ...977..133C},
	pages = {133},
}

@article{speagle_highly_2014,
	title = {A {Highly} {Consistent} {Framework} for the {Evolution} of the {Star}-{Forming} "{Main} {Sequence}" from z {\textasciitilde} 0-6},
	volume = {214},
	issn = {0067-0049},
	url = {https://ui.adsabs.harvard.edu/abs/2014ApJS..214...15S},
	doi = {10.1088/0067-0049/214/2/15},
	urldate = {2026-02-03},
	journal = {The Astrophysical Journal Supplement Series},
	publisher = {IOP},
	author = {Speagle, J. S. and Steinhardt, C. L. and Capak, P. L. and Silverman, J. D.},
	month = oct,
	year = {2014},
	note = {ADS Bibcode: 2014ApJS..214...15S},
	pages = {15},
}

@article{mascia_closing_2023,
	title = {Closing in on the sources of cosmic reionization: first results from the {GLASS}-{JWST} program},
	volume = {672},
	issn = {0004-6361, 1432-0746},
	shorttitle = {Closing in on the sources of cosmic reionization},
	url = {http://arxiv.org/abs/2301.02816},
	doi = {10.1051/0004-6361/202345866},
	urldate = {2026-02-03},
	journal = {Astronomy \& Astrophysics},
	author = {Mascia, S. and Pentericci, L. and Calabro', A. and Treu, T. and Santini, P. and Yang, L. and Napolitano, L. and Roberts-Borsani, G. and Bergamini, P. and Grillo, C. and Rosati, P. and Vulcani, B. and Castellano, M. and Boyett, K. and Fontana, A. and Glazebrook, K. and Henry, A. and Mason, C. and Merlin, E. and Morishita, T. and Nanayakkara, T. and Paris, D. and Roy, N. and Williams, H. and Wang, X. and Brammer, G. and Bradac, M. and Chen, W. and Kelly, P. L. and Koekemoer, A. M. and Trenti, M. and Windhorst, R. A.},
	month = apr,
	year = {2023},
	note = {arXiv:2301.02816 [astro-ph]},
	pages = {A155},
}

@misc{kageura_new_2026,
	title = {A {New} {Constraint} on the {Optical} {Depth} from the {Reionization} {History} {Independent} of {CMB} {Large}-{Scale} {E}-{Mode} {Polarization}},
	url = {http://arxiv.org/abs/2601.09644},
	doi = {10.48550/arXiv.2601.09644},
	urldate = {2026-02-06},
	publisher = {arXiv},
	author = {Kageura, Yuta and Ouchi, Masami and Naokawa, Fumihiro and Umeda, Hiroya and Matsumoto, Akinori and Harikane, Yuichi and Nakane, Minami and Thai, Tran Thi},
	month = jan,
	year = {2026},
	note = {arXiv:2601.09644 [astro-ph]},
}

@article{cain_cosmic_2025,
	title = {The {Cosmic} {Microwave} {Background} {Optical} {Depth} {Constrains} the {Duration} of {Reionization}},
	volume = {987},
	issn = {0004-637X},
	url = {https://ui.adsabs.harvard.edu/abs/2025ApJ...987L..29C},
	doi = {10.3847/2041-8213/ade141},
	urldate = {2026-02-06},
	journal = {The Astrophysical Journal},
	publisher = {IOP},
	author = {Cain, Christopher and Van Engelen, Alexander and Croker, Kevin S. and Kramer, Darby and D'Aloisio, Anson and Lopez, Garett},
	month = jul,
	year = {2025},
	note = {ADS Bibcode: 2025ApJ...987L..29C},
	pages = {L29},
}

@misc{sailer_dispuable_2025,
	title = {Dispu$\tau$able: the high cost of a low optical depth},
	shorttitle = {Dispu$τ$able},
	url = {http://arxiv.org/abs/2504.16932},
	doi = {10.48550/arXiv.2504.16932},
	urldate = {2026-02-06},
	publisher = {arXiv},
	author = {Sailer, Noah and Farren, Gerrit S. and Ferraro, Simone and White, Martin},
	month = apr,
	year = {2025},
	note = {arXiv:2504.16932 [astro-ph]},
}

@article{jhaveri_turning_2025,
	title = {Turning a negative neutrino mass into a positive optical depth},
	volume = {112},
	url = {https://link.aps.org/doi/10.1103/6vd2-rbfn},
	doi = {10.1103/6vd2-rbfn},
	number = {4},
	urldate = {2026-02-06},
	journal = {Physical Review D},
	publisher = {American Physical Society},
	author = {Jhaveri, Tanisha and Karwal, Tanvi and Hu, Wayne},
	month = aug,
	year = {2025},
	pages = {043541},
}

@misc{liu_phantom_2025,
	title = {Phantom {Mirage} from {Axion} {Dark} {Energy}},
	url = {http://arxiv.org/abs/2510.14957},
	doi = {10.48550/arXiv.2510.14957},
	urldate = {2026-02-06},
	publisher = {arXiv},
	author = {Liu, Rayne and Zhu, Yijie and Hu, Wayne and Miranda, Vivian},
	month = oct,
	year = {2025},
	note = {arXiv:2510.14957 [astro-ph]},
}

@article{chen_patchy_2023,
	title = {Patchy {Kinetic} {Sunyaev}–{Zel}’dovich {Effect} with {Controlled} {Reionization} {History} and {Morphology}},
	volume = {943},
	issn = {0004-637X},
	url = {https://doi.org/10.3847/1538-4357/ac8481},
	doi = {10.3847/1538-4357/ac8481},
	language = {en},
	number = {2},
	urldate = {2026-02-06},
	journal = {The Astrophysical Journal},
	publisher = {The American Astronomical Society},
	author = {Chen, Nianyi and Trac, Hy and Mukherjee, Suvodip and Cen, Renyue},
	month = feb,
	year = {2023},
	pages = {138},
}

@article{puchwein_consistent_2019,
	title = {Consistent modelling of the meta-galactic {UV} background and the thermal/ionization history of the intergalactic medium},
	volume = {485},
	issn = {0035-8711},
	url = {https://ui.adsabs.harvard.edu/abs/2019MNRAS.485...47P},
	doi = {10.1093/mnras/stz222},
	urldate = {2026-02-08},
	journal = {Monthly Notices of the Royal Astronomical Society},
	publisher = {OUP},
	author = {Puchwein, Ewald and Haardt, Francesco and Haehnelt, Martin G. and Madau, Piero},
	month = may,
	year = {2019},
	note = {ADS Bibcode: 2019MNRAS.485...47P},
	pages = {47--68},
}

@article{shen_bolometric_2020,
	title = {The {Bolometric} {Quasar} {Luminosity} {Function} at z = 0-7},
	volume = {495},
	issn = {0035-8711, 1365-2966},
	url = {http://arxiv.org/abs/2001.02696},
	doi = {10.1093/mnras/staa1381},
	number = {3},
	urldate = {2026-02-08},
	journal = {Monthly Notices of the Royal Astronomical Society},
	author = {Shen, Xuejian and Hopkins, Philip F. and Faucher-Giguère, Claude-André and Alexander, D. M. and Richards, Gordon T. and Ross, Nicholas P. and Hickox, R. C.},
	month = jul,
	year = {2020},
	note = {arXiv:2001.02696 [astro-ph]},
	pages = {3252--3275},
}

@article{cristiani_spectral_2016,
	title = {The spectral slope and escape fraction of bright quasars at z ∼ 3.8: the contribution to the cosmic {UV} background},
	volume = {462},
	issn = {0035-8711},
	shorttitle = {The spectral slope and escape fraction of bright quasars at z ∼ 3.8},
	url = {https://doi.org/10.1093/mnras/stw1810},
	doi = {10.1093/mnras/stw1810},
	number = {3},
	urldate = {2026-02-08},
	journal = {Monthly Notices of the Royal Astronomical Society},
	author = {Cristiani, Stefano and Serrano, Luisa Maria and Fontanot, Fabio and Vanzella, Eros and Monaco, Pierluigi},
	month = nov,
	year = {2016},
	pages = {2478--2485},
}

@article{iwata_ionizing_2022,
	title = {Ionizing radiation from {AGNs} at z > 3.3 with the {Subaru} {Hyper} {Suprime}-{Cam} {Survey} and the {CFHT} {Large} {Area} {U}-band {Deep} {Survey} ({CLAUDS})},
	volume = {509},
	issn = {0035-8711},
	url = {https://doi.org/10.1093/mnras/stab2742},
	doi = {10.1093/mnras/stab2742},
	number = {2},
	urldate = {2026-02-08},
	journal = {Monthly Notices of the Royal Astronomical Society},
	author = {Iwata, Ikuru and Sawicki, Marcin and Inoue, Akio K and Akiyama, Masayuki and Micheva, Genoveva and Kawaguchi, Toshihiro and Kashikawa, Nobunari and Gwyn, Stephen and Arnouts, Stephane and Coupon, Jean and Desprez, Guillaume},
	month = jan,
	year = {2022},
	pages = {1820--1836},
}

@article{maiolino_jades_2024,
	title = {{JADES}. {The} diverse population of infant {Black} {Holes} at 4<z<11: merging, tiny, poor, but mighty},
	volume = {691},
	issn = {0004-6361, 1432-0746},
	shorttitle = {{JADES}. {The} diverse population of infant {Black} {Holes} at 4<z<11},
	url = {http://arxiv.org/abs/2308.01230},
	doi = {10.1051/0004-6361/202347640},
	urldate = {2026-02-08},
	journal = {Astronomy \& Astrophysics},
	author = {Maiolino, Roberto and Scholtz, Jan and Curtis-Lake, Emma and Carniani, Stefano and Baker, William and Graaff, Anna de and Tacchella, Sandro and Übler, Hannah and D'Eugenio, Francesco and Witstok, Joris and Curti, Mirko and Arribas, Santiago and Bunker, Andrew J. and Charlot, Stéphane and Chevallard, Jacopo and Eisenstein, Daniel J. and Egami, Eiichi and Ji, Zhiyuan and Jones, Gareth C. and Lyu, Jianwei and Rawle, Tim and Robertson, Brant and Rujopakarn, Wiphu and Perna, Michele and Sun, Fengwu and Venturi, Giacomo and Williams, Christina C. and Willott, Chris},
	month = nov,
	year = {2024},
	note = {arXiv:2308.01230 [astro-ph]},
	pages = {A145},
}

@misc{fujimoto_uncover_2024,
	title = {{UNCOVER}: {A} {NIRSpec} {Census} of {Lensed} {Galaxies} at z=8.50-13.08 {Probing} a {High} {AGN} {Fraction} and {Ionized} {Bubbles} in the {Shadow}},
	shorttitle = {{UNCOVER}},
	url = {http://arxiv.org/abs/2308.11609},
	doi = {10.48550/arXiv.2308.11609},
	urldate = {2026-02-08},
	publisher = {arXiv},
	author = {Fujimoto, Seiji and Wang, Bingjie and Weaver, John and Kokorev, Vasily and Atek, Hakim and Bezanson, Rachel and Labbe, Ivo and Brammer, Gabriel and Greene, Jenny E. and Chemerynska, Iryna and Dayal, Pratika and Graaff, Anna de and Furtak, Lukas J. and Oesch, Pascal A. and Setton, David J. and Price, Sedona H. and Miller, Tim B. and Williams, Christina C. and Whitaker, Katherine E. and Zitrin, Adi and Cutler, Sam E. and Leja, Joel and Pan, Richard and Coe, Dan and Dokkum, Pieter van and Feldmann, Robert and Fudamoto, Yoshinobu and Goulding, Andy D. and Khullar, Gourav and Marchesini, Danilo and Maseda, Michael and Nanayakkara, Themiya and Nelson, Erica J. and Smit, Renske and Stefanon, Mauro and Weibel, Andrea},
	month = nov,
	year = {2024},
	note = {arXiv:2308.11609 [astro-ph]},
}

@article{dayal_uncovering_2025,
	title = {{UNCOVERing} the contribution of black holes to reionization},
	volume = {697},
	issn = {0004-6361},
	url = {https://ui.adsabs.harvard.edu/abs/2025A&A...697A.211D},
	doi = {10.1051/0004-6361/202449331},
	urldate = {2026-02-08},
	journal = {Astronomy and Astrophysics},
	publisher = {EDP},
	author = {Dayal, Pratika and Volonteri, Marta and Greene, Jenny E. and Kokorev, Vasily and Goulding, Andy D. and Williams, Christina C. and Furtak, Lukas J. and Zitrin, Adi and Atek, Hakim and Bezanson, Rachel and Chemerynska, Iryna and Feldmann, Robert and Glazebrook, Karl and Labbe, Ivo and Nanayakkara, Themiya and Oesch, Pascal A. and Weaver, John R.},
	month = may,
	year = {2025},
	note = {ADS Bibcode: 2025A\&A...697A.211D},
	pages = {A211},
}

@article{upton_sanderbeck_models_2016,
	title = {Models of the thermal evolution of the intergalactic medium after reionization},
	volume = {460},
	issn = {0035-8711},
	url = {https://ui.adsabs.harvard.edu/abs/2016MNRAS.460.1885U},
	doi = {10.1093/mnras/stw1117},
	urldate = {2026-02-08},
	journal = {Monthly Notices of the Royal Astronomical Society},
	publisher = {OUP},
	author = {Upton Sanderbeck, Phoebe R. and D'Aloisio, Anson and McQuinn, Matthew J.},
	month = aug,
	year = {2016},
	note = {ADS Bibcode: 2016MNRAS.460.1885U},
	pages = {1885--1897},
}

@article{neyer_thesan_2024,
	title = {The thesan project: connecting ionized bubble sizes to their local environments during the {Epoch} of {Reionization}},
	volume = {531},
	issn = {0035-8711},
	shorttitle = {The thesan project},
	url = {https://doi.org/10.1093/mnras/stae1325},
	doi = {10.1093/mnras/stae1325},
	number = {3},
	urldate = {2026-02-08},
	journal = {Monthly Notices of the Royal Astronomical Society},
	author = {Neyer, Meredith and Smith, Aaron and Kannan, Rahul and Vogelsberger, Mark and Garaldi, Enrico and Galárraga-Espinosa, Daniela and Borrow, Josh and Hernquist, Lars and Pakmor, Rüdiger and Springel, Volker},
	month = jul,
	year = {2024},
	pages = {2943--2957},
}

@misc{jecmen_glimpse_2026,
	title = {A {GLIMPSE} into the {UV} {Continuum} {Slopes} of the {Faintest} {Galaxies} in the {Epoch} of {Reionization}},
	url = {https://arxiv.org/abs/2601.19995v1},
	language = {en},
	urldate = {2026-02-18},
	journal = {arXiv.org},
	author = {Jecmen, Michelle C. and Chisholm, John and Atek, Hakim and Kokorev, Vasily and Endsley, Ryan and Chemerynska, Iryna and Furtak, Lukas J. and Pan, Richard and Fujimoto, Seiji and Naidu, Rohan P. and Muñoz, Julian B. and Adamo, Angela and Asada, Yoshihisa and Basu, Arghyadeep and Berg, Danielle A. and Blaizot, Jeremy and Dessauges-Zavadsky, Miroslava and Giovinazzo, Emma and Hsiao, Tiger Yu-Yang and Katz, Harley and Korber, Damien and McKinney, Jed and McQuinn, Kristen B. W. and Oesch, Pascal A. and Schaerer, Daniel},
	month = jan,
	year = {2026},
}

@misc{bera_towards_2025,
	title = {Towards {Reconciling} {Reionization} with {JWST}: {The} {Role} of {Bright} {Galaxies} and {Strong} {Feedback}},
	shorttitle = {Towards {Reconciling} {Reionization} with {JWST}},
	url = {https://arxiv.org/abs/2511.19600v1},
	language = {en},
	urldate = {2026-02-18},
	journal = {arXiv.org},
	author = {Bera, Ankita and Hassan, Sultan and Feldmann, Robert and Davé, Romeel and Finlator, Kristian},
	month = nov,
	year = {2025},
}

@article{robertson_identification_2023,
	title = {Identification and properties of intense star-forming galaxies at redshifts z > 10},
	volume = {7},
	issn = {2397-3366},
	url = {https://ui.adsabs.harvard.edu/abs/2023NatAs...7..611R},
	doi = {10.1038/s41550-023-01921-1},
	urldate = {2026-02-25},
	journal = {Nature Astronomy},
	author = {Robertson, B. E. and Tacchella, S. and Johnson, B. D. and Hainline, K. and Whitler, L. and Eisenstein, D. J. and Endsley, R. and Rieke, M. and Stark, D. P. and Alberts, S. and Dressler, A. and Egami, E. and Hausen, R. and Rieke, G. and Shivaei, I. and Williams, C. C. and Willmer, C. N. A. and Arribas, S. and Bonaventura, N. and Bunker, A. and Cameron, A. J. and Carniani, S. and Charlot, S. and Chevallard, J. and Curti, M. and Curtis-Lake, E. and D'Eugenio, F. and Jakobsen, P. and Looser, T. J. and Lützgendorf, N. and Maiolino, R. and Maseda, M. V. and Rawle, T. and Rix, H.-W. and Smit, R. and Übler, H. and Willott, C. and Witstok, J. and Baum, S. and Bhatawdekar, R. and Boyett, K. and Chen, Z. and de Graaff, A. and Florian, M. and Helton, J. M. and Hviding, R. E. and Ji, Z. and Kumari, N. and Lyu, J. and Nelson, E. and Sandles, L. and Saxena, A. and Suess, K. A. and Sun, F. and Topping, M. and Wallace, I. E. B.},
	month = may,
	year = {2023},
	note = {ADS Bibcode: 2023NatAs...7..611R},
	pages = {611--621},
}

@article{jaskot_ionizing_2025,
	title = {Ionizing {Radiation} {Escape} from {Low}-{Redshift} {Galaxies} and {Its} {Connection} to {Cosmic} {Reionization}},
	volume = {63},
	issn = {0066-4146},
	url = {https://ui.adsabs.harvard.edu/abs/2025ARA&A..63...45J},
	doi = {10.1146/annurev-astro-111324-074935},
	urldate = {2026-02-25},
	journal = {Annual Review of Astronomy and Astrophysics},
	author = {Jaskot, Anne E.},
	month = aug,
	year = {2025},
	note = {ADS Bibcode: 2025ARA\&A..63...45J},
	pages = {45--82},
}

@article{robertson_galaxy_2022,
	title = {Galaxy {Formation} and {Reionization}: {Key} {Unknowns} and {Expected} {Breakthroughs} by the {James} {Webb} {Space} {Telescope}},
	volume = {60},
	issn = {0066-4146},
	shorttitle = {Galaxy {Formation} and {Reionization}},
	url = {https://ui.adsabs.harvard.edu/abs/2022ARA&A..60..121R},
	doi = {10.1146/annurev-astro-120221-044656},
	urldate = {2026-02-25},
	journal = {Annual Review of Astronomy and Astrophysics},
	author = {Robertson, Brant E.},
	month = aug,
	year = {2022},
	note = {ADS Bibcode: 2022ARA\&A..60..121R},
	pages = {121--158},
}

@article{tacchella_redshift-independent_2018,
	title = {A {Redshift}-independent {Efficiency} {Model}: {Star} {Formation} and {Stellar} {Masses} in {Dark} {Matter} {Halos} at z ≳ 4},
	volume = {868},
	issn = {0004-637X},
	shorttitle = {A {Redshift}-independent {Efficiency} {Model}},
	url = {https://doi.org/10.3847/1538-4357/aae8e0},
	doi = {10.3847/1538-4357/aae8e0},
	language = {en},
	number = {2},
	urldate = {2026-03-06},
	journal = {The Astrophysical Journal},
	publisher = {The American Astronomical Society},
	author = {Tacchella, Sandro and Bose, Sownak and Conroy, Charlie and Eisenstein, Daniel J. and Johnson, Benjamin D.},
	month = nov,
	year = {2018},
	pages = {92},
}

@article{zier_adapting_2024,
	title = {Adapting {AREPO}-{RT} for exascale computing: {GPU} acceleration and efficient communication},
	volume = {533},
	issn = {0035-8711},
	shorttitle = {Adapting {AREPO}-{RT} for exascale computing},
	url = {https://ui.adsabs.harvard.edu/abs/2024MNRAS.533..268Z},
	doi = {10.1093/mnras/stae1837},
	urldate = {2026-04-07},
	journal = {Monthly Notices of the Royal Astronomical Society},
	publisher = {OUP},
	author = {Zier, Oliver and Kannan, Rahul and Smith, Aaron and Vogelsberger, Mark and Verbeek, Erkin},
	month = sep,
	year = {2024},
	note = {ADS Bibcode: 2024MNRAS.533..268Z},
	pages = {268--286},
}




\appendix

\section{Thesan-Zoom Potential Indicators} \label{sec:appendix_a}

\begin{figure*}
	\includegraphics[width=\textwidth]{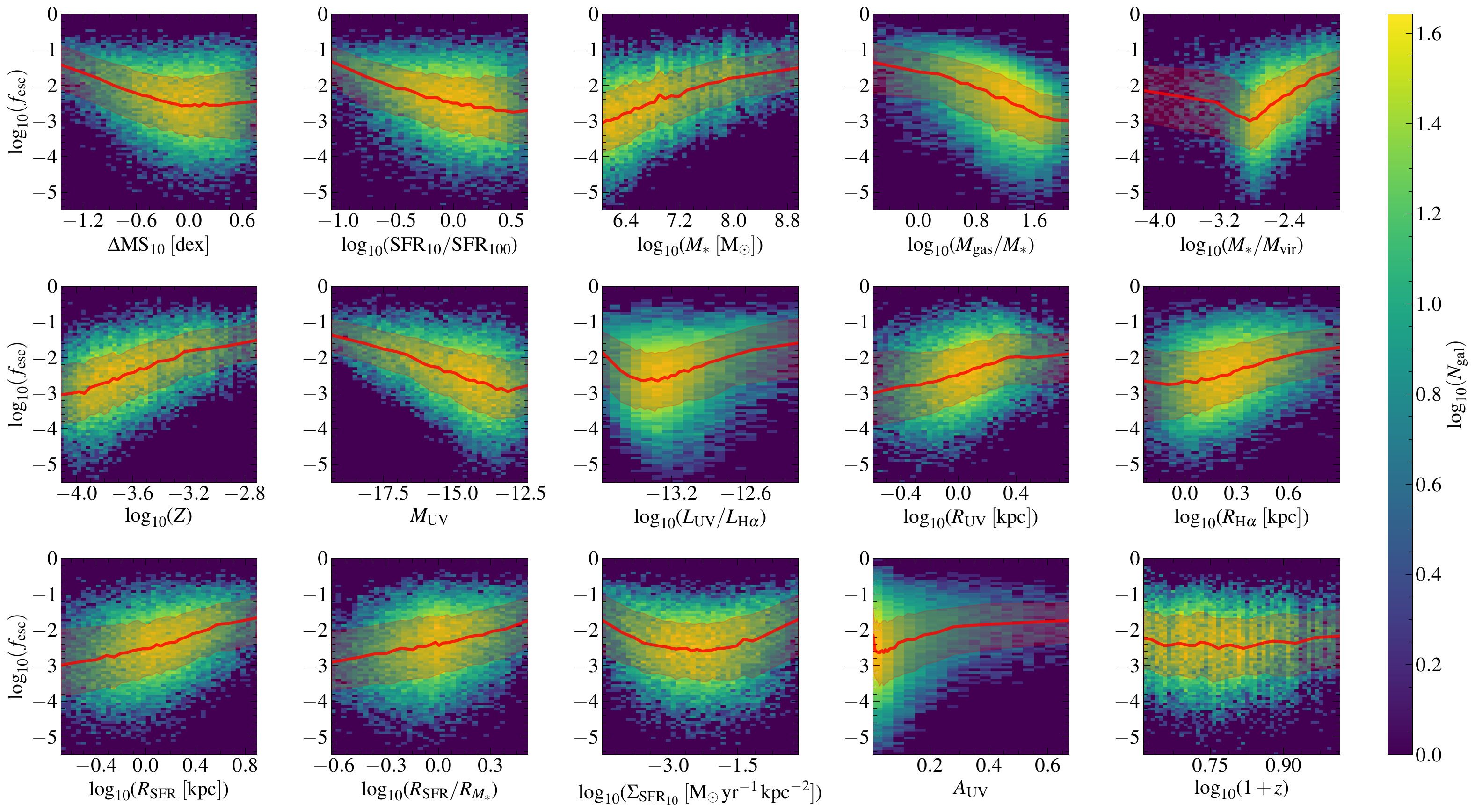}
    \caption{Ionising photon escape fraction ($f_\mathrm{esc}$) versus the 14 identified diagnostics listed in Tab.~\ref{tab:indicators}, as well as $1+z$, for the 35\,512 galaxies selected from the \textsc{thesan-zoom} simulations across $3 < z < 16$. The colour scale indicates the number of galaxies per bin. The red lines trace the median relations, with red shaded regions denoting the 16$^\text{th}$--84$^\text{th}$ percentiles. All the observables exhibit a clear correlation with $f_\mathrm{esc}$, demonstrating that they are all potentially good indicators of $f_\mathrm{esc}$.}
    \label{fig:all_vs_f_esc}
\end{figure*}

\begin{figure*}
	\includegraphics[width=\textwidth]{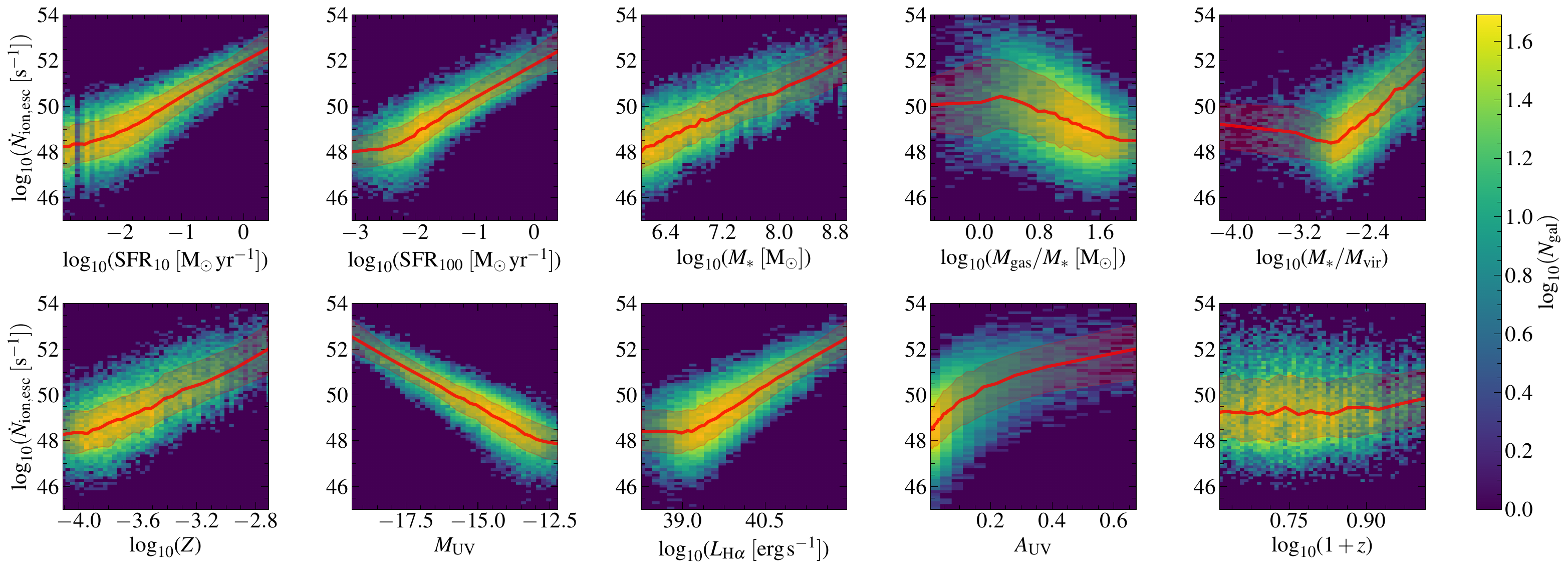}
    \caption{Ionising photon escape rate ($\dot{N}_\mathrm{ion,esc}$) versus the 9 identified diagnostics listed in Tab.~\ref{tab:indicators}, as well as $1+z$, for the 35\,512 galaxies selected from the \textsc{thesan-zoom} simulations simulations across $3 < z < 16$. The colour scale indicates the number of galaxies per bin. The red lines trace the median relations, with red shaded regions denoting the 16$^\text{th}$--84$^\text{th}$ percentiles. All the observables exhibit a clear correlation with $\dot{N}_\mathrm{ion,esc}$, demonstrating that they are all potentially good indicators of $\dot{N}_\mathrm{ion,esc}$.}
    \label{fig:all_vs_n_esc}
\end{figure*}

\begin{figure*}
	\includegraphics[width=\textwidth]{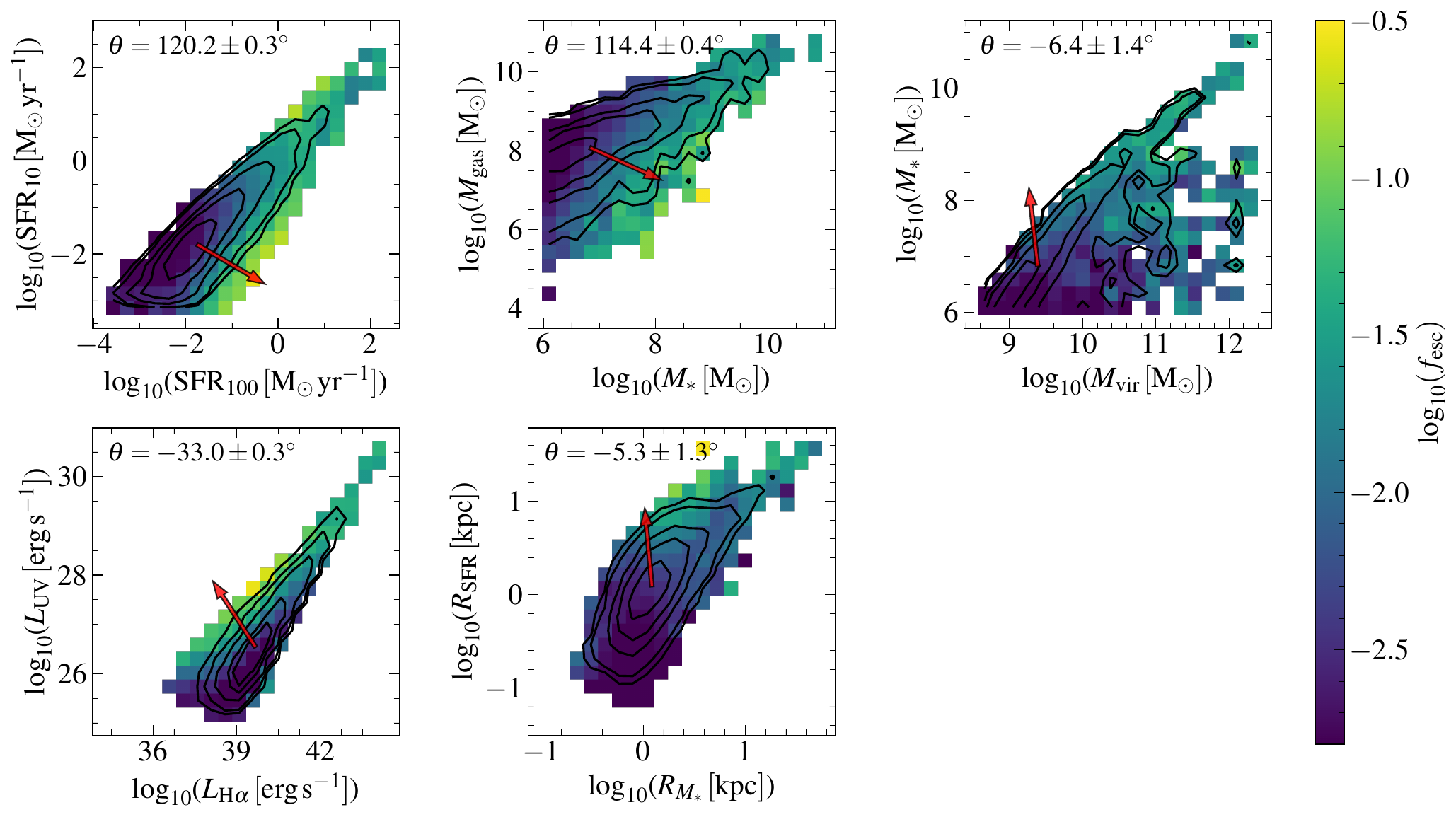}
    \caption{The pairs of observables that form the ionizing photon escape fraction ($f_\mathrm{esc}$) ratio indicators given in Tab.~\ref{tab:indicators} for the \textsc{thesan-zoom} galaxies are plotted against each other, colour-coded by the median $f_\mathrm{esc}$ in each bin. Only bins containing at least five galaxies are plotted. The contours indicate the density of galaxies in the plots, with the innermost contours encompassing 34.0\% of the sample and the outermost contours encompassing 97.5\%. The red arrows indicate the direction of largest positive colour gradient, with their angles, $\theta$, defined clockwise from the positive y-direction. All $\theta$ lie in the regions $(-90,0) \cup (90,180)$, implying that the pairs of observables have correlations of opposite polarity with $f_\mathrm{esc}$. Consequently, $f_\mathrm{esc}$ increases as one observable increases and the other decreases, so their ratios are likely to be good indicators of $f_\mathrm{esc}$.}
    \label{fig:histograms_all_double}
\end{figure*}

We present the complete graphical analysis described in Section \ref{sec:prediction} to identify potential diagnostics of $f_\mathrm{esc}$ and $\dot{N}_\mathrm{ion,esc}$ for all the variables in Tab.~\ref{tab:indicators}, justifying their inclusion  in the list of indicators. Following Fig.~\ref{fig:uv_mag_vs_targets}, the single-variable correlation plots that establish a relationship between each variable and $f_\mathrm{esc}$ or $\dot{N}_\mathrm{ion,esc}$ are shown in Fig.~\ref{fig:all_vs_f_esc} and \ref{fig:all_vs_n_esc} respectively. The ratio analysis, exemplified earlier in Fig.~\ref{fig:histograms_star_mass_vs_gas_mass}, is presented in full in Fig.~\ref{fig:histograms_all_double}. Correlations for all the physically motivated pairs of parameters presented in this figure exhibit opposite polarities, with $f_\mathrm{esc}$ increasing as one parameter increases and the other decreases, explaining why all observable ratios in Tab.~\ref{tab:indicators} are effective diagnostics of $f_\mathrm{esc}$.

\begin{figure*}
	\includegraphics[width=\textwidth]{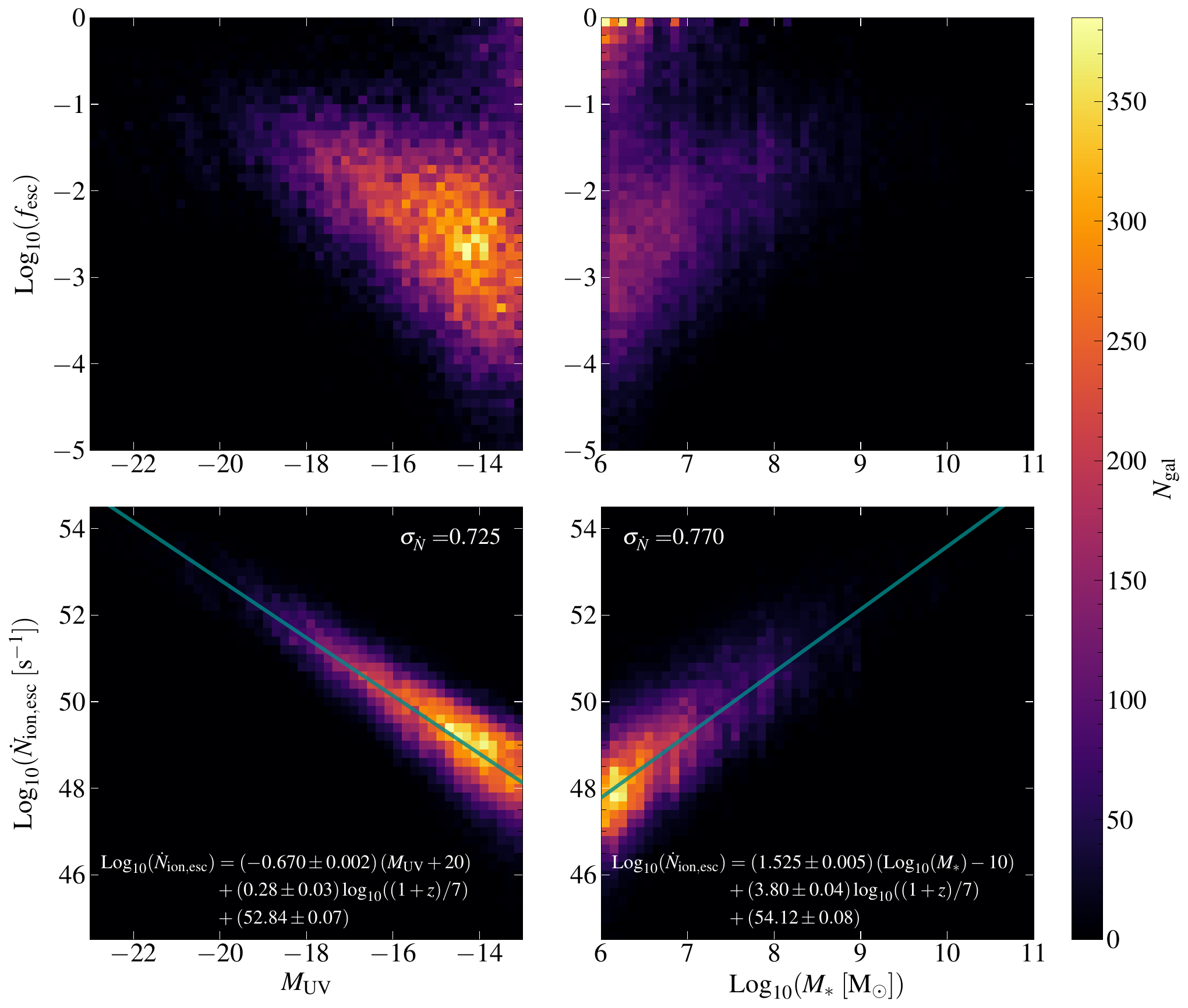}
    \caption{\textit{Top left:} $f_\mathrm{esc}$ as a function of $M_\mathrm{UV}$ for galaxies from the \textsc{thesan-zoom} simulations; \textit{top right:} $f_\mathrm{esc}$ as a function of $M_*$; \textit{bottom left:} $\dot{N}_\mathrm{ion,esc}$ as a function of $M_\mathrm{UV}$; \textit{bottom right:} $\dot{N}_\mathrm{ion,esc}$ as a function of $M_*$. The colour scale indicates the number of galaxies per bin. Like with the predictions of the JWST observational sample in Fig.~\ref{fig:observational_fits}, $\log_{10}(\dot{N}_\mathrm{ion,esc})$ exhibits approximately linear behaviour with both $M_\mathrm{UV}$ and $\log_{10}(M_*)$. Therefore, in both lower panels, we present the orthogonal distance regression linear relations for $\log_{10}(\dot{N}_\mathrm{ion,esc})$, in which we also include a redshift dependence, and plot them in cyan. The scatter about these relations, $\sigma_{\dot{N}}$, computed as half the $16^\mathrm{th} - 84^\mathrm{th}$ percentile range of the residuals, is also shown in both panels. This scatter is slightly smaller for the $M_\mathrm{UV}$ relation, suggesting that $M_\mathrm{UV}$ may be the more important diagnostic of $\dot{N}_\mathrm{ion,esc}$, as we see in Fig.~\ref{fig:importances}.}
    \label{fig:thesan_fits_histogram}
\end{figure*}

The values of $f_\mathrm{esc}$ and $\dot{N}_\mathrm{ion,esc}$ for the \textsc{thesan-zoom} simulations are plotted in Fig.~\ref{fig:thesan_fits_histogram} against $M_\mathrm{UV}$ and $M_*$, analogously with Fig.~\ref{fig:observational_fits} and the observational JWST catalogue. As with JWST, we use orthogonal distance regression to fit linear relationships between $\log_{10}(\dot{N}_\mathrm{ion,esc})$ and $M_\mathrm{UV}$ as well as $\log_{10}(M_*)$, again including a redshift dependence to account for the cosmic evolution of $\dot{N}_\mathrm{ion,esc}$; with the relations presented in their respective panels. Half the $16^\mathrm{th}-84^\mathrm{th}$ percentile range of the residuals of these relations, $\sigma_{\dot{N}}$, is also shown in both panels. Eq~\eqref{thesan_linear_fit} states the $\dot{N}_\mathrm{ion,esc} - M_\mathrm{UV}$ relation found here in the main text. This relation, together with its intrinsic scatter of $\sigma_{\dot{N}} = 0.725$ dex, is used to derive the \textsc{thesan-zoom}-based ionizing emissivity, $\dot{n}_\mathrm{ion}$, as explained in Section.~\ref{sec:ionising_emissivity}.

\section{Optimising The Random Forest} \label{sec:appendix_b}

Here, we explain our processes for optimising the hyperparameters of our random forest (RF) models (see Section~\ref{sec:random_forests}). For each model, we train RFs to target $f_\mathrm{esc}$ using all 14 of the relevant indicators given in Tab.~\ref{tab:indicators}. The Mean Absolute Error (MAE; in dex) reported for each instance of a hyperparameter in this analysis is the mean MAE calculated over 10 separate regressors with the same setup. 

\begin{figure}
	\includegraphics[width=\columnwidth]{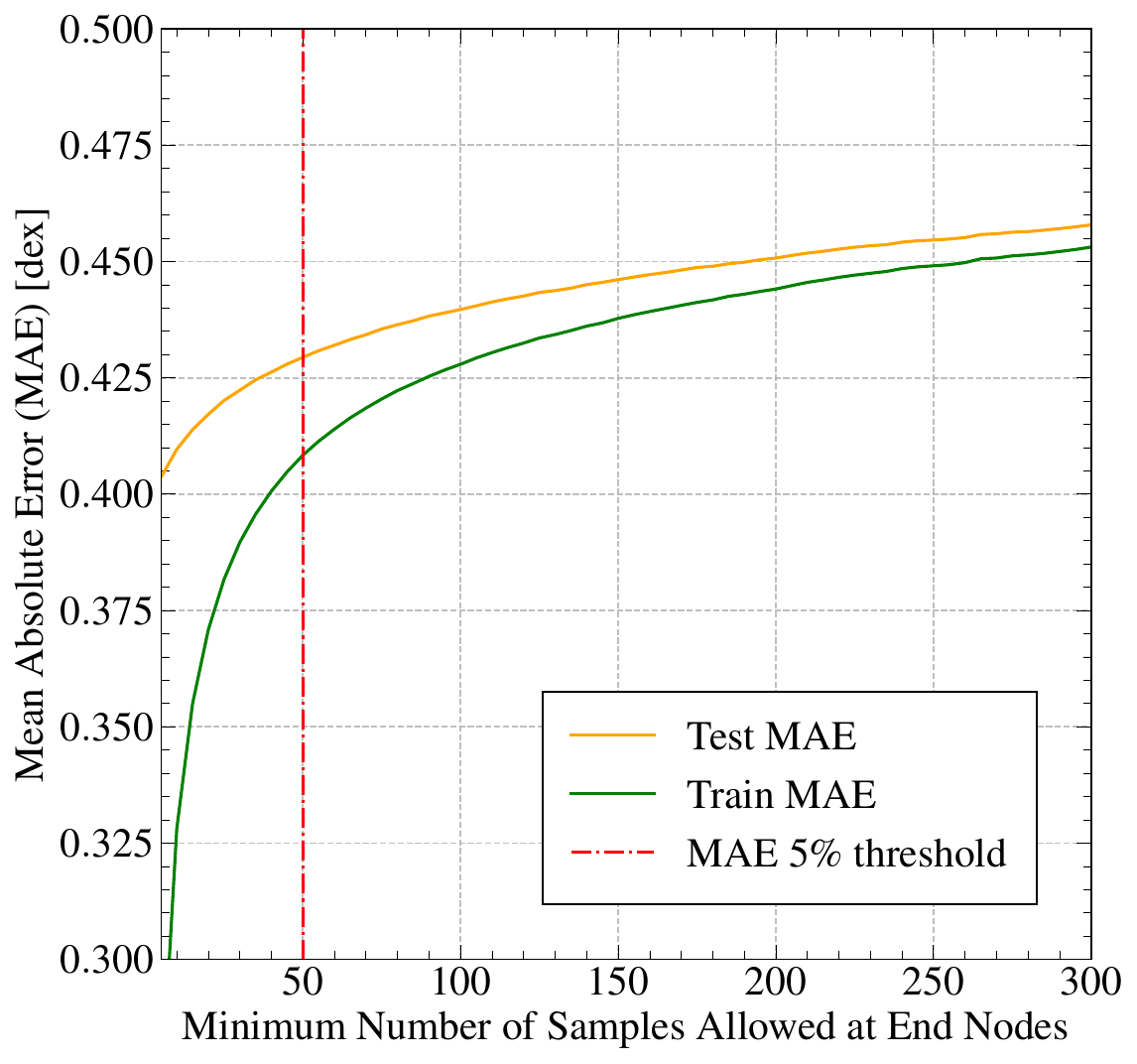}
    \caption{Mean Absolute Error (MAE; in dex) of $f_\mathrm{esc}$ prediction as a function of the minimum number of samples allowed on the end node of a random forest decision tree. The orange line shows the MAE of prediction on the test sample, while the green line shows the MAE of prediction on the training sample. Increasing this hyperparameter reduces overfitting, as evidenced by the decreasing difference in MAE between the two curves. However, larger values also reduce prediction accuracy. We therefore adopt the smallest value for which the difference in MAE between the test and training samples is less than 5\%, which occurs at 50, as indicated by the dashed red line.}
    \label{fig:min_samples_leaf_optimisation}
\end{figure}

The first hyperparameter we address is the minimum number of samples on the final leaf, which dictates the minimum number of samples allowed to exist at an end node of a decision tree. This controls the size a decision tree in the RF can grow to. If too low, the regressor tends to overfit the training sample, however increasing it also decreases the accuracy of the sample (increasing MAE and MSE). We choose the value of the minimum number of samples on the final leaf to be 50, which is the smallest value that yields a difference in MAE between prediction on the testing and training samples of less than 5\% (see Fig.~\ref{fig:min_samples_leaf_optimisation}). 

\begin{figure}
	\includegraphics[width=\columnwidth]{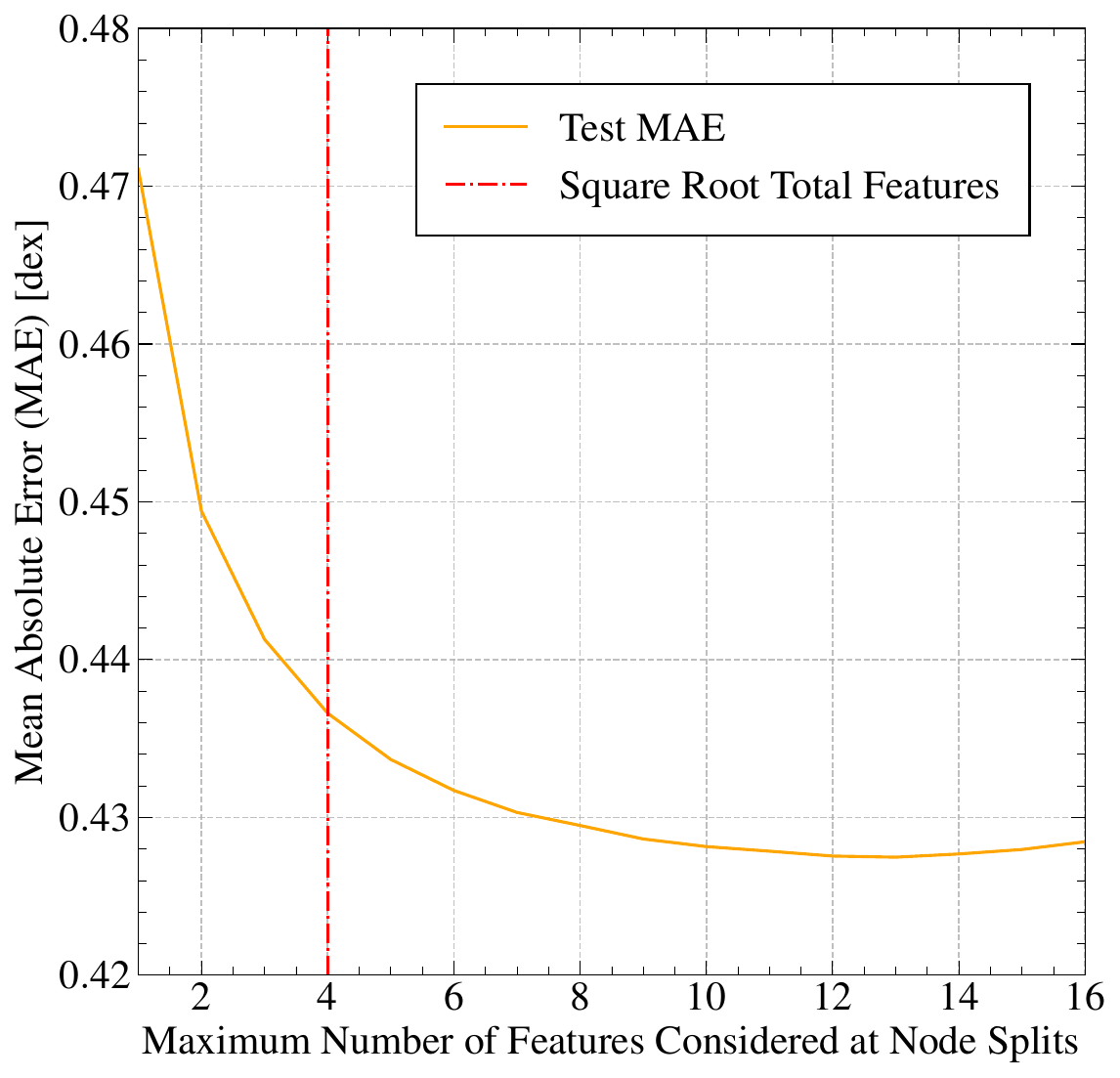}
    \caption{Mean Absolute Error (MAE; in dex) of $f_\mathrm{esc}$ prediction for the test sample as a function of the maximum number of features considered at each decision tree node split in the random forest. We set this hyperparameter equal to the square root of the total number of features (four for the 16 $f_\mathrm{esc}$ random forest features of this work), as indicated by the red dashed line.}
    \label{fig:max_features_optimisation}
\end{figure}

We also optimise the number of features considered when looking for the best split at a node. If set too high, more parameters are evaluated at each split and the model tends to select the most important parameter repeatedly, limiting exploration of other parameters. This removes much of the random nature of the RF and with it the benefits mentioned in Section \ref{sec:random_forests}. We set the maximum features equal to the commonly used square root of the total number of features, for the 16 training features for $f_\mathrm{esc}$ this is a maximum of 4 in this work. Lowering beyond this causes regressor MAE to increase rapidly (see Fig.~\ref{fig:max_features_optimisation}). 

\begin{figure}
	\includegraphics[width=\columnwidth]{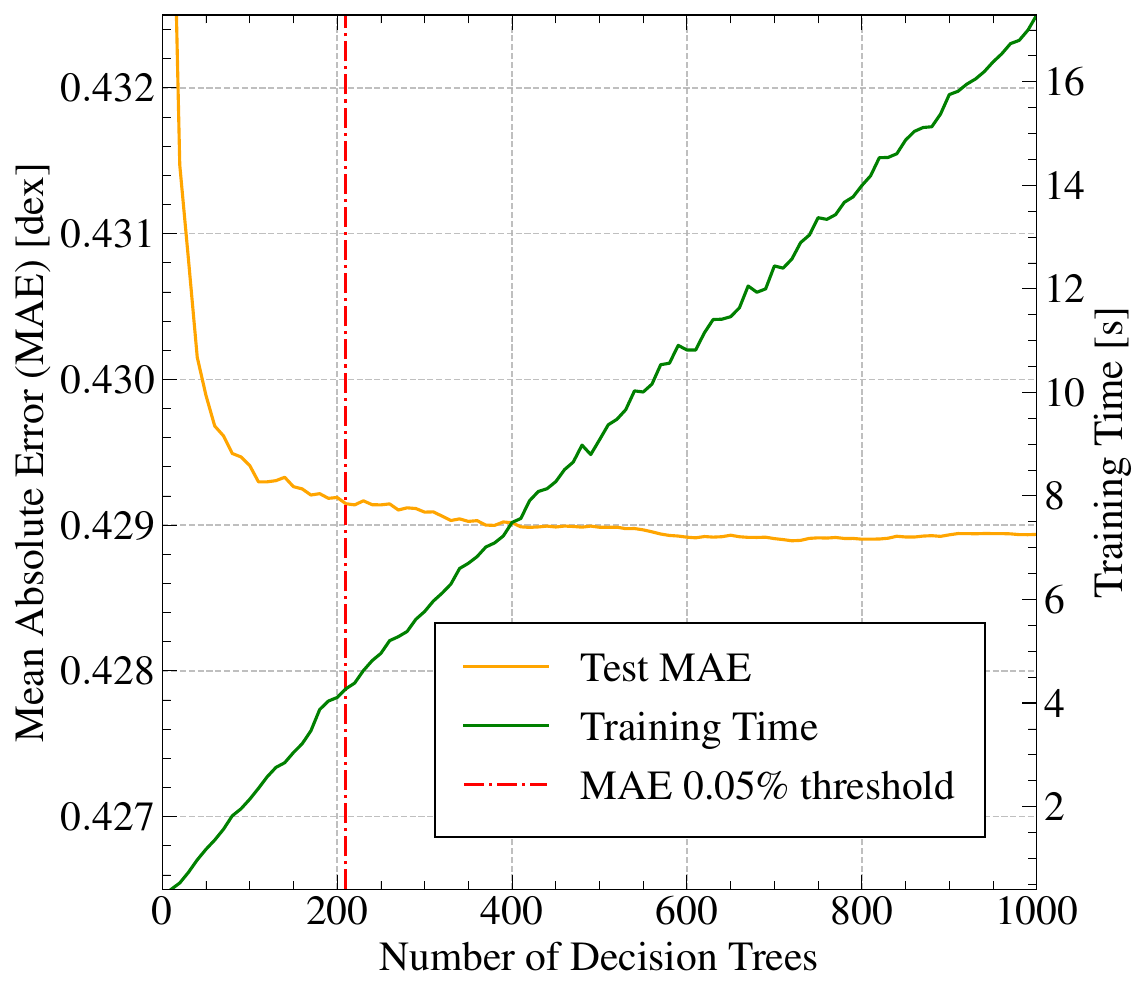}
    \caption{The Orange line shows the Mean Absolute Error (MAE; in dex) of $f_\mathrm{esc}$ prediction for the test sample as a function of the number of decision trees in the random forest. The green line illustrates the linear relationship between model training time and the number of trees. We set this hyperparameter equal to 210, the smallest number of estimators for which the MAE on prediction differs by less than 0.05\% from that of a model trained with 1000 trees. This threshold is indicated by the red dashed line.}
    \label{fig:n_estimators_optimisation}
\end{figure}

The final hyperparameter we address is the number of estimators, which is the number of decision trees in the random forest. A larger number of predictors unsurprisingly decreases errors, although the increase in accuracy falls off exponentially. Unfortunately, training time complexity increases linearly with the number of estimators, expressed as $O(n)$. We therefore conclude that 210 decision trees is an appropriate choice for our models; this is the lowest number of estimators that results in an MAE difference of less than 0.05\% from a model trained with 1000 decision trees (see Fig.~\ref{fig:n_estimators_optimisation}).

\section{Validation of The Ionizing Emissivity Integrations} \label{sec:appendix_c}

\begin{figure*}
    \centering
    \includegraphics[width=\textwidth]{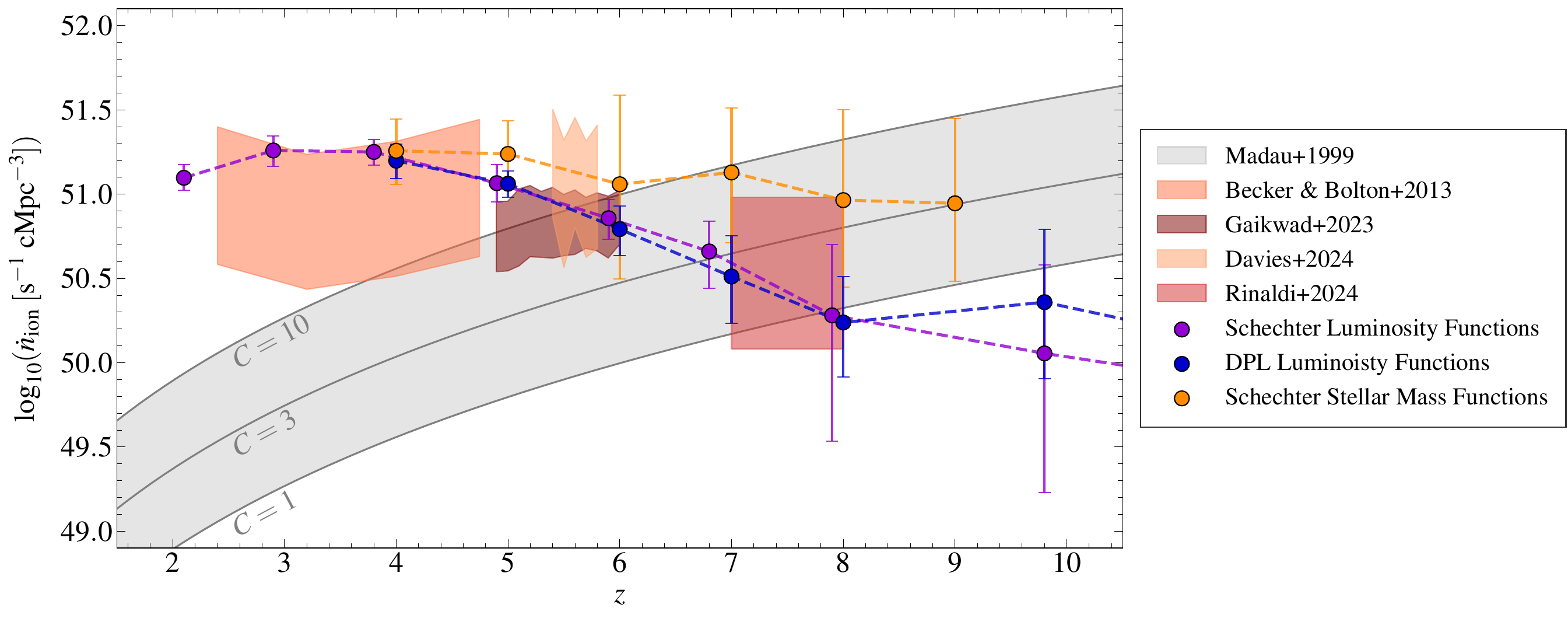}
    \caption{Similar to Fig.~\ref{fig:reionisation_timeline}: comoving cosmic ionizing photon emissivity, $\dot{n}_{\mathrm{ion}}$, as a function of redshift, derived from the \textsc{thesan-zoom} $\dot{N}_\mathrm{ion,esc} - M_\mathrm{UV}$ and $\dot{N}_\mathrm{ion,esc} - M_*$ relations presented in Fig.~\ref{fig:thesan_fits_histogram}. The violet emissivities have been integrated using the Schechter parametrization of the UV luminosity functions (see Eq.~\eqref{schechter_magnitude}) from \citet{bouwens_new_2021} for $z \in [1.5, 8.5]$ and \citet{whitler_z_2025} for $z \in [8.5, 16]$. The blue emissivities have instead been obtained using the Double Power Law (DPL) parametrization of the UV luminosity functions (see Eq.~\eqref{dpl_magnitude}) from \citet{bowler_lack_2020} for $z \in [3.5, 8.5]$ and \citet{whitler_z_2025} for $z \in [8.5, 16]$. Finally, the orange emissivities have been derived using the Schechter parametrization of the stellar mass functions (see Eq.~\eqref{schechter_mass}) from \citet{weibel_galaxy_2024} for $z \in [3.5, 9.5]$. As in Fig.~\ref{fig:reionisation_timeline}, the grey bands indicate the critical ionizing emissivity \citep{madau_radiative_1999} for possible clumping factors $C \in [1, 10]$, and we include the same observational constraints \citep{becker_new_2013, gaikwad_measuring_2023, davies_constraints_2024, rinaldi_midis_2024}.}
    \label{fig:alternative_timeline}
\end{figure*}

In calculating the ionizing emissivities, $\dot{n}_\mathrm{ion}(z)$, of Fig.~\ref{fig:reionisation_timeline} with Eq.~\eqref{integral_magnitude}, we employ the Schechter form of the UV luminosity function (UVLF) \citep{schechter_analytic_1976}:
\begin{align} \label{schechter_magnitude}
    \Phi(M_\mathrm{UV}, z) = 0.4 \, \mathrm{ln(10)} \, \phi^*
    & \cdot 10^{0.4 \, (\alpha + 1 ) \, \Delta M_\mathrm{UV}} \notag  \\
    & \cdot \mathrm{exp} \left( -10^{0.4 \, \Delta M_\mathrm{UV}} \right),
\end{align}
where we have defined $\Delta M_\mathrm{UV} \equiv M_\mathrm{UV}^* - M_\mathrm{UV}$. In this work, the fit parameters $\phi^* \equiv \phi^*(z)$, $M^*_\mathrm{UV} \equiv M^*_\mathrm{UV}(z)$, and $\alpha \equiv \alpha(z)$ from \citet{bouwens_new_2021} are used for $z \in [1.5, 8.5]$ and those from \citet{whitler_z_2025} are used for $z \in [8.5, 16]$. Some sources have found the double power-law (DPL) form of the UVLF to better fit observations at high redshifts of $z \gtrsim 8$ \citep{bowler_lack_2020}:
\begin{equation} \label{dpl_magnitude}
    \Phi(M_\mathrm{UV}, z) = \frac{ 0.4 \, \mathrm{ln(10)} \, \phi^*}{10^{-0.4 \, (\alpha + 1) \, \Delta M_\mathrm{UV}} + 10^{-0.4 \, (\beta + 1) \, \Delta M_\mathrm{UV}}}.
\end{equation}
In Fig.~\ref{fig:alternative_timeline}, we compare the \textsc{thesan-zoom}-based $\dot{n}_\mathrm{ion}(z)$ curve obtained in Section~\ref{sec:ionising_emissivity}, with an alternative \textsc{thesan-zoom}-based curve derived employing the DPL form of the UVLF. The DPL fit parameters $\phi^* \equiv \phi^*(z)$, $M^*_\mathrm{UV} \equiv M^*_\mathrm{UV}(z)$, $\alpha \equiv \alpha(z)$, and $\beta \equiv \beta(z)$ are taken from \citep{bowler_lack_2020} for $z \in [3.5, 8.5]$ and from \citet{whitler_z_2025} for $z \in [8.5, 16]$. The emissivities derived from the Schechter and DPL UV luminosity functions are consistent for $z \leq 8$, the redshift range over which we find the bulk of the ionizing photon budget to be produced; variations in $\dot{n}_\mathrm{ion}(z)$ at earlier times therefore have little effect on the completion of reionization.. However, the DPL curve exhibits an unphysical discontinuous increase from $z \simeq 8$ to $z \simeq 10$, causing it to diverge from the Schechter-based result. We therefore conclude that the emissivity derived from the Schechter fit is more suitable for interpolation and for the analysis of the cosmic hydrogen ionization fraction $Q_\mathrm{HII}$ presented in Section~\ref{sec:ionisation_fraction}.

Eq.~\eqref{integral_magnitude} can be reparametrized for $\dot{N}_\mathrm{ion,esc} - M_*$ relations by adopting galaxy number density functions expressed in terms of stellar mass, $ \Phi(M_\mathrm{*}, z)$, commonly referred to as stellar mass functions (SMFs):
\begin{equation} \label{integral_mass}
    \dot{n}_{\mathrm{ion}}(z) = \int_{M_\mathrm{*,min}}^{M_\mathrm{*,max}} \Phi(M_\mathrm{*}, z) \, \left\langle \dot{N}_\mathrm{ion,esc} \,\middle|\, M_\mathrm{*}, z \right\rangle \, \mathrm{d}M_\mathrm{*}.
\end{equation}
To assess the viability of stellar mass functions, we again compare the \textsc{thesan-zoom}-based $\dot{n}_\mathrm{ion}(z)$ of Fig.~\ref{fig:reionisation_timeline}, with another alternative \textsc{thesan-zoom}-based curve in Fig.~\ref{fig:alternative_timeline}. The latter being derived from the \textsc{thesan-zoom} $\dot{N}_\mathrm{ion,esc} - M_*$ relation presented in the bottom-right panel of Fig.~\ref{fig:thesan_fits_histogram}, employing the Schechter form of the stellar mass function \citep{schechter_analytic_1976}:
\begin{equation} \label{schechter_mass}
    \Phi(y_*, z) = \mathrm{ln}(10) \, \phi' \, 10^{(y_*-\mu) \, (\alpha + 1)} \, \mathrm{exp} \left(-10^{(y_* - \mu)} \right) \, ,
\end{equation}
Where we have defined $y_* \equiv \log_{10}(M_*)$.  We employ the fit parameters $\phi' \equiv \phi'(z)$, $\mu \equiv \log_{10}(M_*'(z))$, and $\alpha \equiv \alpha(z)$ from \citet{weibel_galaxy_2024} for $z \in [3.5, 9.5]$. At high redshifts, the stellar mass derived $\dot{n}_\mathrm{ion}$ values are significantly larger than those derived from UV magnitudes, exhibit large uncertainties, appear less continuous, and do not decline sufficiently to avoid an overabundance of ionising photons, which results in unphysical $Q_\mathrm{HII}$ evolution. This behaviour likely reflects the well-known limitations of stellar mass functions at high redshifts \citep{madau_cosmic_2014}. This approach is therefore not well suited to the reionization analysis presented in Section~\ref{sec:implications}. We therefore adopt Eq.~\eqref{integral_magnitude} for the main results of this study and employ a Schechter parameterisation of the UV luminosity function.

\begin{figure}
    \centering
    \includegraphics[width=\columnwidth]{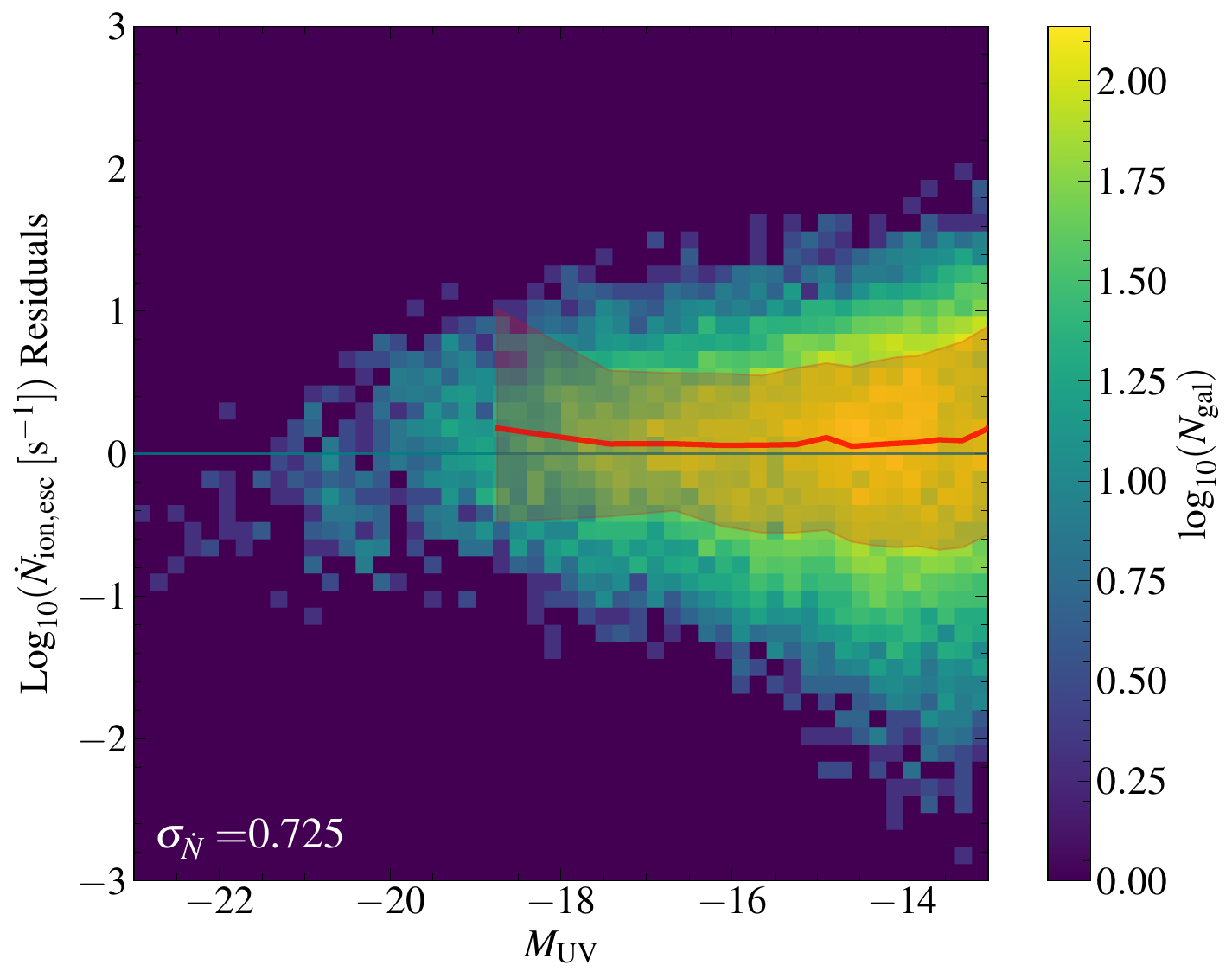}
    \caption{Residuals of ionizing photon escape rate ($\dot{N}_\mathrm{ion,esc}$) as a function of rest-frame UV absolute magnitude ($M_\mathrm{UV}$), relative to the redshift dependent $\dot{N}_\mathrm{ion,esc} - M_\mathrm{UV}$ relation given by Eq.~\eqref{thesan_linear_fit}  for \textsc{thesan-zoom} galaxies. The colour scale indicates the number of galaxies per bin. The red shaded region denotes the $16^\mathrm{th} - 84^\mathrm{th}$ percentile range. We see the width of this range varies only weakly with $M_\mathrm{UV}$ and shows no clear systematic trend, implying that the assumption of constant scatter in the $\dot{N}_\mathrm{ion,esc} - M_\mathrm{UV}$ relation is well motivated. We take this constant scatter to be half of the $16^\mathrm{th} - 84^\mathrm{th}$ percentile range over all the residuals, and report its value in the lower left of this figure.}
    \label{fig:thesan_residuals}
\end{figure}

As described in Section~\ref{sec:ionising_emissivity}, to account for the scatter of $\sigma_{\dot{N}} = 0.725$ dex observed in the $\dot{N}_\mathrm{ion,esc} - M_\mathrm{UV}$ relation in \textsc{thesan-zoom} when computing the ionizing emissivity, we multiply the integrand of Eq.~\eqref{integral_magnitude} by the lognormal correction factor specified in Eq.~\eqref{correction}. In doing this we have implicitly assumed that the scatter in this relation is constant across $M_\mathrm{UV}$. Fig.~\ref{fig:thesan_residuals} shows the residuals of the $\dot{N}_\mathrm{ion,esc} - M_\mathrm{UV}$ relation for the \textsc{thesan-zoom} galaxies as a function of $M_\mathrm{UV}$. In this figure, we see that the width of the $1 \sigma$ percentile range varies only weakly with $M_\mathrm{UV}$ and exhibits no clear systematic trend that would justify parametrizing $\sigma_{\dot{N}}$ as a function of $M_\mathrm{UV}$. This supports our assumption of a constant scatter in the $\dot{N}_\mathrm{ion,esc} - M_\mathrm{UV}$ relation.

\bsp	
\label{lastpage}
\end{document}